\documentclass[manuscript,nonacm]{acmart}
\AtBeginDocument{%
  }

\usepackage{longtable}
\usepackage{booktabs}

\usepackage{tabularx}
\usepackage{multirow}
\usepackage{algorithm}  
\usepackage{algpseudocode}
\usepackage{graphicx}
\usepackage{placeins}
\usepackage{pifont}
  \newcommand{\cmark}{\ding{51}}
  \newcommand{\xmark}{\ding{55}}
  \newcommand{\partialmark}{\ensuremath{\blacktriangle}}%

\usepackage{url}
\usepackage{tcolorbox}

\newcommand{\doubletoprule}{%
  \specialrule{\heavyrulewidth}{0pt}{0.7ex}%
  \specialrule{\lightrulewidth}{0pt}{0pt}}

\newcommand{\doublebottomrule}{%
  \specialrule{\lightrulewidth}{0pt}{0.7ex}%
  \specialrule{\heavyrulewidth}{0pt}{0pt}}

\makeatletter
  \def\@acmdefinitionheadfont{\bfseries}
  \makeatother

\usepackage{xcolor}

  \newcommand{\metric}[2]{%
    #1\,{\scriptsize(%
      \ifdim #2pt > 0pt
        \textcolor{red}{#2}%
      \else
        \ifdim #2pt < 0pt
          \textcolor{green!60!black}{#2}%
        \else
          #2%
        \fi
      \fi
    )}%
  }

\begin{document}

\title{Repairing Broken Access Control in Web Applications with LLMs: An Empirical Benchmark Study}

\author{Mingrui Liu}
\email{liumingrui2024@iie.an.cn}
\affiliation{%
  \institution{Institute of Information Engineering, Chinese Academy of Sciences}
  \city{Beijing}
  \country{China}
}

\author{Liang He}
\correspondingauthor
\email{heliang@iscas.ac.cn}
\affiliation{%
  \institution{Institute of Software, Chinese Academy of Sciences}
  \city{Beijing}
  \country{China}
}

\author{Xinchu Wang}
\email{wangxinchu@iie.an.cn}
\affiliation{%
  \institution{Institute of Information Engineering, Chinese Academy of Sciences}
  \city{Beijing}
  \country{China}
}

\author{Yeting Li}
\email{liyeting@iie.an.cn}
\affiliation{%
  \institution{Institute of Information Engineering, Chinese Academy of Sciences}
  \city{Beijing}
  \country{China}
}

\author{Dongxu Han}
\email{handongxu@iie.an.cn}
\affiliation{%
  \institution{Institute of Information Engineering, Chinese Academy of Sciences}
  \city{Beijing}
  \country{China}
}

\author{Yunpeng Li}
\correspondingauthor
\email{liyunpeng@iie.an.cn}
\affiliation{%
  \institution{Institute of Information Engineering, Chinese Academy of Sciences}
  \city{Beijing}
  \country{China}
}

\author{Yuling Liu}
\email{liuyuling@iie.an.cn}
\affiliation{%
  \institution{Institute of Information Engineering, Chinese Academy of Sciences}
  \city{Beijing}
  \country{China}
}

\renewcommand{\shortauthors}{Liu et al.}

\begin{abstract}
Broken access control (BAC) ranks among the most severe web-application vulnerability classes, yet its automated repair remains largely unexplored. Unlike memory-safety vulnerabilities, BAC is a logic vulnerability whose repair demands a deep understanding of an application's access-control model and business semantics, which traditional automated vulnerability repair techniques handle poorly. Large language models (LLMs) open a promising avenue for this problem, but no standard benchmark exists for studying how well they repair BAC vulnerabilities. We construct \textsc{BAC-Bench}, the first benchmark for BAC repair, consisting of 46 real-world BAC vulnerabilities drawn from six widely used web applications spanning four programming languages. Each vulnerability is equipped with a sandboxed runtime environment, a PoC validation program and a functional test program. We further propose a BAC repair-strategy framework that characterizes any BAC patch along four dimensions, Position, Coverage, Mechanism, and Content. Evaluating eight base LLMs and two commercial LLM agents on \textsc{BAC-Bench}, we find that single-turn base LLMs repair 37.68\% of the vulnerabilities on average, while two minimal scaffolds, \emph{Apply-Feedback} and \emph{Navigate-and-Locate}, raise this average to 66.03\%, with the best base LLM reaching 75.36\%, close to the 78.99\% of the best agent. Notably, some LLM-generated patches surpass the official patches, and one incidentally closed a zero-day vulnerability that had lain dormant for three years. Our findings delineate the current capability boundary of LLM-based BAC repair and offer guidance toward its practical adoption.
\end{abstract}



\maketitle

\section{Introduction}

Broken access control (BAC) has become one of the most damaging classes of software vulnerabilities.
Reports from Mondoo and MITRE document the prevalence of BAC, with CWE-862 Missing Authorization increasing by 56\% from 2024 to 2025 and rising from rank 9 to rank 4 in MITRE's 2025 CWE Top 25, while CWE-284 Improper Access Control increased by 73\% ~\cite{mondoo2026,mitre2025cwetop25}.
OWASP likewise ranks BAC the first web-application risk~\cite{owasp2021,owasp2025} and the first API risk ~\cite{owaspapi2023}.
Its consequences are equally severe.
In December 2024, the Irish Data Protection Commission fined Meta €251 million for an access-control design defect whose leaked access token exposed the profiles of 29 million users~\cite{metadpc2024}.
Meanwhile, the sheer scale of disclosure is overwhelming manual remediation, and AI is accelerating the trend.
A record 48{,}185 CVEs were published in 2025~\cite{gamblin2026cve}, and disclosures in the first four months of 2026 already ran 46\% above projections, prompting FIRST to raise its full-year forecast to approximately 66{,}000 and putting annual disclosures on pace to approach 70{,}000 for the first time in history~\cite{first2026}.
In May 2026, Anthropic reported that Project Glasswing had used a frontier model to uncover more than 10{,}000 high- or critical-severity vulnerabilities in widely deployed software within its first month, yet only 75 of the 530 such vulnerabilities disclosed to maintainers had been patched~\cite{glasswing2026}.
This combination of surging prevalence, severe impact, and lagging repair makes the automated repair of BAC vulnerabilities an urgent priority.

The detection of the BAC has been studied extensively~\cite{dalton2009nemesis,sun2011static,lu2022detecting,shi2025facilitating}.
White-box approaches infer the intended access-control specification from the program itself and flag missing or inconsistent checks, with the detection granularity sharpened from the function level to the object level~\cite{sun2011static,mace2014,huang2024bola}.
Black-box approaches drop the source-code requirement and instead decide violations from externally observable behavior~\cite{authscope,bacscan2025}.
Work that directly targets BAC repair, by contrast, remains scarce. 
FixMeUp, one of the earliest such efforts, mines an access-control template from PHP applications and inserts it where a check is missing, but it relies on developers to hand-annotate sensitive operations and their checks~\cite{fixmeup2013}.
LogicPatcher uses path profiling to place patches for already-detected vulnerabilities~\cite{monshizadeh2016patching}.
DRacv extends this paradigm to both PHP and Java, deriving patches from contrasts between vulnerable and protected pages and deploying each as a standalone guard file ahead of the vulnerable page~\cite{xu2025dracv}.
Yet all of them handle only missing checks, rely on fixed templates or execution-flow patterns, and are designed for traditional architectures, limiting their applicability to the modern web applications in wide use today.

In summary, BAC repairing receive little attention from researchers because of three main reasons. 
\textbf{(1) BAC is a logic vulnerability that lacks modelable syntactic and data-flow signatures.} 
Unlike memory-safety vulnerabilities (e.g., buffer overflow, use-after-free) and injection vulnerabilities (e.g., SQL injection, XSS) that exhibit fixed syntactic patterns or explicit source-to-sink flows, BAC are deeply entangled with business logic and amount to the absence or malfunction of authorization logic, which has neither a fixed syntactic form nor a traceable flow, so pattern-based repair methods do not transfer.
\textbf{(2) Unlike detection, repairing task requires an explicit correctness specification to target.}
Detection need only judge whether an access violates the intended policy, which can be inferred from inconsistencies.
Repairing task, by contrast, need to determine the intended access-control policy, which in practice is usually undocumented or scattered through the business logic, and then realize it faithfully within the concrete context of the application's source code.
Any slight semantic ambiguity or implementation slip, can introduce new logic errors or over-restrict legitimate access.
\textbf{(3) Web access-control implementations are heterogeneous, lacking the standardized mechanisms of smart contracts.}
In smart contracts, access control centers on a few standardized mechanisms (e.g., \texttt{msg.sender}, the \texttt{onlyOwner} modifier), a regularity that recent smart-contract repair leverages by mining common RBAC practices to guide patch generation~\cite{acfix2025}.
Web applications honor no such convention. Different applications adopt RBAC, ABAC, or DAC and realize their checks in disparate forms (session variables, database permission tables, or custom validators), with the logic often spread across many files and call layers, so a repair tool cannot rely on a uniform template or prior knowledge to patch accurately across applications.

The rapid progress of large language models (LLMs) opens a promising path to this problem and lends fresh hope to repairing complex BAC.
Pre-trained on massive code corpora, LLMs command the syntax and semantics of mainstream programming languages and have internalized a wealth of framework idioms, API conventions, and secure-coding practices.
Across downstream tasks ranging from resolving real-world repository issues~\cite{swebench} and detecting vulnerabilities~\cite{ferrag2025securefalcon} to localizing faults and repairing programs~\cite{xia2023apr,zhang2023critical,hossain2024deep} and penetration testing~\cite{pentestgpt}, they already attain near-expert performance on a range of benchmarks, indicating that LLMs are now broadly capable of handling complex, context-coupled code and security tasks.
More importantly, the distinctive strengths of LLMs directly address the core difficulties that have made BAC repair hard. 
\emph{Firstly}, rather than matching a fixed syntactic pattern or template rule, an LLM identifies a missing or misconfigured authorization check through a holistic understanding of code semantics, reasoning from the surrounding business logic and thereby overcoming the \emph{absence of fixed syntactic or data-flow patterns}.
\emph{Secondly}, an LLM can infer implicit access-control intent from code context, variable names, comments, and the implementation of sibling modules in the same project, reconstructing the intended access-control policy \emph{even without an explicit specification} and thereby deciding what check to add and at what granularity to make the fix correct.
\emph{Thirdly}, having encountered RBAC, ABAC, and DAC models and their heterogeneous realizations, including session variables, permission-table lookups, and custom validation routines, throughout pre-training, an LLM generalizes across applications and languages, adapting to each project's checking logic \emph{without a uniform template or task- specific prior}. 

This raises the central empirical question: \textbf{How effective are LLMs at repairing BAC vulnerabilities in web applications?}

Despite a surge of LLM-based repair over the past two years~\cite{appatch,patchagent,pearce2023examining,liu2024exploring}, research efforts on BAC in web applications have been lagged behind.
Prior efforts either target memory-safety vulnerabilities, exploiting a crash or sanitizer signal to complete a locate-generate-and-validate loop, or, in the smart-contract setting, repair access control by mining standardized RBAC practices to guide the model~\cite{acfix2025}, an approach that does not transfer to the diverse landscape of web applications.
LLM-based repair of BAC in web applications thus remains largely unexplored, chiefly for want of a standardized repair benchmark.
LLM-based repair research relies on standardized corpora that pair reproducible vulnerabilities with PoCs and ground- truth patches, yet no such BAC dataset exists in either industry or academia.
The root difficulty is that BAC is a logic vulnerability whose triggering hinges on runtime context, including session state, role configuration, and initial database contents, and which therefore cannot be faithfully reproduced or tested in a standardized manner without a concrete execution environment.
Absent such a benchmark, researchers can neither develop and study repair methods nor compare them fairly on equal footing.

To overcome these challenges, we construct \textsc{BAC-Bench}, the first benchmark for BAC repair, consisting of 46 real-world BAC vulnerabilities drawn from six widely used web applications spanning four programming languages (Java, PHP, Go, and Python).
Each vulnerability is equipped with a vulnerability description, a textual PoC description, the official patch, a sandboxed runtime environment, a PoC validation program, and a functional test program, enabling faithful reproduction of the vulnerability and dynamic validation of repairs.
Building on this benchmark and an in-depth study of real-world BAC patches, we further propose a \emph{BAC repair-strategy framework} that characterizes any BAC patch along four dimensions, Position, Coverage, Mechanism, and Content, and lays bare its internal construction and strategic choices.
We then conduct an extensive evaluation on \textsc{BAC-Bench} with eight base LLMs (DeepSeek-V4 Pro, Qwen3.7-Max, Claude Opus 4.8, GPT-5.5, GLM-5.2, MiMo-V2.5 Pro, Kimi K2.7-Code, and Gemini 3.5 Flash) and two commercial closed-source LLM agents (Claude Code and Codex). Our results show that single-turn base LLMs repair only 37.68\% of the vulnerabilities on average, whereas two minimal scaffolds, \emph{Apply-Feedback} and \emph{Navigate-and-Locate}, raise this average to 66.03\%, with Claude Opus 4.8 reaching 75.36\%, only 3.63 percentage points below the 78.99\% of Claude Code, the best-performing closed-source LLM agent built on the same model. We further find that some LLM-generated patches surpass the official fixes.
On CVE-2023-31066, for example, an LLM-generated patch incidentally closed a latent access-control vulnerability that had gone undiscovered for over three years, which we responsibly disclosed to the Apache Software Foundation and the InLong maintainers and which was assigned CVE-2026-63040~\cite{cve202663040}.
These findings suggest that LLMs can already shoulder part of the BAC repair workload and point to a concrete path toward practical automated BAC repair. All datasets, scaffolds, LLM-generated patches, and experimental scripts are publicly available at \url{https://github.com/TianMaXingK/BAC-BENCH}~\cite{github}.

In summary, this paper makes the following contributions:
\begin{enumerate}
\item \textbf{A real-world benchmark.} We build \textsc{BAC-Bench}, a benchmark of 46 real-world BAC vulnerabilities drawn from six widely used web applications spanning four programming languages (Java, PHP, Go, and Python). Each vulnerability is equipped with a vulnerability description, a textual PoC description, the official patch, a sandboxed runtime environment, a PoC validation program, and a functional test program. We open-source it to facilitate research on BAC repair in web applications.

\item \textbf{A repair-strategy framework.} We propose a BAC repair-strategy framework that characterizes any BAC patch along four dimensions, Position, Coverage, Mechanism, and Content, enabling an in-depth analysis of how a BAC-patch is constructed and why it fails.

\item \textbf{A systematic evaluation.} We systematically evaluate eight base LLMs and two commercial closed-source LLM agents on repairing BAC in web applications, and devise two minimal scaffolds that mitigate the limitations observed in base- LLM repair. Our findings offer guidance for project communities pursuing automated BAC repair.

\end{enumerate}

\section{Background and Preliminary}

\subsection{Broken Access Control in Web Applications}

Broken access control (BAC) occurs when a web application fails to enforce authorization, allowing a request to read or modify resources that it is not entitled to access~\cite{owasp2021}.
Every web application implicitly follows an intended access-control policy set by its designers or mandated by its business domain, and a BAC vulnerability arises precisely when the access decisions realized by the code diverge from this intended policy.
Such divergence takes a family of concrete forms, which are characterized from three complementary perspectives.
The first is the \emph{granularity} at which authorization is missing or incorrect.
Following the OWASP API Security Top~10~\cite{owaspapi2023}, from coarse to fine, \emph{Broken Function-Level Authorization} (BFLA) denotes a sensitive endpoint or function that lacks or mis-implements role-based authorization, \emph{Broken Object-Level Authorization} (BOLA) denotes an attacker manipulating a resource identifier to reach an object owned by another user, a vulnerability widely known as Insecure Direct Object Reference (IDOR), and \emph{Broken Object Property-Level Authorization} (BOPLA) denotes access to an object being legitimate while its individual properties that should be off-limits can still be read or written~\cite{owaspapi2023bopla}.
The second perspective is the \emph{direction} of the privilege violation, which is \emph{horizontal} when a user reaches a peer's resources at the same privilege level and \emph{vertical} when a lower-privilege user acquires higher privileges.
The third perspective is the \emph{operation} the vulnerability enables, which is \emph{read-based} when it breaches confidentiality through data disclosure and \emph{write-based} when it breaches integrity through the unauthorized creation, modification, or deletion of state~\cite{bacscan2025}.
These perspectives are complementary rather than mutually exclusive, and a vulnerability is typically described under all three.
For example, tampering with another user's record through a guessable identifier is a horizontal, write- based BOLA.%

Fundamentally, BAC is a \emph{logic} vulnerability rather than a syntactic or memory-level one~\cite{dalton2009nemesis}.
It exhibits neither a fixed syntactic form nor a traceable data flow, because whether a given access constitutes a violation is inseparable from the application's high-level business logic and domain semantics~\cite{sun2011static}.
An access relation that amounts to a vulnerability in one application may be entirely legitimate behavior in another, as illustrated by OpenEMR's CVE-2022-2730~\cite{cve20222730}, whose fix does not follow the generic IDOR pattern.
Moreover, the intended access-control policy of a web application is rarely documented, and the source code is often its only real specification.
Compounding this, access-control models and implementation styles differ markedly from one application to the next, both the detection and, especially, the repair of BAC are highly complex, and automated tools struggle to generalize across applications~\cite{rolecast2011,mqttactic2024}.%

\subsection{Definition of BAC Vulnerability Repair}

GenProg~\cite{legoues2012genprog} defines a repair as a patch (a set of changes to the program) that makes the patched program pass an entire test suite. The patched program must pass every \emph{positive} test case, which encodes functionality that must be preserved, and at least one \emph{negative} test case, which encodes the bug to be fixed. Monperrus~\cite{monperrus2018} further articulates the oracles underlying repair, dividing them into the \emph{bug oracle}, which detects the unexpected behavior, and the \emph{regression oracle}, which checks that no new bugs are introduced during repair. Building on these notions and on our characterization of BAC above, we define the task of repairing a BAC vulnerability as follows.

Let a web application be denoted $W$, its \emph{request space} $Q$, and a request $q \in Q$. Regardless of whether it is documented, every application's access control in fact follows an \emph{intended policy}, set by the application's designers or mandated by the compliance requirements of its business domain, which we denote
\[
P : Q \rightarrow \{\mathsf{allow}, \mathsf{deny}\}.
\]
In parallel, the application exhibits a \emph{implemented decision} determined entirely by its code,
\[
P_{\mathrm{impl}} : Q \rightarrow \{\mathsf{allow}, \mathsf{deny}\}.
\]
Unlike the abstract, intended $P$, the implemented $P_{\mathrm{impl}}$ is defined by the source code and is therefore observable and testable.

We call the set of requests that the code \emph{allow} but the policy would \emph{deny} the \emph{under-restriction set}, and use it to characterize BAC vulnerabilities. Formally,
\[
\mathrm{Vuln}(W) = \{\, q \in Q \mid P(q) = \mathsf{deny} \;\wedge\;
P_{\mathrm{impl}}(q) = \mathsf{allow} \,\}.
\]

\begin{definition}[BAC vulnerability]\label{def:bac-vulnerability}
An application $W$ contains a BAC vulnerability if and only if
$\mathrm{Vuln}(W) \neq \emptyset$. That is, there exists at least one request $q$ that the intended policy denies but the code admits.
\end{definition}

Dually, the \emph{over-restriction set} is
\[
\mathrm{Over}(W) = \{\, q \in Q \mid P(q) = \mathsf{allow} \;\wedge\;
P_{\mathrm{impl}}(q) = \mathsf{deny} \,\}.
\]
$\mathrm{Over}(W)$ captures functionality or availability defects, in which a legitimate request is wrongly denied. A patch that introduces a new $q' \in \mathrm{Over}(W)$ is over-restrictive, breaking original legitimate functionality.

A CVE record or GitHub issue $d$ describes a single vulnerability. We write $Q_d \subseteq \mathrm{Vuln}(W)$ for the set of requests that exploit the vulnerability reported in $d$.

\begin{definition}[BAC repair]\label{def:bac-repair}
Let a patch $\pi$ be a set of changes to the source-code state $C(W)$, yielding $C(W) \oplus \pi$ and a corresponding application $W_\pi$. Given a vulnerable $W$ (i.e., $\mathrm{Vuln}(W) \neq \emptyset$) and a vulnerability description $d$, the BAC repair task is to find a patch $\pi$ satisfying the following two conditions.
\begin{enumerate}
\item[(1)] \emph{Closure.} The patch denies every request that exploits $d$.
\[
  \forall\, q \in Q_d :\quad P_{\mathrm{impl}}^{W_\pi}(q) = \mathsf{deny}.
\]
\item[(2)] \emph{Preservation.} No legitimate request is newly denied, equivalently,
$\mathrm{Over}(W_\pi) \subseteq \mathrm{Over}(W)$.
\[
  \forall\, q \in Q :\quad P_{\mathrm{impl}}(q) = \mathsf{allow} \;\Rightarrow\; P_{\mathrm{impl}}^{W_\pi}(q) = \mathsf{allow}.
\]
\end{enumerate}
\end{definition}
A patch satisfying both conditions is an \emph{effective patch} for instance $d$, whereas one that fails to block the exploit or breaks legitimate functionality is not.

\section{Overview}
In this section we overview our study, which is designed to answer four research questions:
\par\smallskip
\textbf{RQ1 (Effectiveness).} How effective are LLMs at repairing BAC vulnerabilities in web applications?
\par\smallskip
\textbf{RQ2 (Limitations and mitigation).} What limitations do LLM exhibit when repairing web-application BAC vulnerabilities, and to what extent can these limitations be mitigated?
\par\smallskip
\textbf{RQ3 (Repair strategy).} Which repair-strategy choices underlie the failures of LLM-generated BAC patches?
\par\smallskip
\textbf{RQ4 (Comparison with official patches).} Are official patches always better than LLM-generated patches?
\par\smallskip

\begin{figure}
    \centering
    \includegraphics[width=\linewidth]{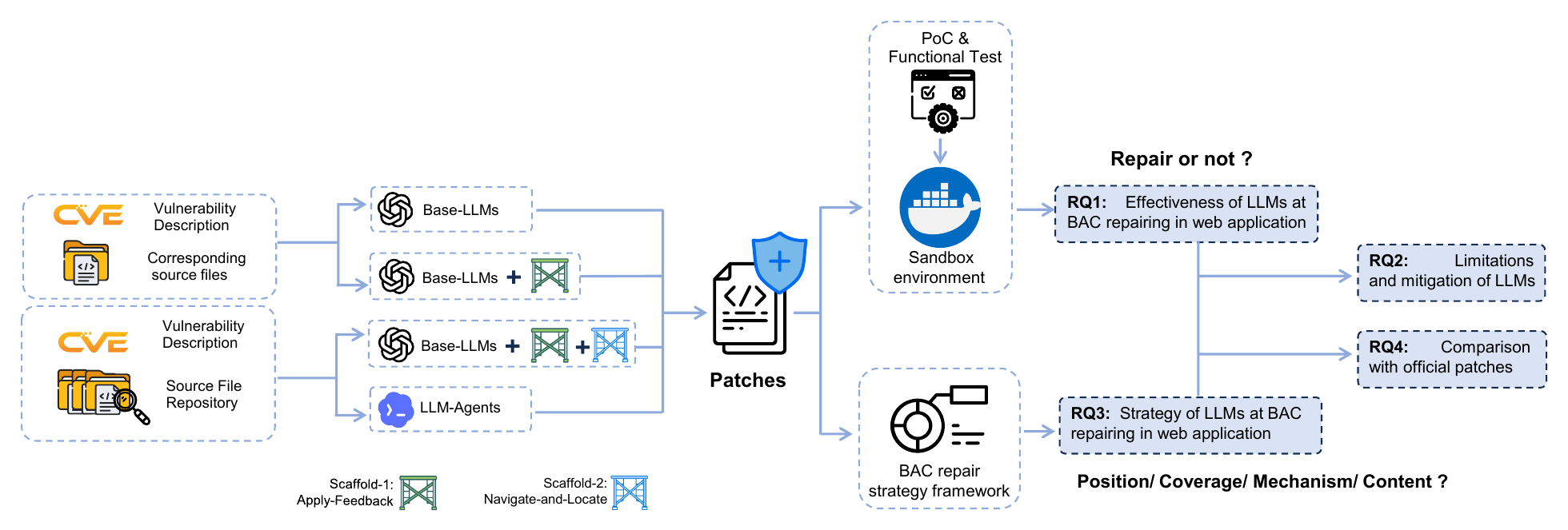}
    \caption{Overview of our study design for evaluating the effectiveness of LLMs in repairing BAC vulnerabilities in web applications.}
    \label{fig:figure-3-1}
\end{figure}

To answer them, we conduct a study depicted in Figure~\ref{fig:figure-3-1}.
\emph{Firstly}, we construct \textsc{BAC-Bench}, a dataset of real-world web BAC spanning multiple languages (Java, PHP, Go, and Python), faithfully reflecting the complex access-control implementation patterns of diverse web applications in the wild.
Using this benchmark, we systematically evaluate a range of state-of-the-art LLMs and compare their behavior, answering RQ1.
\emph{Secondly}, we analyze the failing LLM-generated patches and find that they fall into two broad failure patterns, namely patch-application failures and semantic failures, which stem largely from LLM hallucination and the limited context window, respectively.
We therefore design two minimal scaffolds, \emph{Apply-Feedback} and \emph{Navigate-and-Locate}, to mitigate these failures, answering RQ2.
\emph{Thirdly}, through an in-depth study of real-world BAC patches from web applications, we propose a repair-strategy framework that characterizes any BAC patch along four dimensions, Position, Coverage, Mechanism, and Content, and we apply it to the semantically failing patches to identify the dimensions in which their repair strategies go wrong, answering RQ3.
\emph{Finally}, we examine the official patches and evaluate them side by side with the LLM-generated patches under the same dynamic validation protocol, answering RQ4.

\section{Study Design}

In this section, we present the design of our study: the methodology for constructing the dataset of BAC-bench (\S4.1), the design of validation oracles (\S4.2), the selection of LLM models and the experimental setup (\S4.3), and the construction of BAC repair-strategy framework (\S4.4).

\subsection{Dataset Construction}

\subsubsection{BAC vulnerbility collection}
To investigate whether LLMs can automatically repair BAC, we construct a dataset of vulnerabilities from real-world open-source web applications using two complementary sources.
The first source consists of BAC vulnerabilities made publicly available through prior research, including the vulnerabilities studied by Huang et al.~\cite{huang2024bola} and those discovered by BACFuzz~\cite{bacfuzz}.
The second source is obtained by querying the CVE database for records that were disclosed in or after 2017, affect open-source projects hosted on GitHub, provide an official patch or fixing commit, and contain at least one of the following keywords in the CVE record or its associated references: ``permission check'', ``privilege check'', ``access control'', ``security check'', ``authorization'', ``superuser only'', or ``admin only''.
We then manually review the resulting candidates and remove entries that overlap with the research-derived source or lack a textual description of the exploitation procedure.
The resulting dataset comprises 46 BAC vulnerabilities from six projects, namely OpenEMR, Apache InLong, BookWyrm, Nextcloud, Usememos, and XWiki, which are implemented in four programming languages, namely PHP, Java, Go, and Python. Table~\ref{tab:apps} summarizes the basic characteristics of these applications.

  \begin{table}[t]
    \caption{Web applications in \textsc{BAC-Bench}.}
    \label{tab:apps}

    \centering
    \renewcommand{\arraystretch}{1.25}
    \setlength{\tabcolsep}{12pt}
    \resizebox{0.98\linewidth}{!}{%
      \begin{tabular}{llrll}
        \doubletoprule
        \textbf{Application} &
        \textbf{Language} &
        \textbf{Stars} &
        \textbf{Domain} &
        \textbf{Access-control model} \\
        \midrule

        OpenEMR
          & PHP
          & 5.2k
          & Healthcare Management
          & ACL + RBAC hybrid (phpGACL) \\

        Usememos
          & Go
          & 52.8k
          & Note-taking Platform
          & RBAC  \\

        Apache InLong
          & Java
          & 1.5k
          & Data Integration Framework
          & RBAC + groups \\

        BookWyrm
          & Python
          & 2.6k
          & Social Reading Platform
          & RBAC \\

        XWiki
          & Java
          & 1.2k
          & Knowledge Management
          & ACL + groups \\

        Nextcloud Server
          & PHP
          & 34k
          & Cloud Storage Platform
          & DAC + ACL + RBAC hybrid \\

        \doublebottomrule
      \end{tabular}%
    }
  \end{table}

\subsubsection{Supporting Artifacts}

Each vulnerability entry is equipped with six supporting artifacts, each playing an indispensable role in the evaluation:

\textbf{(a) Vulnerability description.} Taken from the CVE entry: the description, the affected application and version, and the effect of the vulnerability.

\textbf{(b) Textual PoC description.} Taken from the exploitation steps in the CVE's associated links (bug-bounty platforms such as Hunter, GitHub issues, etc.). Only the textual description is provided, without PoC code, which keeps the repair task close to the real-world handling of a security advisory.

\textbf{(c) Official patch and Source files.} The official patch that fixes the vulnerability, together with the original source files it modifies.

\textbf{(d) Sandboxed reproduction environment.} The affected version of the web application, on which the BAC vulnerability is confirmed to trigger, is deployed in a Docker container, providing a deterministic, isolated runtime.

\textbf{(e) PoC validation program.} Instantiates the vulnerability-blocking oracle of closure condition of Definition~\ref{def:bac-repair}, and its decision mechanism is detailed in \S4.2.1.

\textbf{(f) Functional test program.} Instantiates the functionality-preservation oracle of preservation condition of Definition~\ref{def:bac-repair}, and its decision mechanism is detailed in \S4.2.2.

Notably, (e) and (f) are constructed by security experts with the assistance of large language models, and each is validated against objective ground truth before inclusion in the dataset. Specifically, every PoC validation program is executed in the sandbox and confirmed to trigger the disclosed vulnerability on the unpatched original application, and every functional test program is confirmed to pass on the unpatched original application. The validity and fidelity of these artifacts thus derive from actual dynamic execution against ground truth, rather than from their authorship.

\subsection{Oracle Design}
To gauge the true effectiveness of LLM-generated BAC patches, the PoC validation program~(e) and the functional test program~(f) must themselves be valid and faithful. As a logic vulnerability, BAC lacks the universal sanitizer oracle available for memory-safety bugs. Whereas AddressSanitizer automatically captures out-of-bounds accesses and use-after-free~\cite{serebryany2012asan}, the exploitation of a BAC vulnerability triggers neither a crash nor an explicit error. Our oracle design must therefore employ clear decision criteria and satisfy three requirements: it must block the exploit (\S4.2.1), preserve legitimate functionality (\S4.2.2), and remain replayable (\S4.2.3).

\subsubsection{Vulnerability-Blocking Oracle}
We design separate vulnerability-blocking oracles for unauthorized reads and unauthorized writes.

\textbf{Unauthorized read.} Response-similarity oracles are brittle and prone to false positives due to their fixed similarity thresholds. We instead use a more precise unique-marker content match (honeytoken~\cite{spitzner2003honeytokens}). A high-entropy, unguessable unique marker is pre-seeded in the victim's protected resource, and we then assert that the unauthorized attacker's response contains this marker byte-for-byte. Because the marker is unique and unguessable, its appearance unambiguously proves that an unauthorized read has occurred.

\textbf{Unauthorized write.} Unauthorized writes, which include UPDATE, INSERT, and DELETE operations, typically leave no observable trace in the direct response, causing response-similarity oracles to misjudge modification-based violations at a very high rate~\cite{bacscan2025}. The classic differential-response oracle~\cite{authscope} can decide only unauthorized reads and cannot automatically decide unauthorized writes. Since our benchmark is white-box with a controllable database, we adopt the more direct database-state inspection as the primary oracle. After the attacker issues an unauthorized write carrying a unique marker, we directly query the victim's corresponding database row and assert that the marker has been persisted (for UPDATE/DELETE, we additionally assert that the original value has been overwritten or removed). We complement this with a victim-side re-read through the application's own interface, verifying that the returned content reflects the attacker's marker. The former observes the committed, persisted state, whereas the latter confirms that the unauthorized write is semantically visible through the application itself.

\subsubsection{Functionality-Preservation Oracle}

A failed PoC attack alone is insufficient to conclude that a repair succeeds. The patch must also preserve the application's original legitimate functionality, as required by the preservation condition of Definition~\ref{def:bac-repair}. In practice, a substantial fraction of patches trade functionality preservation for ``passing the PoC'' and some even bluntly deny all user access. Although such patches block the attack, they severely violate the application's business semantics and constitute typical instances of \emph{over-restriction}, a class of incorrect patches~\cite{qi2015analysis}. Moreover, a project's built-in test suite is largely oriented toward general functionality and rarely touches the specific endpoints, operations, and roles involved in the vulnerability at hand. Hence, passing the project suite does not establish that the affected functionality is preserved~\cite{smith2015overfitting}.

Our functionality-preservation oracle therefore adopts a \emph{functional test program} authored for each BAC. The test targets exactly the functionality affected by the vulnerability and asserts that a legitimate subject that should retain access still succeeds in its legitimate operation both before and after patching. Its decision mechanism follows the same marker-based and database-state checks as the vulnerability-blocking oracle~(\S4.2.1), with the polarity reversed. If the project ships its own functional tests, we run them together with our targeted test. Otherwise, only ours is run.

\subsubsection{Replayability}

Mutating operations (writes, modifications, or deletions) alter shared application state, making any test program that performs them \emph{destructive}, i.e., a \emph{polluter}~\cite{gyori2015poldet}. Once the mutation footprint of a destructive PoC intersects the verification objects on which another PoC (or its own replay) depends, the latter's oracle verdict varies with the execution order and becomes a flaky test~\cite{luo2014flaky}. To eliminate this test-order dependency, we make every destructive PoC \emph{hermetic} (self-contained)~\cite{winters2020swe}. At the start of its execution, the PoC establishes the data it depends on, creating the resource it will later delete and seeding the initial value whose unauthorized modification it will check. Before it exits, it reverts every change it has made to the shared application and restores the pre-run state. Each PoC is thereby \emph{idempotent}, leaving no residual state and yielding the same verdict regardless of the execution order or the number of repeated runs. PoCs are therefore individually deterministic and mutually non-interfering, which ensures \emph{observation independence}.

\subsection{Models and Experimental Setup}

\subsubsection{Model Selection}

This work aims to systematically assess the current ability of LLMs to repair BAC vulnerabilities. Accordingly, our criteria for selecting the evaluation targets include advancedness, representativeness, and comparability. We particularly consider SWE-bench Verified~\cite{swebench,openai2024swebenchverified}, a benchmark that evaluates the coding and program-repair ability of LLMs, and GPQA Diamond~\cite{rein2023gpqa}, a graduate-level reasoning benchmark that targets complex reasoning problems. Our final selection comprises two commercial closed-source general-purpose LLM agents (Claude Code and Codex) and eight base LLMs (DeepSeek-V4 Pro, Qwen3.7-Max, Claude Opus 4.8, GPT-5.5, GLM-5.2,Gemini 3.5 Flash, MiMo-V2.5 Pro and Kimi K2.7-Code). Table~\ref{tab:models} lists the relevant parameters of these models.

  \begin{table}[t]
    \caption{LLM-based systems evaluated on our \textsc{BAC-Bench}. The DeepSeek entry refers to the preview version used in our experiments. \emph{Release} gives the release or publication date of the listed version.}
    \label{tab:models}

    \centering
    \renewcommand{\arraystretch}{1.25}
    \setlength{\tabcolsep}{15pt}
    \resizebox{0.92\linewidth}{!}{%
      \begin{tabular}{llllcl}
        \doubletoprule
        \rule{0pt}{2.8ex}%
        \textbf{Category} &
        \textbf{Model} &
        \textbf{Developer} &
        \textbf{Release} &
        \textbf{Access} \\
        \midrule

        \multirow{2}{*}{\shortstack[l]{LLM Agents}}
          & Claude Code (2.1.160) & Anthropic    & 2026-06-02 & Closed \\
          & Codex (0.139.0)       & OpenAI       & 2026-06-09 & Closed \\
        \midrule

        \multirow{8}{*}{Base LLMs}
          & Claude Opus 4.8  & Anthropic    & 2026-05-28 & Closed \\
          & GPT-5.5          & OpenAI       & 2026-04-23 & Closed \\
          & Qwen3.7-Max      & Alibaba      & 2026-05-20 & Closed \\
          & Gemini 3.5 Flash & Google       & 2026-05-19 & Closed \\
          & DeepSeek-V4 Pro  & DeepSeek     & 2026-04-23 & Open   \\
          & GLM-5.2          & Z.ai         & 2026-06-13 & Open   \\
          & MiMo-V2.5 Pro    & Xiaomi       & 2026-04-23 & Open   \\
          & Kimi K2.7-Code   & Moonshot AI  & 2026-06-12 & Open   \\

        \doublebottomrule
      \end{tabular}%
    }
  \end{table}

\subsubsection{Common setup}

All LLMs are run under a unified, isolated harness to ensure comparability and evaluation validity.

\begin{itemize}
\item \textbf{Input.} For each vulnerability instance, the model receives the vulnerability description, the textual PoC description (without PoC code), and the source code of the web application (the source granularity varies by category, as detailed below). 
\item \textbf{Contamination control.} To evaluate the models' intrinsic repair ability, we disable all network access for the evaluated systems, preventing them from retrieving the official patch, post-fix versions, or any other ground-truth information, and we audit the run log after each inference to confirm that no such information was accessed.
\item \textbf{Session isolation.} Each run starts a fresh session with an empty working directory.
\item \textbf{Repetition.} Under identical conditions, we independently repeat the repair of each (model, vulnerability) pair three times to assess stability.
\end{itemize}

The input granularity and the number of interaction turns differ by model category. Base LLMs are queried in a single-turn session, with the source input restricted to the source files modified by the official patch~\cite{pearce2023examining}. They must produce a patch in a single response, which we automatically extract from the LLM's answer via a predefined regular-expression match and materialize as a patch file. The two agents (Claude Code and Codex) instead run their own multi-turn loops, autonomously reading across the entire repository and emitting a patch file. We use them as-is, without altering their internal procedures, and constrain only the outer sandbox (network isolation, session isolation, and log auditing). In addition, we implement two multi-turn \emph{scaffolds}~(\S4.3.3) that place base LLMs on an equal footing with the mature LLM agents, i.e., on exactly the same input.

\subsubsection{Two Scaffolds for Base LLMs}
\label{subsubsec:scaffolds}
Beyond single-turn inference, we design two mutually independent scaffolds for base LLMs. Both are minimal, transparent, and deliberately simple, built solely to reveal the core capabilities underlie effective LLM-based BAC repair.

\paragraph{(1) Apply-Feedback (Algorithm~\ref{alg:apply-feedback})}

LLM-generated patches often fail to apply because their unified-diff metadata, source context, or file paths are inconsistent with the target repository. Such patches cannot reach the subsequent semantic evaluation. To mitigate this application-stage failure, we introduce the \emph{Apply-Feedback} scaffold.

The procedure first constructs the initial dialogue context $H_0=\textsc{InitialPrompt}(d,p,S)$ from the vulnerability description $d$, the PoC description $p$, and the source files $S$. At turn $k$, the LLM receives the current context $H_k$ and returns a raw textual response $o_k$. The scaffold extracts a candidate patch $\pi_k=\textsf{ExtractPatch}(o_k)$ and invokes $\textsf{ApplyCheck}(\pi_k,R)$ to test whether the candidate can be applied to repository $R$. The application check uses \texttt{git apply -{}-check} or \texttt{git apply -{}-recount -{}-check}, and returns a Boolean result $\mathit{ok}_k$ together with the diagnostic $e_k$. If $\mathit{ok}_k$ is true, the candidate patch is returned immediately. Otherwise, the model output and the feedback generated from $e_k$ are appended to the dialogue context, and the LLM is asked to revise its patch. The process starts at $k=0$ and continues through $k=K$. Thus, \(K\), which we set to 5 in our experiments, denotes the maximum number of self-healing attempts after the initial generation, and the LLM is invoked at most \(K+1\) times. If no candidate passes the application check, the most recent candidate is returned.

  \begin{algorithm}
    \caption{Apply-Feedback}
    \label{alg:apply-feedback}
    \begin{algorithmic}[1]
      \Require vulnerability description $d$, PoC text $p$, source files $S$,
      repository $R$, self-heal budget $K$
      \Ensure patch $\pi$
      \State $H_0 \gets \textsc{InitialPrompt}(d,p,S)$
      \State $\pi_{\mathrm{last}} \gets \emptyset$
      \For{$k = 0 \ldots K$}
        \State $o_k \gets \textsf{LLM}(H_k)$
        \State $\pi_k \gets \textsf{ExtractPatch}(o_k)$
        \State $\pi_{\mathrm{last}} \gets \pi_k$
        \State $(\mathit{ok}_k,e_k)
          \gets \textsf{ApplyCheck}(\pi_k,R)$
        \If{$\mathit{ok}_k = \mathsf{true}$}
          \State \Return $\pi_k$
        \EndIf
        \If{$k < K$}
          \State $H_{k+1}
            \gets H_k \oplus o_k \oplus \textsc{Feedback}(e_k)$
        \EndIf
      \EndFor
      \State \Return $\pi_{\mathrm{last}}$
    \end{algorithmic}
  \end{algorithm}

In Algorithm~\ref{alg:apply-feedback}, $H_k$ denotes the dialogue context available at the beginning of turn $k$, and $o_k$ denotes the raw textual response returned by the LLM in that turn. The candidate patch extracted from this response is denoted by $\pi_k=\textsf{ExtractPatch}(o_k)$. The procedure $\textsf{ApplyCheck}(\pi_k,R)$ returns the Boolean applicability result $\mathit{ok}_k$ and the corresponding diagnostic $e_k$. If no unified-diff patch can be extracted from $o_k$, $\textsf{ExtractPatch}(o_k)$ returns the empty patch. The variable $\pi_{\mathrm{last}}$ stores the most recent candidate and is returned if none of the candidates passes the application check.

\paragraph{(2) Navigate-and-Locate (Algorithm~\ref{alg:navigate-locate})}

Because of their input-size and context limitations, LLMs cannot generally receive an entire repository as input. In the single-turn setting, we therefore follow Pearce et al.~\cite{pearce2023examining} and provide the base LLM with only the source files modified by the official patch. This simplification has three drawbacks. First, it does not reflect a realistic repository-level repair scenario and may implicitly reveal the likely repair location. Second, it omits files on which the target files depend, limiting the model's access to the program semantics needed for a correct repair. Third, it restricts the space of possible repairs, even though a BAC vulnerability may admit multiple valid fixes. In particular, if the official patch is incomplete or defective, the required repair may involve files outside the provided context.

To address these issues, we build the \emph{Navigate-and-Locate} scaffold, thereby restoring the real-world repair scenario, testing the model's end-to-end ability to locate and repair a vulnerability within a code repository, and aligning the inputs of base LLMs and commercial agents so that they are evaluated on exactly the same input and their true capability gap can be isolated. The scaffold initially provides the repository directory tree and allows the LLM to request source files on demand. At each exploration turn, the LLM receives the current dialogue context $H_m$ and returns a raw response $o_m$. The function $\textsf{ExtractPatch}(o_m)$ checks whether the response contains a patch. If no patch is found, $\textsf{ParseFileRequest}(o_m)$ identifies the files requested by the model, and $\textsf{Retrieve}(F_m,R)$ retrieves their contents from repository $R$. The retrieved contents are appended to the dialogue context for the next turn. Exploration continues until the model emits a patch, signals that it has sufficient context, or reaches the exploration budget $M$. If no patch is produced during exploration, the scaffold issues one final request for a repair patch based on the files provided so far. The resulting patch is then passed to the downstream application, compilation, and semantic-validation pipeline.

  \begin{algorithm}
    \caption{Navigate-and-Locate}
    \label{alg:navigate-locate}
    \begin{algorithmic}[1]
      \Require vulnerability description $d$, PoC text $p$, repository $R$,
      exploration budget $M$
      \Ensure patch $\pi$
      \State $H_0 \gets
        \textsc{InitialPrompt}(d,p,\text{directory tree of }R)$
      \For{$m = 0 \ldots M-1$}
        \State $o_m \gets \textsf{LLM}(H_m)$
        \State $\pi_m \gets \textsf{ExtractPatch}(o_m)$
        \If{$\pi_m \neq \emptyset$}
          \State \Return $\pi_m$
        \EndIf
        \State $F_m \gets \textsf{ParseFileRequest}(o_m)$
        \If{$F_m \neq \emptyset$}
          \State $H_{m+1} \gets
            H_m \oplus o_m \oplus \textsf{Retrieve}(F_m,R)$
        \Else
          \State $H_{m+1} \gets H_m \oplus o_m$
          \State \textbf{break}
        \EndIf
      \EndFor
      \State $H_{\mathrm{final}} \gets
        H_{m+1} \oplus
        \textsc{FinalPatchRequest}$
      \State $o_{\mathrm{final}} \gets
        \textsf{LLM}(H_{\mathrm{final}})$
      \State $\pi_{\mathrm{final}} \gets
        \textsf{ExtractPatch}(o_{\mathrm{final}})$
      \State \Return $\pi_{\mathrm{final}}$
    \end{algorithmic}
  \end{algorithm}

In Algorithm~\ref{alg:navigate-locate}, $H_m$ denotes the dialogue context available at exploration turn $m$, and $o_m$ denotes the corresponding raw LLM response. The function $\textsf{ExtractPatch}(o_m)$ extracts a unified-diff patch from the response and returns the empty patch if no patch is found. The function $\textsf{ParseFileRequest}(o_m)$ identifies the set $F_m$ of repository files requested by the model, while $\textsf{Retrieve}(F_m,R)$ returns their contents from repository $R$. The retrieved contents are appended to form \(H_{m+1}\). The exploration budget \(M\), which we set to 10 in our experiments, limits the number of exploratory LLM turns.

The \emph{Apply-Feedback} and \emph{Navigate-and-Locate} scaffolds are decoupled, the former mitigating the model's hallucination problem and the latter resolving the source-file compromise forced by the LLM's limited context. They can also be composed into a single pipeline, which we release as a reproducible baseline for \textsc{BAC-Bench}.

\subsection{BAC Repair-Strategy Framework}
\label{sec:framework}

\subsubsection{Conceptual Dimensions}

Evaluating an access-control patch, whether it originates from a human developer, a traditional program-repair tool, or an LLM, typically reduces to an \emph{effectiveness oracle} that deems a patch a successful repair if it blocks the proof-of-concept (PoC) exploit and preserves normal functionality under regression testing (Definition~\ref{def:bac-repair} in \S2)~\cite{pearce2023examining,swebench}. This oracle is objective and indispensable, yet opaque. A binary pass/fail verdict treats the patch as a black box, explaining neither \emph{how} the patch is constructed nor \emph{why} a rejected patch fails. To open this black box, we introduce the \emph{BAC repair-strategy framework}, a compact and principled coordinate system that characterizes the internal construction and the strategic choices of a patch for a broken-access-control (BAC) vulnerability.

Through an in-depth study of a broad set of BAC patches, we observe that repairing a BAC defect essentially amounts to inserting an authorization decision into the program. In the reference architecture standardized by XACML and the earlier IETF policy framework~\cite{xacml30,rfc2753}, an authorization action is decomposed into a Policy Enforcement Point (PEP), which specifies where a request is intercepted and the decision is applied, and a Policy Decision Point (PDP), which specifies what the decision logic is. Refining this anatomy, we contend that any authorization insertion can be characterized by four questions: where it is enforced, how many request paths it covers, how it is implemented, and what authorization decision it makes. This characterization motivates a four-dimensional framework for BAC repair strategies, comprising \emph{Position}, \emph{Coverage}, \emph{Mechanism}, and \emph{Content}, which locate a patch within the complete repair space.

\textbf{Position} describes where an access-control check is enforced in a web application, including its logical architectural layer, its placement along the vulnerable execution path, and whether that location serves as a centralized chokepoint. The choice of architectural layer captures the \emph{depth} of the repair, while the degree of centralization shapes its potential reach and granularity. A check placed at an upstream, centralized point may govern multiple downstream vulnerability units, whereas a check embedded in operation-specific business logic can provide more focused protection. Neither placement is inherently preferable. The appropriate choice depends on the structure of the vulnerability and the desired balance between broad reach and operation-specific precision. Position thus concerns the placement and potential scope of a repair. The protection actually realized across vulnerability units and attack paths is captured by Coverage.

\textbf{Coverage} captures the scope of protection both across vulnerability units sharing the same root cause within a single BAC and across attack paths reaching each unit. The units may correspond, for example, to the read, update, and delete operations on the same protected object, or to a group of sibling endpoints implementing related functionality. A single unit may in turn be reachable through multiple attack paths, such as distinct request routes, request sequences, or input conditions that eventually trigger the same vulnerable operation. Coverage therefore records the breadth of a repair, reflecting whether a patch protects all vulnerability units sharing the root cause and blocks every attack path through which those units can be reached.

\textbf{Mechanism} specifies how the check is implemented, whether by reusing a canonical authorization helper already present in the codebase, by hand-rolling a bespoke check function in the application code, or by adding an extra filtering predicate to the SQL query at the data layer. These choices carry different trade-offs. Reusing a helper keeps the codebase consistent and is less error-prone, hand-rolling a dedicated check adds complexity and is more prone to omissions, and a query-level predicate is a fine-grained modification that demands precise handling.

\textbf{Content} concerns what the check decides, which is tightly coupled to the concrete business scenario. It can be viewed as an explicit rendering of the \emph{intended policy} the application designs for that scenario, shaped by the access-control model it follows, whether ownership/DAC, role/RBAC, attribute/ABAC, or relationship/ReBAC~\cite{sandhu1996rbac,nist_abac,fong2011rebac}. Its granularity ranges from coarse, role-level discrimination to fine, per-object ownership.

\subsubsection{Operationalization and Binary Mapping}

The conceptual framework is not directly observable clearly without additional coding rules.
To apply it consistently to thousands of repository-level patches, we operationalize each dimension as an observable variable with explicit and reproducible boundaries.
The operationalization uses two stages.
First, each patch receives a fine-grained label that preserves its concrete repair strategy and failure mode.
Then, the fine-grained label is mapped to a binary outcome for quantitative comparison across models, applications, vulnerabilities, and experimental conditions. Table~\ref{tab:framework-operationalization} summarizes this coding scheme.

  \begin{table}[t]
    \centering
    \caption{Operationalization of the four-dimensional BAC repair-strategy framework.}
    \label{tab:framework-operationalization}
    \renewcommand{\arraystretch}{1.18}
    \setlength{\tabcolsep}{6pt}

    \resizebox{1\linewidth}{!}{%
      \begin{tabular}{
        p{0.16\linewidth}
        p{0.45\linewidth}
        @{\hspace{16pt}}
        p{0.27\linewidth}
      }

      \specialrule{\heavyrulewidth}{0pt}{0.3ex}
      \specialrule{\lightrulewidth}{0pt}{0pt}

      \rule{0pt}{2.8ex}%
      \textbf{Dimension} &
      \textbf{Fine-grained categories and criterion} &
      \textbf{Binary mapping } \\

      \midrule

      \textbf{Position} &
      Presentation, control, business, data, and cross-cutting authorization
      layers. A repair point is accurate only when it lies on the vulnerable
      execution path and is placed at a time that can constrain the sensitive
      operation. &
      \textbf{Success:} All required repair points are accurate.
      \newline
      \textbf{Failure:} At least one repair point is inaccurate. \\

      \addlinespace[2pt]
      \cmidrule(lr){1-3}
      \addlinespace[2pt]

      \textbf{Coverage} &
      The ratio \(k/n\), where \(n\) is the number of vulnerability units
      frozen before inspecting candidate patches and \(k\) is the number of
      units receiving a structural security intervention. &
      \textbf{Success:} \(k=n\), coverage complete.
      \newline
      \textbf{Failure:} \(k<n\), coverage incomplete. \\

      \addlinespace[2pt]
      \cmidrule(lr){1-3}
      \addlinespace[2pt]

      \textbf{Mechanism} &
      \ding{172} Standard mechanism reuse.
      \ding{173} Reliable project-native mechanism.
      \ding{174} Fragile substitute mechanism.
      \ding{175} Mechanism implementation failure. &
      \textbf{Success:} \ding{172} or \ding{173}.
      \newline
      \textbf{Failure:} \ding{174} or \ding{175}. \\

      \addlinespace[2pt]
      \cmidrule(lr){1-3}
      \addlinespace[2pt]

      \textbf{Content} &
      Correct, partially correct, and incorrect. &
      \textbf{Success:} \emph{Correct} only.
      \newline
      \textbf{Failure:} \emph{Partially correct} or \emph{incorrect}. \\

      \specialrule{\lightrulewidth}{0pt}{0.3ex}
      \specialrule{\heavyrulewidth}{0pt}{0pt}

      \end{tabular}%
    }
  \end{table}

\textbf{Position} identifies the logical architectural location at which a patch introduces or modifies an authorization enforcement decision. Drawing on established presentation-domain-data layering \cite{fowler2002eaa} and recent empirical evidence on controller-service-repository architectures \cite{ozkan2023ddd}, and informed by our analysis of BAC patches in web applications, we classify repair locations into five logical categories: Presentation, Control, Business, Data, and Cross-Cutting Authorization.
The presentation layer handles rendering and final output. The control layer receives external requests, parses inputs, establishes request context, and selects an entry point.
The business layer executes domain operations and workflow logic.
The data layer reads, filters, updates, and persists application data.
The cross-cutting authorization layer contains ACL, RBAC, policy, or other centralized authorization infrastructure.
The authorization layer is treated as a cross-cutting concern rather than as a fifth sequential runtime layer.
A Position is accurate only when the patch modifies the actual execution path of the vulnerable request and the check is placed at a time that can constrain the sensitive operation or reliably compensate for its effect.
For a multi-point patch, any path-level or temporal misalignment causes the Position result to be classified as a failure.

\textbf{Coverage} is quantified using the ratio \(k/n\).
Before inspecting candidate patches, we manually enumerate and freeze $n$ of each BAC based on the vulnerability description, disclosure material, and target-version source code.
To make large-scale annotation tractable and reproducible, we consider only the attack path that is explicitly documented in the vulnerability description for each BAC and do not separately enumerate or score alternative attack paths.
The values of \(n\) for all vulnerabilities are reported in Appendix~\ref{app:vuln-list}.
Here, \(k\) is the number of frozen units that receive a structural security intervention from the patch.
Coverage is classified as complete when \(k=n\) and incomplete when \(k<n\).
This dimension does not evaluate whether the authorization condition on a covered path is correct.

\textbf{Mechanism} is classified into four fine-grained categories.
Standard mechanism reuse refers to the correct use of an existing central ACL, authorization helper, security-aware query, or other project security facility.
A reliable project-native mechanism does not directly reuse a standard helper but derives the decision from trusted principals, authoritative resources, stable APIs, and fail-closed control flow in a manner compatible with the application architecture.
These two categories are mapped to mechanism success.
A fragile substitute mechanism relies on proxy attributes, unstable internal state, or an inappropriate mechanism family.
Mechanism implementation failure includes incorrect APIs, types, queries, cache handling, exception propagation, or control flow that prevents the decision from being reliably enforced.
These last two categories are mapped to mechanism failure.

\textbf{Content} is classified as \emph{correct}, \emph{partially correct}, or \emph{incorrect}.
A patch is \emph{correct} when it expresses the target allowed set, blocks unauthorized access, and preserves legitimate functionality.
It is \emph{partially correct} when it blocks the core unauthorized access and retains basic legitimate authorization, but exhibits a bounded deviation in administrator exceptions, action granularity, relationship direction, boundary inputs, or the complete allowed set.
It is \emph{incorrect} when the subject, resource, action, core relationship, or security effect is wrong, or when the patch does not express a meaningful authorization decision.
Under the strict binary mapping used in this study, only \emph{correct} counts as Content success. Both \emph{partially correct} and \emph{incorrect} count as Content failures.
We retain the three-way labels in the detailed analysis because they distinguish bounded semantic deviations from fundamentally incorrect repairs.

We use the BAC repair-strategy framework to conduct an in-depth analysis of patches for BAC vulnerabilities.
Recent studies suggest that LLMs can serve as reliable labeling instruments for software-engineering artifacts when guided by explicit coding rules~\cite{ahmed2025llmannotation}.
In patch evaluation, rubric-guided LLM judgments have achieved agreement with human consensus comparable to that of human raters~\cite{shi2025llmjudge}, while LLM-generated labels for commit classification have approached inter-human agreement~\cite{wan2026commitsuite}.
Guided by the four-dimensional criteria in Table~\ref{tab:framework-operationalization}, we therefore use an LLM to analyze each BAC patch and assign fine-grained labels, which are subsequently converted into binary outcomes using the deterministic mappings in the same table.
To assess labeling reliability, three PhD students specializing in cybersecurity independently relabel a stratified random sample using the same coding rules while blinded to the LLM labels and to each other's judgments.
We report both human inter-rater agreement and the agreement between the LLM labels and the human consensus using percent agreement and Cohen's~$\kappa$ in \S5.3.

\section{Study Result}
\subsection{RQ1: How effective are LLM at repairing BAC vulnerabilities in web application?}
 
  \begin{table}[!t]
    \centering
    \caption{Repair Effectiveness of LLM Agents and Single-Turn Base LLMs on BAC-BENCH. Each percentage denotes the proportion of these 46 vulnerability instances for which the generated patch is effective after execution in the corresponding sandboxed environment. The Mean column reports the arithmetic average over the three independent runs, while each macro-average row reports the unweighted arithmetic mean across models within the corresponding category for that run. The Mean value of a macro-average row is then computed from its three run-level macro-averages. A patch is counted as finally effective only if it is successfully applied, successfully compiled when compilation is applicable, resistant to the PoC exploit, and compatible with the functional tests.}
    \label{tab:Base LLMs effectiveness}

    \small
    \setlength{\tabcolsep}{15pt}
    \renewcommand{\arraystretch}{1.15}

    \resizebox{1\linewidth}{!}{%
      \begin{tabular}{llrrrr}
        \doubletoprule

        \rule{0pt}{2.8ex}
        \textbf{Model} &
        \textbf{Condition} &
        \textbf{Run 1} &
        \textbf{Run 2} &
        \textbf{Run 3} &
        \textbf{Mean} \\
        \midrule

        Claude Code(Opus 4.8)
          & LLM Agent & 84.78\% & 71.74\% & 80.43\% & \textbf{78.99\%} \\
        Codex(GPT-5.5)
          & LLM Agent & 78.26\% & 76.09\% & 76.09\% & 76.81\% \\
        \midrule
        \textbf{Macro-average}
        & LLM Agent & 81.52\% & 73.92\% & 78.26\% & \textbf{77.90\%} \\
        \midrule
        DeepSeek-V4 Pro
          & Base LLM & 36.96\% & 43.48\% & 28.26\% & 36.23\% \\
        Qwen3.7-Max
          & Base LLM & 36.96\% & 41.30\% & 36.96\% & 38.41\% \\
        Claude Opus 4.8
          & Base LLM & 30.43\% & 34.78\% & 26.09\% & 30.43\% \\
        GPT-5.5
          & Base LLM & 63.04\% & 63.04\% & 56.52\% & \textbf{60.87\%} \\
        Gemini 3.5 Flash
          & Base LLM & 30.43\% & 26.09\% & 34.78\% & 30.43\% \\
        GLM-5.2
          & Base LLM & 41.30\% & 43.48\% & 39.13\% & 41.30\% \\
        MiMo-V2.5 Pro
          & Base LLM & 30.43\% & 19.57\% & 28.26\% & 26.09\% \\
        Kimi K2.7-Code
          & Base LLM & 30.43\% & 41.30\% & 41.30\% & 37.68\% \\
        \midrule
        \textbf{Macro-average}
        & Base LLM & 37.50\% & 39.13\% & 36.41\% & \textbf{37.68\%} \\

        \doublebottomrule
      \end{tabular}%
    }
  \end{table}

Table~\ref{tab:Base LLMs effectiveness} reports the final repair effectiveness of eight base LLMs and two closed-source LLM agents on \textsc{BAC-Bench}. Claude Code achieves the highest mean effectiveness among all evaluated models, reaching \(78.99\%\), followed by Codex at \(76.81\%\). The two LLM agents obtain an average effectiveness of \(77.90\%\), exceeding the eight base LLMs' average of \(37.68\%\) by \(40.22\) percentage points. Each model was evaluated in three independent runs under the same experimental configuration. To quantify run-to-run variability, let \(E_{m,r}\) denote the final effectiveness of model \(m\) in run \(r\), and define the within-model spread as
\begin{equation}
R_m =
\max_{r \in \{1,2,3\}} E_{m,r}
-
\min_{r \in \{1,2,3\}} E_{m,r}.
\label{eq:run-spread}
\end{equation}

The mean value of \(R_m\) across the eight base LLMs is \(8.69\) percentage points, indicating non-negligible stochastic variation in patch generation. We therefore use the arithmetic mean across the three runs as the primary estimate of each model's effectiveness.
The difference between LLM agents and base LLMs is particularl pronounced under the matched Opus-based comparison.
Claude Code based on Opus 4.8  achieves \(78.99\%\), whereas the same model evaluated directly in the single-turn base LLM setting achieves only \(30.43\%\), yielding a difference of \(48.56\) percentage points.
Unlike single-turn patch generation, LLM agents can iteratively search the repository, retrieve additional source files, invoke tools, apply patches, inspect diagnostics, and revise their hypotheses.
This reasoning--action--observation loop enables them to use execution feedback during repair, whereas the base LLM condition generates a patch in a single turn.

The eight base LLMs exhibit substantial performance heterogeneity. GPT-5.5 achieves the highest mean effectiveness at \(60.87\%\), outperforming the second-best base model, GLM-5.2 at \(41.30\%\), by \(19.57\) percentage points.
MiMo-V2.5 Pro obtains the lowest mean effectiveness at \(26.09\%\).
The remaining models achieve mean effectiveness values between \(30.43\%\) and \(38.41\%\).
Among the eight base LLMs, the four closed-source models achieve a mean effectiveness of \(40.04\%\), compared with \(35.33\%\) for the four open-weight models.
The difference is \(4.71\) percentage points.
This result indicates a modest aggregate advantage for the closed-source models on BAC repair, but does not establish their general superiority.
GLM-5.2 achieves \(41.30\%\), exceeding Claude Opus 4.8 and Gemini 3.5 Flash, both at \(30.43\%\), and slightly outperforming Qwen3.7-Max at \(38.41\%\).
In addition, open-weight models offer practical advantages for private deployment, task-specific adaptation, andlong-term reproducibility.

\begin{tcolorbox}
\textbf{Answer to RQ1:} LLM agents can repair a substantial fraction of real-world BAC vulnerabilities, reaching a mean effectiveness of \(77.90\%\), whereas single-turn base LLMs reach only \(37.68\%\), indicating that repository-level interaction is important for effective LLM-based BAC repair.
\end{tcolorbox}

\subsection{RQ2: What limitations do LLM exhibit when repairing web-application BAC vulnerabilities, and to what extent can these limitations be mitigated ?}

To answer RQ2, we analyze at which stage an ineffective patch fails.
Application of a patch is a prerequisite for the two behavioral tests, yet passing \texttt{git apply} verifies only format compliance rather than whether the patch repairs the vulnerability.
The PoC and functional tests instead evaluate the intended access-control behavior and preserved functionality required by Definition~\ref{def:bac-repair}.
We therefore distinguish \emph{Patch-Application Failures}, which occur at the application stage, from \emph{BAC Semantic Failures}, which surface in the PoC or functional tests.
\S5.2.1 examines the former and their mitigation through the \emph{Apply-Feedback} scaffold, while \S5.2.2 examines the latter and the mitigation provided by \emph{Navigate-and-Locate}.

\subsubsection{Patch-Application Failures and Apply-Feedback Mitigation}

\begin{table}[t]
    \centering
    \small
    \setlength{\tabcolsep}{12pt}
    \renewcommand{\arraystretch}{1.15}
    \caption{Stage-wise success rates of base LLMs in BAC repair.}
    \label{tab:stage-wise}
    
    \resizebox{1\linewidth}{!}{%

    \begin{tabular}{llcccc}
    \doubletoprule
    \rule{0pt}{2.8ex}
    \textbf{Model} &
    \textbf{Condition} &
    \textbf{Applicability} &
    \textbf{PoC} &
    \textbf{Functional} &
    \textbf{Effective} \\
    \midrule

    DeepSeek-V4 Pro
    & Base LLM
    & 60.14\% & 39.86\% & 49.28\% & 36.23\% \\

    \midrule

    Qwen3.7-Max
    & Base LLM
    & 62.32\% & 42.75\% & 47.83\% & 38.41\% \\

    \midrule

    Claude Opus 4.8
    & Base LLM
    & 42.03\% & 32.61\% & 38.41\% & 30.43\% \\
    \midrule

    GPT-5.5
    & Base LLM
    & 86.96\% & 68.12\% & 73.19\% & 60.87\% \\

    \midrule

    Gemini 3.5 Flash
    & Base LLM
    & 69.57\% & 39.13\% & 44.93\% & 30.43\% \\

    \midrule

    GLM-5.2
    & Base LLM
    & 64.49\% & 43.48\% & 55.07\% & 41.30\% \\

    \midrule

    MiMo-V2.5 Pro
    & Base LLM
    & 57.97\% & 33.33\% & 42.03\% & 26.09\% \\
    \midrule

    Kimi K2.7-Code
    & Base LLM
    & 56.52\% & 42.75\% & 44.93\% & 37.68\% \\

    \midrule
    \textbf{Macro-average}
    & Base LLM
    & \textbf{62.50\%} & 42.75\% & 49.46\% & \textbf{37.68\%} \\

    \doublebottomrule
    \end{tabular}%
    }
    
  \end{table}

Table~\ref{tab:stage-wise} shows that patch application is a substantial failure point for single-turn base LLM repair.
Across the eight base LLMs, the macro-average applicability is only \(62.50\%\), meaning that \(37.50\%\) of the generated patches fail before reaching PoC or functional evaluation. The problem is particularly severe for Claude Opus 4.8, whose applicability is \(42.03\%\).
An inapplicable patch cannot be materialized in the target repository, so its intended BAC repair remains unobservable through the downstream behavioral tests.

\begin{figure}
    \centering
    \includegraphics[width=0.92\linewidth]{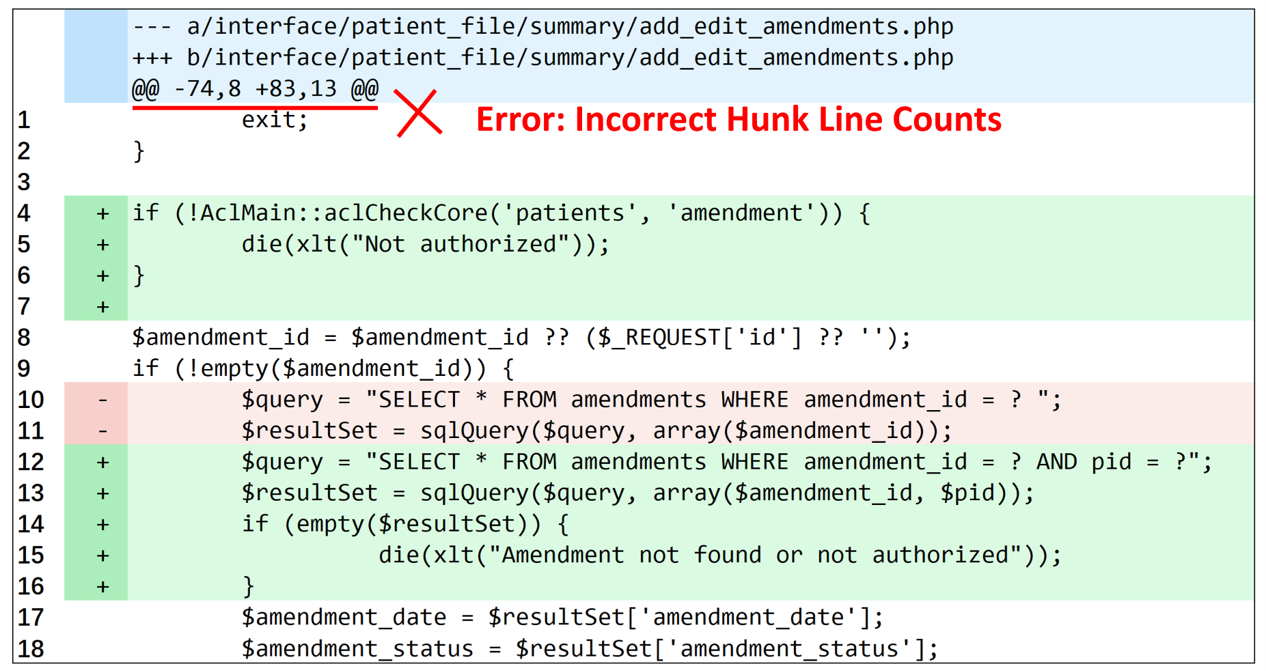}
    \caption{Case A: a patch generated by Qwen3.7-Max under Base condition for OpenEMR's CVE-2022-2730. The patch declares incorrect old and new hunk line counts, causing a hunk-header line-count mismatch.}
    \label{fig:figure-5-1}
\end{figure}

We manually inspected patches that failed to apply and identified several recurring patch-application failure patterns.
These included hunk-header range/count mismatches, malformed hunk-header syntax, hunk-context mismatches, and patch-path resolution failures.
There were also cases in which generated patches contained explanatory prose, ellipses, or message headers as non-comment text.
Case A (Figures~\ref{fig:figure-5-1}) and Case B (Figures~\ref{fig:figure-5-2}) respectively present representative examples, and Case C and Case D are provided in Appendix~\ref{app:patch-failures}.

\begin{figure}
    \centering
    \includegraphics[width=0.8\linewidth]{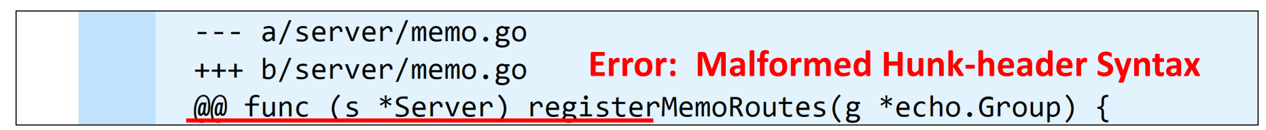}
    \caption{Case B: a patch generated by Opus4.8 under Base condition for Usememos's CVE-2022-4684. The patch uses a hunk header without the required old and new line ranges, resulting in malformed hunk-header syntax.}
    \label{fig:figure-5-2}
\end{figure}

As shown in Figure~\ref{fig:figure-5-1}, Case A exemplifies the hunk-header range/count mismatch pattern.
The Qwen3.7-Max patch for OpenEMR's CVE-2022-2730 declares "\(\texttt{\@\@ -74,8 +83,13 \@\@}\)" in its hunk header.
The header specifies 8 old-file lines and 13 new- file lines, whereas the hunk body actually contains 9 old-file lines and 16 new-file lines.
Although the hunk contains a plausible authorization check and a query restriction, this mismatch causes \texttt{git apply} to reject the file as a corrupt patch.
The case shows that a locally meaningful repair remains unevaluable when its diff metadata is invalid.
Case B corresponding to Figure~\ref{fig:figure-5-2} shows malformed hunk-header syntax.
The Opus4.8 patch for Usememos' CVE-2022-4684 writes the hunk header as \(\texttt{\@\@ func (s *Server) registerMemoRoutes(...)}\), omitting the old-file and new-file ranges required by the unified-diff format. Consequently, \texttt{git apply} reports \texttt{unrecognized input}.
Although the output conveys a human-readable source location, it is not a well-formed patch that \texttt{git apply} can consume.

Taken together, these cases show that many Patch-Application Failures stem not from misguided repair strategies but from hallucinated diff metadata.
When reconstructing a diff over hundreds of lines, an LLM may misremember hunk ranges, contextual lines, or path prefixes that \texttt{git apply} rejects deterministically.
This failure mode is therefore an application-stage bottleneck distinct from BAC authorization reasoning, and it should be recoverable once the model sees the tool's diagnostics.
To test this, we evaluate the \emph{Apply-Feedback} scaffold (Algorithm~\ref{alg:apply-feedback}), which applies the generated patch to the target repository and feeds the diagnostics from \texttt{git apply} back to the base LLM for revision.

\begin{figure}
    \centering
    \includegraphics[width=\linewidth]{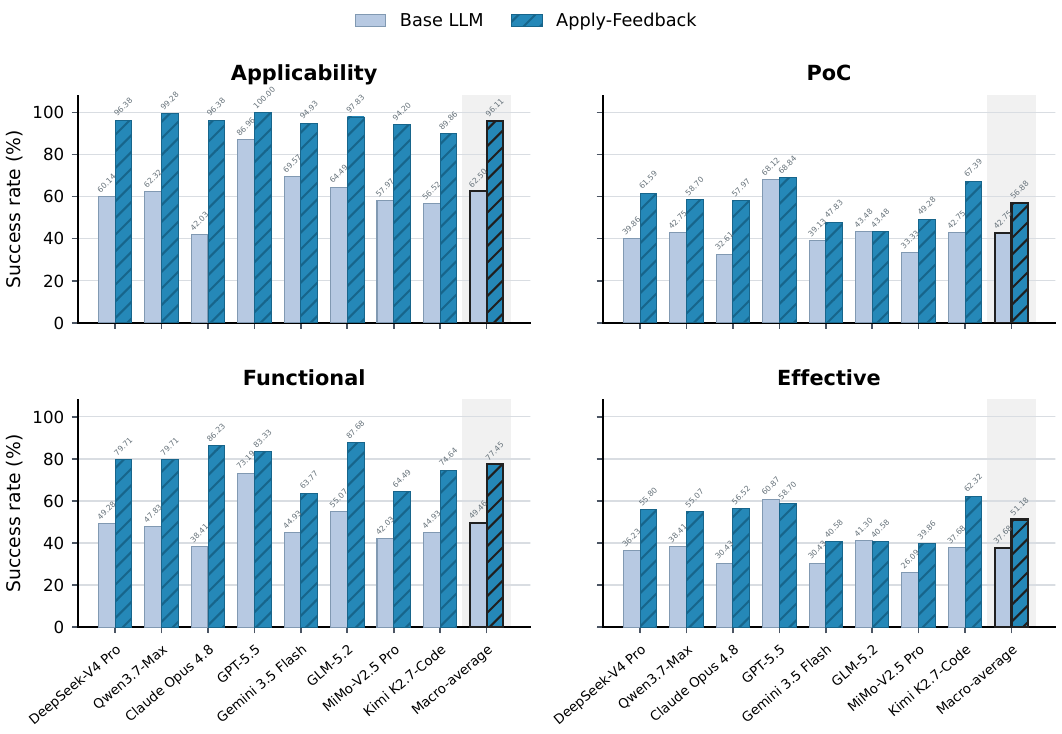}
    \caption{Repair performance under Base and Apply-Feedback conditions.}
    \label{fig:apply-feedback}
\end{figure}

Figure~\ref{fig:apply-feedback} reports the effect of the \emph{Apply-Feedback} scaffold.
The macro-average applicability rate increases from \(62.50\%\) under the Base LLM condition to \(96.11\%\) under Apply-Feedback, an improvement of \(33.61\) percentage points.
All eight models show substantial gains. Claude Opus 4.8 exhibits the largest increase, from \(42.03\%\) to \(96.38\%\), while GPT-5.5 reaches \(100.00\%\) applicability.
These results show that Base LLMs can produce applicable repository patches in most cases when they receive direct feedback from the patch-application tool.

The higher applicability is accompanied by higher downstream behavioral-test rates.
The macro-average PoC pass rate rises from \(42.75\%\) to \(56.88\%\), and the functional-test pass rate rises from \(49.46\%\) to \(77.45\%\).
These increases partly reflect the fact that more patches become eligible for behavioral evaluation once the application barrier is reduced.
The macro-average final effectiveness nevertheless increases by only \(13.50\) percentage points, from \(37.68\%\) to \(51.18\%\), which is substantially smaller than the gain of \(33.61\) percentage points in applicability.
This gap indicates that Apply-Feedback primarily mitigates Patch-Application Failures.
It does not by itself eliminate BAC Semantic Failures or resolve the underlying difficulty of constructing semantically correct BAC repairs.

\subsubsection{Semantic Failures and Navigate-and-Locate Mitigation}

After Apply-Feedback substantially reduces Patch-Application Failures, this subsection focuses on the remaining BAC Semantic Failures exposed by the dynamic validation tests.
Under the Apply-Feedback condition, the macro-average applicability rate reaches \(96.11\%\), whereas the PoC-based security-test and functional-test pass rates are \(56.88\%\) and \(77.45\%\), respectively.
The macro-average final effectiveness is only \(51.18\%\), which is \(44.93\) percentage points lower than the applicability rate.
This substantial gap motivate our analysis of the semantic limitations of LLM-generated BAC patches.

We manually audited patches that passed application-stage validation but failed at least one of the two dynamic validation tests.
The main failure patterns were semantic-contract misunderstandings, incomplete authorization-state reconstruction, enforcement-point mislocalization, and symptom-level repairs.
These patterns show that BAC repair requires repository-level reasoning about data definitions, call relationships, request handlers, authorization state, and state-transition logic rather than local code generation alone.
Case E (Figures~\ref{fig:figure-5-5}) presents a representative example, and Case F and Case G are provided in Appendix~\ref{app:semantic-failures}.

\begin{figure}
    \centering
    \includegraphics[width=0.94\linewidth]{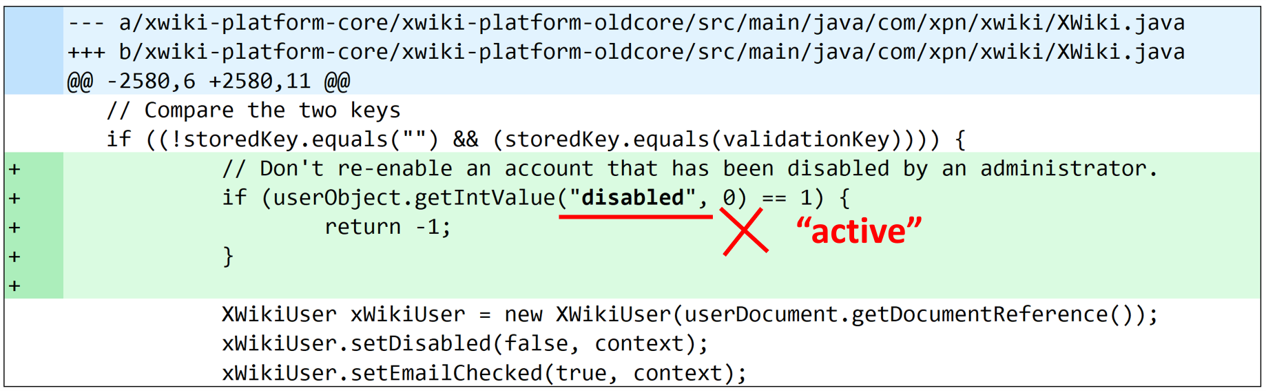}
    \caption{Case E: a patch generated by Kimi K2.7-Code under the Base LLM condition for XWiki's CVE-2021-32620. The patch checks the nonexistent \texttt{disabled} property instead of XWiki's \texttt{active} property, causing the account-state authorization check to fail.}
    \label{fig:figure-5-5}
\end{figure}

Figure~\ref{fig:figure-5-5} presents Case E, an instance of the semantic-contract misunderstanding pattern.
The Kimi K2.7-Code patch for XWiki's CVE-2021-32620 checks \texttt{userObject.getIntValue("disabled", 0) == 1} before re-enabling the account.
In the target XWiki version, however, the account state is represented by the \texttt{active} property, and an account is disabled when \texttt{active = 0}.
The relevant contract is defined by \texttt{XWikiUser.ACTIVE\_PROPERTY} in \texttt{XWikiUser.java}.
That file is not modified by the official patch and was therefore not included in the source context provided under the Base LLM and Apply-Feedback conditions.
Consequently, the model had no direct access to the actual field definition when generating the repair and had to infer the property name from incomplete context.
The resulting patch uses a nonexistent or semantically incorrect field, so its check does not take effect as intended.
This case illustrates how source-context restrictions can prevent an LLM from recovering the correct data-field contract.

  \begin{table}[t]
    \centering
    \small
    \setlength{\tabcolsep}{12pt}
    \renewcommand{\arraystretch}{1.15}

    \caption{Incremental effect of the \emph{Navigate-and-Locate} scaffold combined with \emph{Apply-Feedback}. Each cell reports the mean rate under the combined condition, followed in parentheses by the change in percentage points relative to the \emph{Apply-Feedback}-only condition.}
    \label{tab:navigation-increment}

    \resizebox{1\linewidth}{!}{%
      \begin{tabular}{lcccc}
        \doubletoprule

        \rule{0pt}{2.8ex}%
        \textbf{Model} &
        \textbf{Applicability} &
        \textbf{PoC} &
        \textbf{Functional} &
        \textbf{Effective} \\
        \midrule

        DeepSeek-V4 Pro
        & \metric{91.30\%}{-5.07}
        & \metric{65.22\%}{+3.62}
        & \metric{79.71\%}{+0.00}
        & \metric{62.32\%}{+6.52} \\

        Qwen3.7-Max
        & \metric{100.00\%}{+0.72}
        & \metric{76.09\%}{+17.39}
        & \metric{78.26\%}{-1.45}
        & \textbf{\metric{71.01\%}{+15.94}} \\

        Claude Opus 4.8
        & \metric{100.00\%}{+3.62}
        & \metric{77.54\%}{+19.57}
        & \metric{91.30\%}{+5.07}
        & \textbf{\metric{75.36\%}{+18.84}} \\

        GPT-5.5
        & \metric{86.23\%}{-13.77}
        & \metric{81.16\%}{+12.32}
        & \metric{76.81\%}{-6.52}
        & \textbf{\metric{73.19\%}{+14.49}} \\

        Gemini 3.5 Flash
        & \metric{87.68\%}{-7.25}
        & \metric{73.91\%}{+26.09}
        & \metric{73.19\%}{+9.42}
        & \metric{64.49\%}{+23.91} \\

        GLM-5.2
        & \metric{97.83\%}{+0.00}
        & \metric{69.57\%}{+26.09}
        & \metric{81.88\%}{-5.80}
        & \metric{63.77\%}{+23.19} \\

        MiMo-V2.5 Pro
        & \metric{98.55\%}{+4.35}
        & \metric{69.57\%}{+20.29}
        & \metric{74.64\%}{+10.14}
        & \metric{60.14\%}{+20.29} \\

        Kimi K2.7-Code
        & \metric{85.51\%}{-4.35}
        & \metric{65.94\%}{-1.45}
        & \metric{71.01\%}{-3.62}
        & \metric{57.97\%}{-4.35} \\

        \midrule

        \textbf{Macro-average}
        & \textbf{\metric{93.39\%}{-2.72}}
        & \textbf{\metric{72.37\%}{+15.49}}
        & \textbf{\metric{78.35\%}{+0.91}}
        & \textbf{\metric{66.03\%}{+14.86}} \\

        \doublebottomrule
      \end{tabular}%
    }
  \end{table}

These cases expose the limitations of restricting the Base LLM's source context to the pre-patch source files touched by the official patch, thereby hindering the LLM's reconstruction of the relevant application logic and authorization semantics.
To assess whether repository-level source exploration can mitigate these BAC Semantic Failures, we evaluate the \emph{Navigate-and-Locate} scaffold introduced in \S~\ref{subsubsec:scaffolds} and formalized in Algorithm~\ref{alg:navigate-locate}.
The scaffold exposes the repository tree and allows the LLM to request source files on demand.
We compose it with \emph{Apply-Feedback} and report the results for the resulting combined condition in Table~\ref{tab:navigation-increment}.
The macro-average final effectiveness increases from \(51.18\%\) under Apply-Feedback to \(66.03\%\), a gain of \(14.86\) percentage points.
Relative to the Base LLM condition, the combined scaffolds improve final effectiveness by \(28.35\) percentage points.
The largest improvement occurs for the PoC-based security test, whose macro-average pass rate rises from \(56.88\%\) to \(72.37\%\), an increase of \(15.49\) percentage points.
In contrast, the functional-test pass rate increases only from \(77.45\%\) to \(78.35\%\), a change of \(0.91\) percentage points.
The combined condition also has a slightly lower macro-average applicability rate than Apply-Feedback, decreasing from \(96.11\%\) to \(93.39\%\).

Seven of the eight models improve under the combined condition. Claude Opus 4.8 achieves the highest final effectiveness at \(75.36\%\), followed by GPT-5.5 at \(73.19\%\) and Qwen3.7-Max at \(71.01\%\).
Claude Opus 4.8's \(75.36\%\) is only \(3.63\) percentage points below the \(78.99\%\) achieved by Claude Code, a closed-source agent based on Opus 4.8, while GPT-5.5's \(73.19\%\) is \(3.62\) percentage points below the \(76.81\%\) achieved by Codex, a closed-source agent based on GPT-5.5.
Claude Opus 4.8 is particularly illustrative and exhibits the largest absolute gain among the eight Base LLMs.
Its final effectiveness rises from \(30.43\%\) in the file-restricted Base LLM condition to \(75.36\%\) under the combined scaffold, an increase of \(44.93\) percentage points.
This result shows not only that providing sufficient repository context and interaction can substantially improve an LLM's understanding of the BAC vulnerability, the relevant application logic, and its access-control semantics, thereby enabling more effective repository-level repairs, but also that the two lightweight scaffolds introduced in this work can substantially improve Base LLM repair effectiveness and mitigate the limitations.

Grouping the models into closed-source and open-weight categories, the combined condition also reveals a consistent advantage for the evaluated closed-source models.
The four evaluated closed-source Base LLMs achieve a mean final effectiveness of \(71.01\%\), compared with \(61.05\%\) for the four evaluated open-weight Base LLMs, a difference of \(9.96\) percentage points.

\begin{tcolorbox}
\textbf{Answer to RQ2:} LLM-based BAC repair is limited primarily by patch-application errors and insufficient repository context for reconstructing the repository-level authorization semantics.
The \emph{Apply-Feedback} scaffold raises the mean applicability rate from $62.50\%$ to $96.11\%$ and the mean final effectiveness from $37.68\%$ to $51.18\%$.
Further composing it with the \emph{Navigate-and-Locate} scaffold increases the mean final effectiveness to $66.03\%$, with the best-performing model, Claude Opus 4.8, reaching $75.36\%$. These results show that lightweight scaffolding can substantially mitigate both limitations.
\end{tcolorbox}

\subsection{RQ3: Which repair-strategy choices underlie the failures of LLM-generated BAC patches?}

\subsubsection{Failure-Dimension Analysis}

The PoC-based security test and the functional test can determine whether a patch exhibits the desired behavior on the tested scenarios, but they cannot explain why a semantically failing patch fails or which aspect of its repair strategy is responsible.
We therefore apply the BAC Repair-Strategy Framework defined in \S~\ref{sec:framework} to characterize the observable failure dimensions of LLM-generated patches.

\begin{table}[!t]
\centering
\scriptsize
\setlength{\tabcolsep}{6pt}
\renewcommand{\arraystretch}{1.15}
\caption{Failure prevalence across the four dimensions of the BAC repair-strategy framework among all LLM-generated patches that pass application and compilation when applicable. Only the \(810\) dynamically failing patches are analyzed under the framework, whereas the \(1{,}925\) dynamically successful patches are treated as successful in all four dimensions by convention for aggregate prevalence estimation. Percentages use the corresponding \(N\) as the denominator and are pooled across the three runs. Because a patch may fail in multiple dimensions, the four prevalence values are not mutually exclusive and need not sum to \(100\%\). AF denotes \emph{Apply-Feedback}, and NL denotes \emph{Navigate-and-Locate}.}
\label{tab:all-eligible-failure-profiles}
\resizebox{0.96\linewidth}{!}{%
\begin{tabular}{llrrrrr}
\doubletoprule
\rule{0pt}{2.8ex}%
\textbf{Model} &
\textbf{Condition} &
\(\boldsymbol{N}\) &
\textbf{Position} &
\textbf{Coverage} &
\textbf{Mechanism} &
\textbf{Content} \\
\midrule

Claude Code & LLM Agent
& 133 & 3.01\% & 9.02\% & 8.27\% & 14.29\% \\

Codex & LLM Agent
& 132 & 2.27\% & 3.03\% & 5.30\% & 11.36\% \\

\midrule

DeepSeek-V4 Pro & Base LLM
& 72 & 5.56\% & 16.67\% & 13.89\% & 26.39\% \\

Qwen3.7-Max & Base LLM
& 73 & 5.48\% & 10.96\% & 20.55\% & 24.66\% \\

Claude Opus 4.8 & Base LLM
& 55 & 1.82\% & 14.55\% & 9.09\% & 21.82\% \\

GPT-5.5 & Base LLM
& 114 & 7.02\% & 8.77\% & 19.30\% & 14.91\% \\

Gemini3.5 Flash & Base LLM
& 72 & 5.56\% & 13.89\% & 26.39\% & 30.56\% \\

GLM-5.2 & Base LLM
& 80 & 2.50\% & 5.00\% & 20.00\% & 22.50\% \\

MiMo-V2.5 Pro & Base LLM
& 67 & 8.96\% & 20.90\% & 32.84\% & 31.34\% \\

Kimi K2.7-Code & Base LLM
& 69 & 1.45\% & 1.45\% & 15.94\% & 21.74\% \\

\midrule
\multicolumn{2}{l}{\textbf{All Base LLMs}}
& \textbf{602} & \textbf{4.98\%} & \textbf{11.13\%}
& \textbf{19.93\%} & \textbf{23.59\%} \\

\midrule

DeepSeek-V4 Pro & AF
& 117 & 9.40\% & 15.38\% & 23.08\% & 30.77\% \\

Qwen3.7-Max & AF
& 117 & 5.98\% & 10.26\% & 29.91\% & 28.21\% \\

Claude Opus 4.8 & AF
& 121 & 18.18\% & 27.27\% & 13.22\% & 33.88\% \\

GPT-5.5 & AF
& 131 & 9.16\% & 11.45\% & 29.77\% & 19.85\% \\

Gemini3.5 Flash & AF
& 100 & 6.00\% & 16.00\% & 30.00\% & 38.00\% \\

GLM-5.2 & AF
& 128 & 9.38\% & 21.09\% & 44.53\% & 44.53\% \\

MiMo-V2.5 Pro & AF
& 103 & 11.65\% & 23.30\% & 29.13\% & 38.83\% \\

Kimi K2.7-Code & AF
& 111 & 3.60\% & 8.11\% & 16.22\% & 19.82\% \\

\midrule
\multicolumn{2}{l}{\textbf{All AF}}
& \textbf{928} & \textbf{9.27\%} & \textbf{16.59\%}
& \textbf{27.16\%} & \textbf{31.57\%} \\

\midrule

DeepSeek-V4 Pro & AF + NL
& 117 & 5.98\% & 7.69\% & 13.68\% & 20.51\% \\

Qwen3.7-Max & AF + NL
& 116 & 0.86\% & 1.72\% & 8.62\% & 12.93\% \\

Claude Opus 4.8 & AF + NL
& 130 & 3.08\% & 8.46\% & 5.38\% & 16.15\% \\

GPT-5.5 & AF + NL
& 117 & 0.85\% & 1.71\% & 1.71\% & 5.98\% \\

Gemini3.5 Flash & AF + NL
& 112 & 3.57\% & 0.89\% & 11.61\% & 15.18\% \\

GLM-5.2 & AF + NL
& 122 & 6.56\% & 9.02\% & 17.21\% & 26.23\% \\

MiMo-V2.5 Pro & AF + NL
& 117 & 7.69\% & 7.69\% & 23.93\% & 26.50\% \\

Kimi K2.7-Code & AF + NL
& 109 & 6.42\% & 7.34\% & 10.09\% & 22.94\% \\

\midrule
\multicolumn{2}{l}{\textbf{All AF + NL}}
& \textbf{940} & \textbf{4.36\%} & \textbf{5.64\%}
& \textbf{11.49\%} & \textbf{18.30\%} \\

\midrule
\multicolumn{2}{l}{\textbf{All}}
& \textbf{2735} & \textbf{6.00\%} & \textbf{10.60\%}
& \textbf{18.21\%} & \textbf{23.44\%} \\

\doublebottomrule
\end{tabular}%
}
\end{table}

Table~\ref{tab:all-eligible-failure-profiles} reports the four-dimensional failure prevalence among all \(2{,}735\) LLM-generated patches that pass application and build validation.
Across all of these patches, Content has the highest failure prevalence at \(23.44\%\), followed by Mechanism at \(18.21\%\), Coverage at \(10.60\%\), and Position at \(6.00\%\).
This ordering indicates that the dominant limitation lies in reconstructing or expressing the complete authorization policy required by the target business scenario and in implementing that policy through a reliable, project-compatible mechanism, rather than in identifying a plausible repair location.

Under the Base condition, Content is the most prevalent failure dimension for six of the eight LLMs, whereas Mechanism is more prevalent for GPT-5.5 and MiMo-V2.5 Pro, and Position is the least prevalent failure dimension.
The model-level profiles reveal distinct failure tendencies. GPT-5.5 has the lowest Content failure prevalence among the Base LLMs at \(14.91\%\), although its Mechanism failure prevalence remains \(19.30\%\).
This profile indicates a relative advantage in reconstructing the intended authorization relationship, but not necessarily in implementing it through the application's existing security facilities.
Kimi K2.7-Code has the lowest Position and Coverage failure prevalences, both at \(1.45\%\), suggesting that its patches less frequently target unrelated locations or omit structurally related vulnerability units.
Its Mechanism and Content failure prevalences remain \(15.94\%\) and \(21.74\%\), respectively, showing that broad structural coverage does not guarantee a correct authorization policy or a reliable enforcement mechanism.
Claude Opus 4.8 has the lowest Mechanism failure prevalence among the Base LLMs at \(9.09\%\), together with a low Position failure prevalence of \(1.82\%\), but its Content failure prevalence remains \(21.82\%\).
MiMo-V2.5 Pro has the highest Coverage, Mechanism, and Content failure prevalences among the Base LLMs, at \(20.90\%\), \(32.84\%\), and \(31.34\%\), respectively.
These profiles show that the models do not fail uniformly. Some more often reconstruct a plausible authorization policy but implement it unreliably, whereas others more frequently omit related vulnerability units or fail to recover the required policy semantics.

The experimental conditions further clarify the roles of the two scaffolds.
Under the Base LLM condition, the failure prevalences are \(4.98\%\) for Position, \(11.13\%\) for Coverage, \(19.93\%\) for Mechanism, and \(23.59\%\) for Content.
Under \emph{Apply-Feedback}, the corresponding values increase to \(9.27\%\), \(16.59\%\), \(27.16\%\), and \(31.57\%\).
This increase should not be interpreted as evidence that \emph{Apply-Feedback} worsens semantic repair.
The scaffold increases the number of patches that pass application and build validation from \(602\) to \(928\), thereby admitting patches that would otherwise have been excluded by Patch-Application Failures.
The newly admitted patches can contain weaker semantic repairs, which raises the observed prevalence of semantic defects.

Adding \emph{Navigate-and-Locate} yields a clear improvement relative to \emph{Apply-Feedback} alone.
The combined condition reduces the failure prevalence from \(9.27\%\) to \(4.36\%\) for Position, from \(16.59\%\) to \(5.64\%\) for Coverage, from \(27.16\%\) to \(11.49\%\) for Mechanism, and from \(31.57\%\) to \(18.30\%\) for Content.
Mechanism shows the largest absolute reduction of \(15.67\) percentage points, corresponding to a \(57.69\%\) relative reduction.
Coverage exhibits the largest relative reduction, decreasing by \(10.96\) percentage points, or \(66.02\%\) relative to its \emph{Apply-Feedback} baseline.
Equivalently, the \emph{Apply-Feedback} prevalence is \(2.36\) times the combined-condition prevalence for Mechanism and \(2.94\) times that for Coverage. Position and Content decrease by \(52.93\%\) and \(42.05\%\), respectively.
These results suggest that repository exploration helps models identify and structurally address a more complete set of vulnerable paths, including other exploitable sites that share the same root cause, while also enabling them to select and correctly reuse the application's existing authorization facilities.
The lower Content prevalence further suggests improved reconstruction of the target authorization policy, although Content remains the most prevalent failure dimension under the combined condition.

The combined condition also narrows the gap between the Base LLMs and the closed-source agents.
Claude Code reports failure prevalences of \(3.01\%\), \(9.02\%\), \(8.27\%\), and \(14.29\%\) for Position, Coverage, Mechanism, and Content, respectively, while Codex reports \(2.27\%\), \(3.03\%\), \(5.30\%\), and \(11.36\%\). The combined Base LLM condition reaches \(4.36\%\), \(5.64\%\), \(11.49\%\), and \(18.30\%\) for the same dimensions.
In particular, GPT-5.5 under the combined condition has failure prevalences of \(0.85\%\) for Position, \(1.71\%\) for Coverage, \(1.71\%\) for Mechanism, and \(5.98\%\) for Content, closely approaching the Codex profile.
Claude Opus 4.8 likewise approaches Claude Code in Content and Mechanism, with corresponding prevalences of \(16.15\%\) and \(5.38\%\).
These results indicate that the two lightweight scaffolds recover part of the repository-level interaction available to stronger LLM agents, although they do not eliminate the remaining difficulty of reconstructing complete BAC authorization semantics.

\subsubsection{Dynamic Failure Patterns}

\begin{table}[t]
    \centering
    \small
    \setlength{\tabcolsep}{12pt}
    \renewcommand{\arraystretch}{1.15}

    \caption{Conditional failure prevalence across the four dimensions of the BAC repair-strategy framework within each dynamic-failure category. \cmark\ and \xmark\ denote passing and failing a test, respectively. Percentages in each row use the corresponding category size \(N\) as the denominator, and a patch may fail in multiple
    dimensions.}
    \label{tab:dynamic-failure-types}

    \resizebox{1\linewidth}{!}{%
      \begin{tabular}{lrrrrr}
        \doubletoprule

        \rule{0pt}{2.8ex}%
        \textbf{Dynamic-failure type} &
        \textbf{\(N\)} &
        \textbf{Position} &
        \textbf{Coverage} &
        \textbf{Mechanism} &
        \textbf{Content} \\
        \midrule

        PoC \xmark,\ functional \cmark
        & 588
        & 23.98\%
        & 45.92\%
        & 61.05\%
        & 81.12\% \\

        PoC \cmark,\ functional \xmark
        & 169
        & 8.88\%
        & 9.47\%
        & 55.62\%
        & 72.19\% \\

        PoC \xmark,\ functional \xmark
        & 53
        & 15.09\%
        & 7.55\%
        & 84.91\%
        & 79.25\% \\

        \doublebottomrule
      \end{tabular}%
    }
  \end{table}

We next group the \(810\) dynamically failing patches by their dynamic-failure pattern: PoC failure with functional-test pass, PoC pass with functional-test failure, and failure of both tests, as illustrated in Table~\ref{tab:dynamic-failure-types}.
Content and Mechanism exhibit the highest failure proportions among the four dimensions, consistent with the corresponding pattern reported in Table~\ref{tab:all-eligible-failure-profiles}.
The PoC-pass and functional-failure category has the lowest Position and Coverage failure proportions, at \(8.88\%\) and \(9.47\%\), respectively. These patches often reach the correct repair location and cover the relevant vulnerability paths, but impose an authorization restriction that is too strong or otherwise inconsistent with legitimate application behavior. Their Content failure proportion remains high at \(72.19\%\), indicating that blocking the tested exploit may be achieved at the cost of rejecting legitimate subjects or actions.

Moreover, the PoC-failure and functional-pass category has the highest Coverage failure proportion, at \(45.92\%\).
This pattern is consistent with patches that omit one or more vulnerable paths, actions, or sibling resource branches.
Finally, the category in which both tests fail has the highest Mechanism failure proportion, at \(84.91\%\).
This suggests that an incorrect authorization facility, API contract, query, cache interaction, exception path, or control flow can simultaneously fail to block the attack and disrupt normal functionality.

\subsubsection{Annotation Reliability}

  \begin{table}[t]
      \centering
      \small
      \setlength{\tabcolsep}{16pt}
      \renewcommand{\arraystretch}{1.15}

      \caption{Reliability of the LLM-assisted static audit. The table
      reports, for each dimension, the agreement between the labels
      assigned by the LLM and the majority-vote consensus of three
      blinded cybersecurity PhD annotators on a stratified sample of
      108 patches. Each dimension contains 108 patch-dimension cells,
      yielding 432 cells overall. \emph{Corr.} denotes the number of
      LLM-assigned labels that disagree with the majority consensus.}

      \label{tab:audit-reliability}

      \resizebox{0.72\linewidth}{!}{%
        \begin{tabular}{lrrr}
          \doubletoprule

          \rule{0pt}{2.8ex}%
          \textbf{Dimension} &
          \textbf{Agreement} &
          \textbf{Cohen's \(\boldsymbol{\kappa}\)} &
          \textbf{Corr.} \\
          \midrule

          Position
          & 98.1\%
          & 0.955
          & 2 \\

          Coverage
          & 100.0\%
          & 1.000
          & 0 \\

          Mechanism
          & 94.4\%
          & 0.874
          & 6 \\

          Content
          & 92.6\%
          & 0.781
          & 8 \\

          \midrule

          \textbf{Overall}
          & \textbf{96.3\%}
          & \textbf{0.926}
          & \textbf{16} \\

          \doublebottomrule
        \end{tabular}%
      }
  \end{table}

The four-dimensional failure prevalences reported above rest on labels produced by the LLM-assisted static audit (\S4.4).
To validate this labeling instrument, three PhD students specializing in cybersecurity independently labeled a stratified random sample of 108 patches, corresponding to 13.3\% of the 810 dynamically failing patches and covering all six applications, all ten evaluated systems, and all three experimental conditions.
The auditors followed the coding rules in Table~\ref{tab:framework-operationalization} and were blinded to the LLM-assigned labels and to one another.
For each patch-dimension cell, the human consensus was determined by majority vote. Table~\ref{tab:audit-reliability} reports the agreement between the LLM-assigned labels and the majority-vote consensus.
Overall, the LLM labels agree with the human consensus on \(96.3\%\) of the audited cells (Cohen's \(\kappa = 0.926\)), with 16 disagreements among 432 cells.

\begin{tcolorbox}
\textbf{Answer to RQ3:} LLM-generated BAC patches fail across all four dimensions of the repair-strategy framework, with Content and Mechanism being the most prevalent failure dimensions, followed by Coverage and Position. The results indicate that the main difficulty lies in reconstructing the intended authorization policy and implementing it through a reliable, project-compatible mechanism, rather than merely locating a plausible repair site.
\end{tcolorbox}

\subsection{RQ4: Are Official Patches Always Better than LLM-Generated Patches?}

Although the preceding analysis identifies the dominant failure dimensions of LLM-generated BAC patches, it does not establish that official patches are always semantically correct or uniformly superior.
We therefore compare official and LLM-generated patches under the same dynamic validation protocol and using the BAC repair-strategy framework. Our audit identified defects in several official patches, as summarized in Table~\ref{tab:defective-official-patches}.
We next examine three representative cases, namely, CVE-2022-4505, CVE-2023-2946, and CVE-2023-31066, which illustrate distinct ways in which some LLM-generated patches outperform the corresponding official patches.

  \begin{table}[t]
    \centering
    \small
    \setlength{\tabcolsep}{15pt}
    \renewcommand{\arraystretch}{1.18}

    \caption{Official patches with defects. \cmark\ and \xmark\ denote passing and failing a test, respectively, whereas \partialmark\ indicates that a better-performing patch exists}
    \label{tab:defective-official-patches}

    \resizebox{0.82\linewidth}{!}{%
      \begin{tabular}{llcc}
        \doubletoprule

        \multirow{2}{*}{\textbf{Vulnerability}} &
        \multirow{2}{*}{\textbf{Application}} &
        \multicolumn{2}{c}{\textbf{Official patch}} \\
        \cmidrule(lr){3-4}

        & &
        \textbf{PoC test} &
        \textbf{Functional test} \\
        \midrule

        CVE-2022-4505
        & OpenEMR
        & \cmark
        & \textbf{\xmark} \\

        CVE-2023-2946
        & OpenEMR
        & \cmark
        & \textbf{\xmark} \\

        \addlinespace[2pt]

        CVE-2022-4684
        & Usememos
        & \textbf{\xmark}
        & \cmark \\

        CVE-2022-4802
        & Usememos
        & \textbf{\xmark}
        & \textbf{\xmark} \\

        CVE-2022-4803
        & Usememos
        & \textbf{\xmark}
        & \cmark \\

        \addlinespace[2pt]

        CVE-2023-31066
        & Apache InLong
        & \partialmark
        & \cmark \\

        \doublebottomrule
      \end{tabular}%
    }
  \end{table}

\begin{figure}[t]
      \centering
      \includegraphics[width=0.94\linewidth]{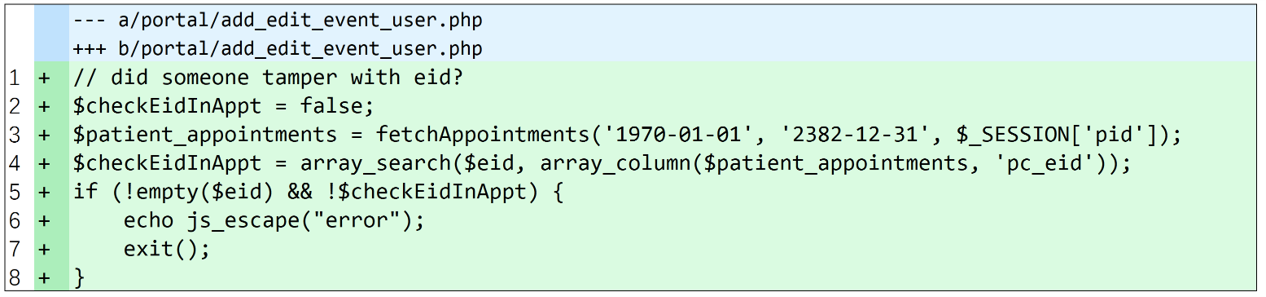}
      \caption{Excerpt illustrating the core logic of OpenEMR's official patch for CVE-2022-4505 (commit \texttt{235b1910}), released on August 20, 2022~\cite{openemr4505initial}. The patch checks whether the requested appointment belongs to the authenticated patient's appointment set, but incorrectly interprets the index returned by \texttt{array\_search()} as a Boolean value.}
      \label{fig:figure-5.4-1}
  \end{figure}

  \begin{figure}[t]
      \centering
      \includegraphics[width=0.7\linewidth]{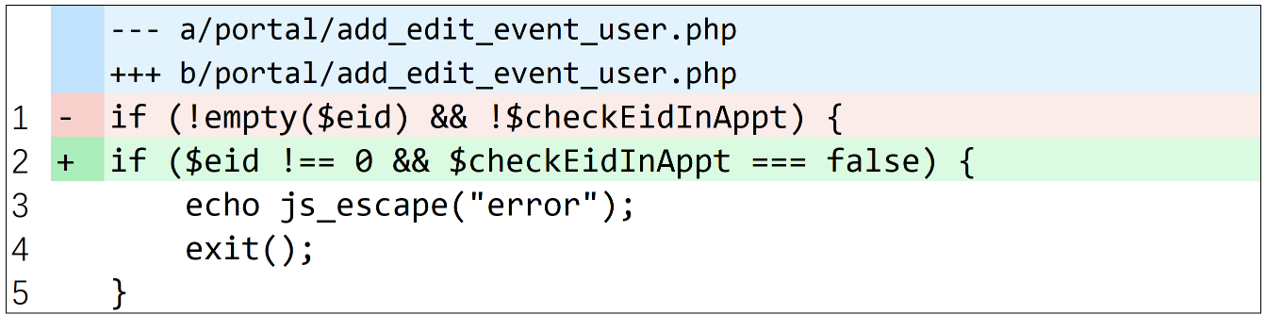}
      \caption{Excerpt illustrating OpenEMR's follow-up repair for CVE-2022-4505 (commit \texttt{210a63b4}), released on August 8, 2024~\cite{openemr4505followup}. The strict comparison with \texttt{false} distinguishes a valid match at index \(0\) from the absence of a match. }
      \label{fig:figure-5.4-2}
  \end{figure}

  \begin{figure}[t]
      \centering
      \includegraphics[width=0.94\linewidth]{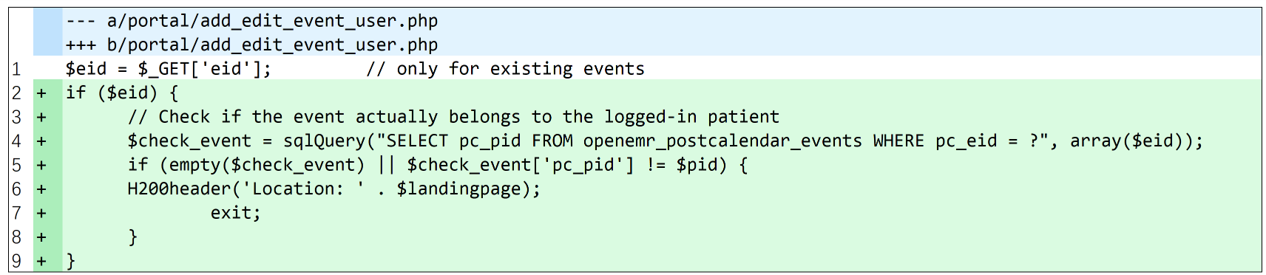}
      \caption{Excerpt illustrating an effective patch generated by Gemini 3.5 Flash for CVE-2022-4505, which validates appointment ownership directly through an SQL query.}
      \label{fig:figure-5.4-3}
  \end{figure}

\textbf{CVE-2022-4505.}
In OpenEMR's CVE-2022-4505, a portal patient can gain unauthorized access to another patient's appointment by modifying the request parameter \texttt{eid}.
Figure~\ref{fig:figure-5.4-1} shows an excerpt from the official patch (commit \texttt{235b1910}), released on August 20, 2022 ~\cite{openemr4505initial}.
The patch uses the patient identifier stored in \texttt{\$\_SESSION['pid']} to retrieve the patient's appointments through the existing project facility \texttt{fetchAppointments()}, and then uses PHP's built-in \texttt{array\_search()} function to determine whether the requested appointment belongs to that set.
This implements the intended owner-based authorization policy.
However, the patch directly interprets the index returned by \texttt{array\_search()} as a Boolean value.
If a legitimate appointment is located at index \(0\), the function returns \(0\), which is interpreted as false and causes the request to be rejected. The official patch therefore blocks the reported unauthorized access but can also deny legitimate owner access.
This implementation defect remained in the official repair for nearly two years before OpenEMR corrected it on August 8, 2024 by strictly comparing the return value with \texttt{false}, as illustrated by the follow-up patch (commit \texttt{210a63b4}) in Figure~\ref{fig:figure-5.4-2}~\cite{openemr4505followup}.

The BAC repair-strategy framework distinguishes the intended authorization policy from its implementation.
The original official patch has an accurate Position, complete Coverage, and correct Content, but its Mechanism fails because it violates the return-value contract of \texttt{array\_search()}.
Some LLM-generated patches avoid this particular pitfall through alternative mechanisms. For example, the patch generated by Gemini 3.5 Flash in Figure~\ref{fig:figure-5.4-3} relies on neither \texttt{fetchAppointments()} nor \texttt{array\_search()}.
Instead, it issues a parameterized SQL query to retrieve the authoritative \texttt{pc\_pid} of the appointment identified by \texttt{eid} and compares it directly with the authenticated patient identifier.
It rejects nonexistent appointments and ownership mismatches before loading the appointment, thereby implementing the same owner-based policy without an index-valued intermediate result. The patch therefore constitutes a reliable project-native Mechanism and passes both dynamic tests.
This case demonstrates that an LLM-generated patch can preserve the intended authorization semantics while avoiding a mechanism-level defect in the corresponding official repair.

\begin{figure}[t]
  \centering
  \includegraphics[width=0.86\linewidth]{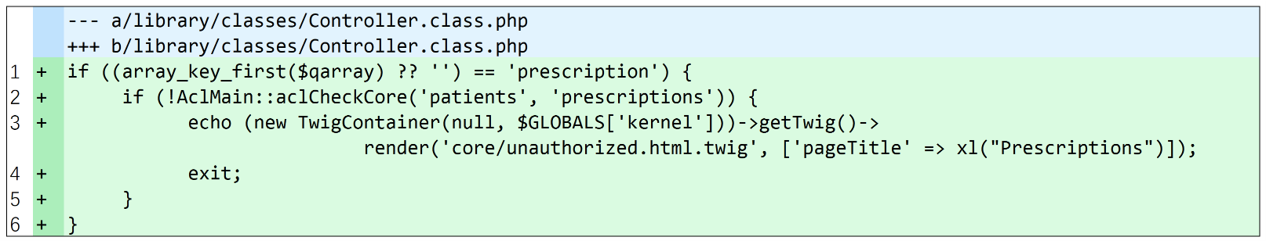}
  \caption{Excerpt from the original official patch for CVE-2023-2946
    (commit \texttt{81832ac}), released on March 14, 2023
    ~\cite{openemr2946initial}. The patch incorrectly invokes the central ACL facility with the ACO name \texttt{prescriptions} rather than its configured value \texttt{rx}.}
  \label{fig:figure-5.4-4}
\end{figure}

\begin{figure}[t]
  \centering
  \includegraphics[width=0.8\linewidth]{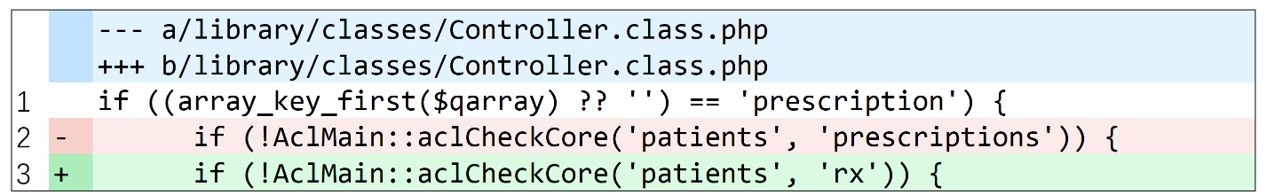}
  \caption{Excerpt illustrating the follow-up official patch for CVE-2023-2946 (commit \texttt{a8ed9cd}), released on May 12, 2023 ~\cite{openemr2946followup}. The correction replaces the invalid ACO name \texttt{prescriptions} with the configured value \texttt{rx}.}
  \label{fig:figure-5.4-5}
\end{figure}

\begin{figure}[t]
  \centering
  \includegraphics[width=0.8\linewidth]{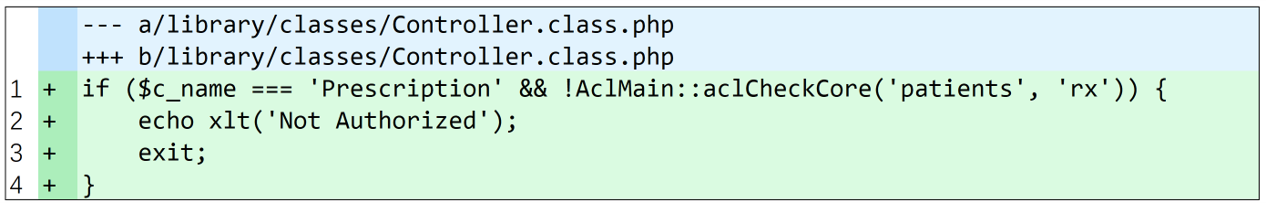}
  \caption{Excerpt illustrating an effective Kimi K2.7-Code-generated patch for CVE-2023-2946. The patch guards the prescription controller using the correct \texttt{('patients', 'rx')} ACL coordinate.}
  \label{fig:figure-5.4-6}
\end{figure}

\textbf{CVE-2023-2946.}
In OpenEMR's CVE-2023-2946, the prescription module lacks an appropriate authorization gate, allowing any authenticated user, including low-privilege Front Office and Receptionist users, to view, create, or modify prescriptions for arbitrary patients.
Figure~\ref{fig:figure-5.4-4} shows an excerpt from the official patch (commit \texttt{81832ac}), released on March 14, 2023 ~\cite{openemr2946initial}.
The patch attempts to enforce the prescription permission through OpenEMR's phpGACL-based authorization mechanism by invoking the central \texttt{AclMain::aclCheckCore()} facility.
In terms of authorization \emph{Content}, it targets the intended prescription access-control object (ACO) in the \texttt{patients} section.
However, the patch confuses the ACO's \texttt{name} with its lookup \texttt{value} and invokes \texttt{AclMain::aclCheckCore('patients', 'prescriptions')}. OpenEMR maintains its ACL configuration across multiple tables, including \texttt{gacl\_aco\_map}.
In this table, \texttt{prescriptions} is the ACO's \texttt{name}, whereas \texttt{rx} is its \texttt{value}. The configured section/value pair for the prescription ACO is therefore \texttt{('patients', 'rx')}.
The pair supplied by the official patch, \texttt{('patients', 'prescriptions')}, does not identify this configured ACO.
Because \texttt{AclMain::aclCheckCore()} resolves the ACO using its configured \texttt{value}, the call returns false for legitimate physicians. The official patch therefore blocks the reported unauthorized access but also denies legitimate prescription operations.

An OpenEMR community user reported this regression on May 11, 2023, under the title ``Physician Can't Fill Up Prescriptions''~\cite{openemr2946forum}. OpenEMR released a corrective patch the following day by replacing \texttt{prescriptions} with \texttt{rx}, as illustrated by the follow-up patch (commit \texttt{a8ed9cd}) in Figure~\ref{fig:figure-5.4-5}~\cite{openemr2946followup}.
Under the BAC repair-strategy framework, the official patch targets the intended authorization Content and places the check at a relevant Position, but its Mechanism fails because it violates the identifier contract of the existing phpGACL facility.
Several effective LLM-generated patches avoid this error by invoking \texttt{AclMain::aclCheckCore()} with the correct \texttt{('patients', 'rx')} section/value pair.
For example, the Kimi K2.7-Code-generated patch excerpted in Figure~\ref{fig:figure-5.4-6} inserts a concise guard before loading the prescription controller and uses the correct ACO \texttt{value},which passes both dynamic tests.
This case demonstrates that some LLM-generated patches can correctly reuse a project's standard authorization mechanism while avoiding an identifier-contract error in the corresponding official repair.

\textbf{CVE-2023-31066.}
Apache InLong's CVE-2023-31066 is an object-level broken access-control vulnerability in \texttt{StreamSource} management.
A user can tamper with the source identifier in a deletion request and illegally \texttt{delete} a \texttt{StreamSource} belonging to another user.
The same missing group-level authorization check affects the corresponding \texttt{stop} and \texttt{start/restart} paths.
In the vulnerable service implementation, these methods resolve the target source from a caller-controlled identifier and perform the state-changing operation without verifying whether the caller is authorized to operate on the source's owning InLong group.
The intended policy permits the group's designated \texttt{inCharges} users and administrators to perform these operations.
To address this vulnerability, the official patch (commit \texttt{231dbd5a}), released on April 4, 2023~\cite{inlong31066initial} adds group-level authorization checks to the three affected service paths and is effective.

\begin{figure}[t]
  \centering
  \includegraphics[width=0.94\linewidth]{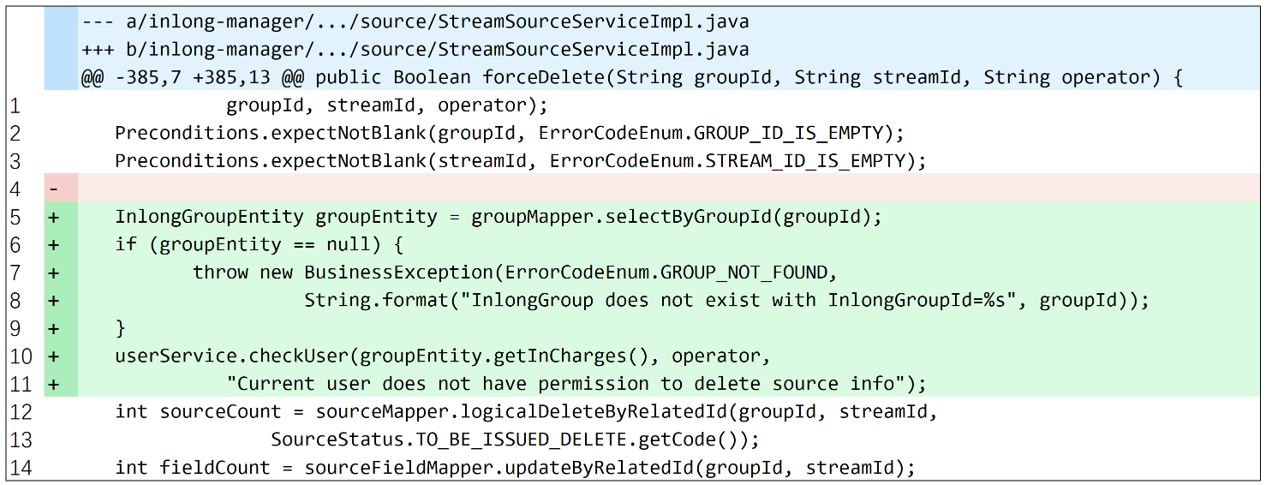}
  \caption{Excerpt from an effective patch generated by DeepSeek-V4 Pro for CVE-2023-31066, showing its authorization check for the sibling \texttt{forceDelete} path.}
  \label{fig:figure-5.4-7}
\end{figure}

\begin{figure}[t]
  \centering
  \includegraphics[width=0.94\linewidth]{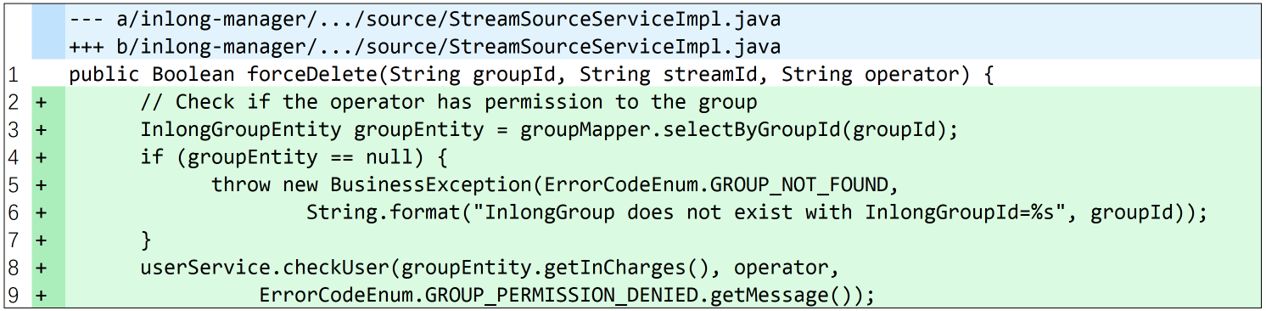}
  \caption{Excerpt illustrating the official patch for the sibling \texttt{forceDelete} path associated with CVE-2023-31066, released on June 22, 2026~\cite{inlong31066followup}.}
  \label{fig:figure-5.4-8}
\end{figure}

Applying the \emph{Coverage} dimension of the BAC repair-strategy framework to the LLM-generated patches revealed that many of them also protected the sibling \texttt{forceDelete} path, as illustrated by the effective patch generated by DeepSeek-V4 Pro in Figure~\ref{fig:figure-5.4-7}.
Both \texttt{delete} and \texttt{forceDelete} perform deletion-related state changes on \texttt{StreamSource} records, but \texttt{delete} accepts a source identifier and targets one source, whereas \texttt{forceDelete} accepts a \texttt{groupId}/\texttt{streamId} pair and affects all sources associated with the specified group and stream.
A manual source audit followed by an extended PoC confirmed that \texttt{forceDelete} lacked the same group-level authorization check as the \texttt{delete} path.
A separate exploit against the latest available InLong release at the time of our study, version 1.6.0, further confirmed that the path remained vulnerable.
We reported the finding to the Apache Software Foundation and the InLong maintainers.
The finding was subsequently assigned a new CVE, and the maintainers released a patch (commit \texttt{e92cb3c}) for \texttt{forceDelete} on June 22, 2026, as illustrated in Figure~\ref{fig:figure-5.4-8} ~\cite{inlong31066followup}.

In effect, while repairing a BAC vulnerability disclosed in 2023, some LLM-generated patches incidentally protected a previously unknown sibling path with the same root cause, without any explicit instruction to inspect that path.
The BAC vulnerability in \texttt{forceDelete} had remained undiscovered for more than three years across multiple InLong releases, despite the earlier identification and remediation of closely related vulnerable paths.
This case shows that some LLMs can explore vulnerable code regions and related execution paths with sufficient depth and breadth to uncover less conspicuous vulnerable paths that may escape maintainers' attention, thereby exhibiting strong performance in the \emph{Coverage} dimension and achieving broader vulnerability coverage than the original official repair in some cases.

For the other three official patches with defects, the official patch for CVE-2022-4684 places the repair at an inaccurate Position and therefore performs only a superficial repair that fails to block the PoC exploit.
The official patch for CVE-2022-4802 has two independent defects.
First, it invokes a security-helper function that is itself buggy in the target source version, causing the patch to fail the functional test.
Second, its Coverage is incomplete, leaving part of the vulnerable behavior unprotected and causing the PoC test to fail.
The official patch for CVE-2022-4803 likewise has incomplete Coverage and therefore fails to block the PoC exploit.
In contrast, for each of these vulnerabilities, there are effective LLM-generated patches that passed both the PoC and functional tests.

\begin{tcolorbox}
\textbf{Answer to RQ4:} Although LLM-generated patches remain limited in their ability to repair BAC vulnerabilities in web applications, some repairs do outperform their official counterparts. In some cases, LLMs demonstrate greater precision and thoroughness than the corresponding maintainer-developed repairs through a more accurate understanding and use of project-native security facilities and a more comprehensive identification and repair of sibling paths that share the same root cause.
\end{tcolorbox}

\FloatBarrier

\section{Discussion}

\subsection{Reason for Choosing Base LLMs}

Given the remarkable ability of agentic systems to solve complex security tasks~\cite{wang2026cybergym,ji2026firmagent,li2026detecting,ullah2026cve,zhang2026fixing}, our work, focusing on the base LLMs, is driven by following reason.
Specifically, as the LLM agents become more capable, their architectures also become too complex to be explored and analyzed.
It remains unclear how large the performance gap between base LLMs and LLM agents is on a specific downstream task, which agent components contribute most to this gap, and which limitations of the underlying models these components alleviate.
\textit{Our evaluation of eight base LLMs and two widely used commercial LLM agents on \textsc{BAC-Bench} helps demystify the role of agentic support.}
With just two lightweight scaffolds, \emph{Apply-Feedback} and \emph{Navigate-and-Locate}, Opus 4.8 improves its repair rate from 30.43\% in the single-turn setting to 75.36\%, only 3.63 percentage points below the 78.99\% achieved by Claude Code using the same underlying model.
Across all eight base LLMs, these scaffolds increase the average repair rate from 37.68\% in the single-turn setting to 66.03\%.
These results indicate that the key to BAC repair lies in mitigating the model's hallucination of diff metadata during patch generation and granting the model repository-level access to the codebase so that it can locate the vulnerability, confirm the root cause, and construct the repair. Together, these two forms of support enable base LLMs to approach the repair capabilities of LLM agents.
Our experimental setup thus enables this finding, providing empirical guidance and insight for designing minimal BAC vulnerability repair harnesses with improved effectiveness and efficiency in future work.

\subsection{About an Ideal Patch}

Although our BAC repair-strategy framework helps pinpoint the dimensions along which a patch fails, it has not fully solved everything.
Basically, compared with the relatively standardized repair patterns for memory-safety vulnerabilities, BAC admits a wider range of repair strategies. 
For a BAC vulnerability caused by a missing access-control check, viable repairs may range from adding a filtering predicate to an SQL query at the data layer, through implementing a dedicated authorization function from scratch at the business layer, to reusing an existing authorization helper at the control layer.
This diversity raises the question of what constitutes a better BAC patch, which motivated us to develop a four-dimensional framework for identifying desirable solutions within the broad BAC repair space.
Choosing among these strategies is non-trivial because repair placement can affect both the vulnerability units and the attack paths covered.
A single BAC may manifest as several vulnerability units sharing the same root cause, and each reachable through multiple attack paths.
For example, an access-control check placed in operation-specific business logic at the business layer can fix only one unit yet block every attack path reaching it, whereas a repair at an endpoint at the presentation layer may protect several units, but only along paths through that endpoint (such as Usememos's CVE-2022-4684).
These differences in repair location and protection scope are captured by the framework's \emph{Position} and \emph{Coverage} dimensions, while \emph{Mechanism} and \emph{Content} assess the security facilities used and the correctness of the implemented authorization policy, respectively.
An Ideal patch therefore requires a carefully chosen repair location that provides robust protection with minimal code changes, reuses existing security facilities wherever possible.

\subsection{Future Work}

In our large-scale analysis of LLM-generated patches, we adopted a coarse-grained assessment scheme that assigns binary correctness labels along each dimension of the framework.
On the one hand, our results indeed reveal a valuable finding: \textit{failures along the Content dimension remain the most common, suggesting that the hardest part of BAC repair is reconstructing the intended authorization policy expressed in the application's business logic.} 
This finding guides our future work on BAC repair.
Specifically, we plan to investigate policy-inference techniques that recover the intended access-control policy of the vulnerable scenario from source code, which is often the only material available in a web application for this purpose, by drawing on evidence such as sibling endpoints, data models, and project conventions, so that generated checks implement the intended policy more faithfully. 
On the other hand, we still regard binary correctness labeling as only one practical way to apply the BAC repair-strategy framework to large numbers of patches, reflecting a trade-off between analytical detail and evaluation feasibility.
As a methodology for characterizing and analyzing BAC patches, the framework can provide richer guidance and support deeper analysis when applied to a specific patch in a concrete web application without such simplification.
We therefore plan to further investigate and refine the framework to better analyze BAC patches, evaluate BAC repair tools more objectively and comprehensively, and guide improvements to these tools in the future.

\section{Threats to Validity}

\textbf{External Threats} mainly lie in the scope of the benchmark and the studied systems. \textsc{BAC-Bench} covers 46 BAC vulnerabilities from six applications in four languages, so our findings may not generalize to BAC vulnerabilities in web applications built with other programming languages or access-control architectures. To mitigate this threat, the studied applications span RBAC, ACL, and hybrid access-control models, and the vulnerabilities cover Broken Function-Level Authorization (BFLA) and Broken Object-Level Authorization (BOLA), horizontal and vertical violations, and read- and write-based unauthorized operations. To mitigate the second threat, we conducted a systematic survey of state-of-the-art LLMs and selected the most advanced and representative models at the time of our experiments, spanning both closed-source and open-weight models and including the strongest LLM agents.

\textbf{Internal Threats} mainly concern the fairness of the comparison across the evaluated systems and the reliability of the labeling procedure. All systems are evaluated in the same isolated sandbox, with identical inputs and the same validation pipeline, and the combined-scaffold condition further places the base LLMs on exactly the same input as the agents, so that the reported performance differences reflect the systems' capabilities rather than unequal experimental conditions. Our scaffolds are model-agnostic by design, and the full replication package, including all prompts, configurations, and experimental scripts, is publicly released~\cite{github}. To account for the non-deterministic nature of LLMs, we explicitly document the configuration settings of every evaluated system and repeat each (system, vulnerability) pair three times, reporting all runs and their spread. In addition, most vulnerabilities in \textsc{BAC-Bench} were disclosed around 2022 and may thus appear in the models' pre-training corpora, but the sampled LLM-generated patches differ substantially from the official fixes with almost no exact matches, and one patch even repaired a zero-day vulnerability CVE-2026-63040~\cite{cve202663040}, that no model could have known in advance, suggesting that the measured effectiveness is not reducible to memorization. Finally, the four-dimensional labels are assigned by an LLM-assisted audit, and a blinded audit of a stratified random sample of 108 patches found the labels to agree with the human consensus at 96.3\% ($\kappa = 0.926$), supporting the reliability of the labels used in our analysis.

\textbf{Construct Threats} primarily stem from the operationalization of repair effectiveness. The first threat concerns the test instruments. A patch that blocks the attack may still break legitimate functionality, so the PoC exploit alone cannot establish the effectiveness of a patch. The built-in test suites of web applications may not fill this gap, as they are largely oriented toward general functionality and rarely touch the specific endpoints, operations, and roles involved in a given vulnerability. To mitigate this threat, we develop a functional test program for each vulnerability, and we run the project's own test suite whenever available to complement its limited breadth. Together, the PoC validation program and the functional test program instantiate closure and preservation in Definition~\ref{def:bac-repair}. The second threat concerns the decision mechanism of the oracles, which may misjudge a violation. To mitigate this threat, we use a unique-marker content match for unauthorized reads and database-state inspection complemented by an application-level re-read for unauthorized writes, rather than response-based oracles, which are brittle for reads and misjudge writes at a high rate. The third threat concerns the implementation of the oracle programs, which are drafted by the LLM. To ensure their correctness and fidelity to the oracle design (\S4.2), every program is manually reviewed and validated through actual execution in the runtime sandbox, confirming that each PoC program triggers the disclosed vulnerability on the unpatched application and that each functional test passes on it.

\section{Related Work}

\subsection{Detecting Broken Access Control} 
The central task of the detection of BAC is to identify missing or inconsistent authorization checks, and existing approaches fall broadly into white-box and black-box paradigms.
White-box approaches analyze the application's source code toward this end. Because web applications seldom ship an explicit access-control specification, AutoISES infers security specifications from code automatically and flags their violations~\cite{tan2008autoises}, and Sun et al. infer the intended per-role accesses from the program itself, treating any successful forced browsing as a likely vulnerability~\cite{sun2011static}.
RoleCast locates missing security checks even when the form of the checks is unknown~\cite{rolecast2011}, and MACE extracts authorization context from database operations and verifies that it is enforced consistently, catching both horizontal and vertical privilege escalation~\cite{mace2014}. BolaRay proposes four authorization models common to database-backed applications and detects object-level violations against them~\cite{huang2024bola}.
The same missing-check problem has also been studied beyond traditional web applications, in the Linux kernel through precise permission-check analysis~\cite{zhang2019pex}, in distributed cloud systems~\cite{lu2022detecting}, in Kubernetes applications where excessive RBAC permissions are detected~\cite{gu2025epscan}, and in modern Java web applications through accurate permission-check identification~\cite{shi2025facilitating}.

Black-box approaches drop the source-code requirement and instead decide whether a violation has occurred from externally observable behavior.
AuthScope discovers vulnerable authorizations in online services through differential response analysis~\cite{authscope}. Observing that response-similarity oracles can catch only unauthorized reads, BACScan introduces a feedback-driven oracle that infers cross-operation dependencies and detects unauthorized reads and writes alike~\cite{bacscan2025}.
BACFuzz further observes SQL-layer behavior during fuzzing to expose the otherwise silent violations~\cite{bacfuzz}.
In sum, BAC detection is relatively mature in both paradigms, whereas the automated repair of BAC vulnerabilities has received far less attention.

\subsection{Repairing Broken Access Control} 
Work that directly automates BAC repair remains scarce and concentrated in the template-based paradigm.
FixMeUp, the earliest such effort, mines an access-control template from PHP applications and inserts it where a check is missing~\cite{fixmeup2013}.
The repair guarantees that the patched code implements the same access-control policy as the template, but not that the resulting program is correct, and the authors observe that web applications provide no standard framework for access control and that each application implements its policies in its own way.
LogicPatcher, developed by the same research group as a repair follow-up to the MACE detector, extends detection to repair and uses path profiling to place patch code at a precise location, addressing parameter-tampering and authorization logic vulnerabilities~\cite{monshizadeh2016patching}.
DRacv models RBAC web applications with a fine-grained navigation graph and mounts each patch as a standalone guard file that behaves like a firewall ahead of the vulnerable page rather than modifying the vulnerable code~\cite{xu2025dracv}.
Beyond web applications, ACFix mines common RBAC practices from smart contracts and uses them to guide LLMs toward context-aware repair of access-control vulnerabilities~\cite{acfix2025}.

These works share two limitations.
First, the repair pattern is single.
Nearly all of them reduce repair to inserting a missing check, which hardly covers object-level violations, property-level violations, or other more complex forms.
Second, their applicability and scale are limited.
All three systems are designed for traditional web application architectures, leaving their effectiveness on contemporary web applications unclear. ACFix targets smart contracts, where access control follows standardized idioms such as \texttt{msg.sender}, the \texttt{onlyOwner} modifier, and OpenZeppelin-style RBAC. Web applications offer no such regularity, so the paradigm of mining uniform practices and reapplying them does not transfer directly.
In summary, as a logic vulnerability whose intended policy is rarely documented and whose realization varies across applications, BAC is inherently difficult to repair, which may explain why automated BAC repair has received only limited attention despite its prevalence.

\subsection{LLM-Based Automated Program Repair}

LLM-based automated program repair has advanced rapidly, evolving from neural machine translation models to conversational LLMs and, most recently, to repository-level repair agents.
The first generation translates defective functions into fixed ones end to end, exemplified by SequenceR~\cite{chen2021sequencer} and CURE~\cite{jiang2021cure} for general bugs and by VRepair~\cite{chen2023vrepair} and VulRepair~\cite{fu2022vulrepair} for security vulnerabilities, the latter two marking the vulnerable lines explicitly. The LLM generation shifts to zero-shot and conversational repair.
AlphaRepair performs cloze-style infilling at the defective location~\cite{xia2023apr}, Pearce et al. systematically examine the zero-shot repair of security vulnerabilities~\cite{pearce2023examining}, and ChatRepair iterates patches through multi-turn dialogue~\cite{xia2024chatrepair}.

SWE-bench then raised the task to the repository level, asking a model to edit a codebase so that a described issue is resolved~\cite{swebench}, and spawned a generation of repair agents.
SWE-agent introduces an agent-computer interface that lets the model navigate entire repositories and execute tests~\cite{yang2024sweagent}.
AutoCodeRover combines structured code search over the program's abstract syntax with spectrum-based fault localization~\cite{zhang2024autocoderover}.
Agentless takes the opposite direction and shows that a fixed three-phase pipeline of localization, repair, and patch validation remains competitive~\cite{xia2025agentless}. Security-oriented agents have followed. PatchAgent integrates fault localization, patch generation, and validation into one autonomous agent that mimics human expertise, repairing over 90\% of 178 real-world vulnerabilities~\cite{patchagent}.
APPATCH couples semantics-aware code pruning with adaptive prompting and repairs vulnerabilities, including zero-days, without requiring a triggering input, although it assumes that the vulnerability location is known~\cite{appatch}.
The finalist teams in DARPA's AI Cyber Challenge developed more comprehensive end-to-end systems that automatically discover and repair vulnerabilities in real-world software~\cite{zhang2026aixccsok,atlantis2025,osscrs2026}.

\section{Conclusion}

Broken access control (BAC) remains one of the most damaging classes of web-application vulnerabilities, yet its automated repair is still largely underexplored.
This paper presented a systematic empirical study of the effectiveness of large language models (LLMs) in repairing BAC vulnerabilities in web applications. We constructed \textsc{BAC-Bench}, a benchmark of 46 real-world BAC vulnerabilities drawn from six widely used web applications spanning four programming languages.
Each vulnerability is equipped with a documented description, a textual PoC description, the official patch, a sandboxed runtime environment, a PoC validation program, and a functional test program.
We further propose a four-dimensional repair-strategy framework of Position, Coverage, Mechanism, and Content, which characterizes how BAC patches are constructed and explains why patches rejected by the dynamic tests fail.
Our evaluation shows that base LLMs repair only 37.68\% of the vulnerabilities on average in the single-turn setting, and that two minimal scaffolds, \emph{Apply-Feedback} and \emph{Navigate-and-Locate}, raise this average to 66.03\%, with the highest-performing base LLM, Claude Opus 4.8, reaching 75.36\%, only 3.63 percentage points below the 78.99\% of Claude Code, the best-performing closed-source LLM agent (likewise built on Opus 4.8).
Inferring the intended authorization policy, captured by the Content dimension, remains the dominant failure mode. Notably, some LLM-generated patches surpass the official fixes, and one even incidentally closed a zero-day vulnerability that had lain dormant for three years in Apache InLong.
We responsibly disclosed this vulnerability to the Apache Software Foundation and the InLong maintainers, and it was assigned CVE-2026-63040.
These results delineate the current capability boundary of LLM-based BAC repair and point to an important next step toward practical automated BAC repair, namely, inferring the precise authorization policy intended for each application's specific business scenario. All artifacts are publicly available to facilitate future research.

\section{Data Availability}
We released all the experimental data and source code on the project homepage for replication, future research, and practical use~\cite{github}.

\begin{acks}

We would like to thank the anonymous reviewers for their insightful comments that helped improve the quality of the paper. This work was supported by the National Natural Science Foundation of China (No. 62372437), and Beijing High Innovation Plan (No. 202504841069). The authors gratefully acknowledge the financial support provided by these programs.

\end{acks}

\bibliographystyle{ACM-Reference-Format}
\bibliography{references}


\begin{thebibliography}{85}


\ifx \showCODEN    \undefined \def \showCODEN     #1{\unskip}     \fi
\ifx \showISBNx    \undefined \def \showISBNx     #1{\unskip}     \fi
\ifx \showISBNxiii \undefined \def \showISBNxiii  #1{\unskip}     \fi
\ifx \showISSN     \undefined \def \showISSN      #1{\unskip}     \fi
\ifx \showLCCN     \undefined \def \showLCCN      #1{\unskip}     \fi
\ifx \shownote     \undefined \def \shownote      #1{#1}          \fi
\ifx \showarticletitle \undefined \def \showarticletitle #1{#1}   \fi
\ifx \showURL      \undefined \def \showURL       {\relax}        \fi
\providecommand\bibfield[2]{#2}
\providecommand\bibinfo[2]{#2}
\providecommand\natexlab[1]{#1}
\providecommand\showeprint[2][]{arXiv:#2}
\makeatletter
\@ifundefined{NAT@parse@date}{}{\let\NAT@parse@date@orig\NAT@parse@date}
\@ifundefined{NAT@parse@date}{}{\def\NAT@parse@date#1#2#3#4#5#6@@{\NAT@parse@date@orig#1#2#3#4#5#6@@\def\NAT@tempyear{0000}\def\NAT@tempexlab{{?}}\ifx\NAT@year\NAT@tempyear\ifx\NAT@exlab\NAT@tempexlab\def\NAT@date{[n.\,d.]}\else\edef\NAT@date{[n.\,d.]\NAT@exlab}\fi\fi}}
\makeatother

\bibitem[Ahmed et~al\mbox{.}(2025)]%
        {ahmed2025llmannotation}
\bibfield{author}{\bibinfo{person}{Toufique Ahmed}, \bibinfo{person}{Premkumar Devanbu}, \bibinfo{person}{Christoph Treude}, {and} \bibinfo{person}{Michael Pradel}.} \bibinfo{year}{2025}\natexlab{}.
\newblock \showarticletitle{Can {LLMs} Replace Manual Annotation of Software Engineering Artifacts?}. In \bibinfo{booktitle}{\emph{Proceedings of the 22nd IEEE/ACM International Conference on Mining Software Repositories (MSR)}}.
\newblock
\href{https://doi.org/10.1109/MSR66628.2025.00086}{doi:\nolinkurl{10.1109/MSR66628.2025.00086}}


\bibitem[{Anthropic}(2026)]%
        {glasswing2026}
\bibfield{author}{\bibinfo{person}{{Anthropic}}.} \bibinfo{year}{2026}\natexlab{}.
\newblock \bibinfo{title}{{Project Glasswing: An Initial Update}}.
\newblock \bibinfo{howpublished}{\url{https://www.anthropic.com/research/glasswing-initial-update}}.
\newblock


\bibitem[{Apache InLong Project}(2023)]%
        {inlong31066initial}
\bibfield{author}{\bibinfo{person}{{Apache InLong Project}}.} \bibinfo{year}{2023}\natexlab{}.
\newblock \bibinfo{title}{[INLONG-7774][Manager] Add permission verification for StreamSource}.
\newblock \bibinfo{howpublished}{GitHub commit}.
\newblock
\urldef\tempurl%
\url{https://github.com/apache/inlong/commit/231dbd5aadb35a4c770ffd83c6f963b3dcf1913e}
\showURL{%
Retrieved August 27, 2026 from \tempurl}


\bibitem[{Apache InLong Project}(2026)]%
        {inlong31066followup}
\bibfield{author}{\bibinfo{person}{{Apache InLong Project}}.} \bibinfo{year}{2026}\natexlab{}.
\newblock \bibinfo{title}{[INLONG-12144][Manager] Perform permission verification for the deletion operation of StreamSource}.
\newblock \bibinfo{howpublished}{GitHub commit}.
\newblock
\urldef\tempurl%
\url{https://github.com/apache/inlong/commit/e92cb3c2930025f49ffefc93bf93a4ac5f57d4a9}
\showURL{%
Retrieved August 27, 2026 from \tempurl}


\bibitem[{Apache Software Foundation}(2026)]%
        {cve202663040}
\bibfield{author}{\bibinfo{person}{{Apache Software Foundation}}.} \bibinfo{year}{2026}\natexlab{}.
\newblock \bibinfo{title}{{CVE-2026-63040}: Apache {InLong}: Missing authorization in {StreamSource} {forceDelete}}.
\newblock \bibinfo{howpublished}{Apache security advisory}.
\newblock
\urldef\tempurl%
\url{https://lists.apache.org/thread/sbqrk88cjv3r9rnqfqgn31ox4711offy}
\showURL{%
\tempurl}
\newblock
\shownote{Published 2026-08-20}.


\bibitem[Chen et~al\mbox{.}(2023)]%
        {chen2023vrepair}
\bibfield{author}{\bibinfo{person}{Zimin Chen}, \bibinfo{person}{Steve Kommrusch}, {and} \bibinfo{person}{Martin Monperrus}.} \bibinfo{year}{2023}\natexlab{}.
\newblock \showarticletitle{Neural Transfer Learning for Repairing Security Vulnerabilities in {C} Code}.
\newblock \bibinfo{journal}{\emph{IEEE Transactions on Software Engineering}} \bibinfo{volume}{49}, \bibinfo{number}{1} (\bibinfo{year}{2023}).
\newblock


\bibitem[Chen et~al\mbox{.}(2021)]%
        {chen2021sequencer}
\bibfield{author}{\bibinfo{person}{Zimin Chen}, \bibinfo{person}{Steve Kommrusch}, \bibinfo{person}{Michele Tufano}, \bibinfo{person}{Louis-No{\"e}l Pouchet}, \bibinfo{person}{Denys Poshyvanyk}, {and} \bibinfo{person}{Martin Monperrus}.} \bibinfo{year}{2021}\natexlab{}.
\newblock \showarticletitle{{SequenceR}: Sequence-to-Sequence Learning for End-to-End Program Repair}.
\newblock \bibinfo{journal}{\emph{IEEE Transactions on Software Engineering}} \bibinfo{volume}{48}, \bibinfo{number}{1} (\bibinfo{year}{2021}).
\newblock


\bibitem[Chin et~al\mbox{.}(2026)]%
        {osscrs2026}
\bibfield{author}{\bibinfo{person}{Andrew Chin}, \bibinfo{person}{Dongkwan Kim}, \bibinfo{person}{Yu-Fu Fu}, \bibinfo{person}{Fabian Fleischer}, \bibinfo{person}{Youngjoon Kim}, \bibinfo{person}{HyungSeok Han}, \bibinfo{person}{Cen Zhang}, \bibinfo{person}{Brian~Junekyu Lee}, \bibinfo{person}{Hanqing Zhao}, {and} \bibinfo{person}{Taesoo Kim}.} \bibinfo{year}{2026}\natexlab{}.
\newblock \showarticletitle{OSS-CRS: Liberating AIxCC Cyber Reasoning Systems for Real-World Open-Source Security}.
\newblock \bibinfo{journal}{\emph{arXiv preprint arXiv:2603.08566}} (\bibinfo{year}{2026}).
\newblock


\bibitem[Dalton et~al\mbox{.}(2009)]%
        {dalton2009nemesis}
\bibfield{author}{\bibinfo{person}{Michael Dalton}, \bibinfo{person}{Christos Kozyrakis}, {and} \bibinfo{person}{Nickolai Zeldovich}.} \bibinfo{year}{2009}\natexlab{}.
\newblock \showarticletitle{Nemesis: Preventing authentication \& [and] access control vulnerabilities in web applications}.
\newblock  (\bibinfo{year}{2009}).
\newblock


\bibitem[{Data Protection Commission (Ireland)}(2024)]%
        {metadpc2024}
\bibfield{author}{\bibinfo{person}{{Data Protection Commission (Ireland)}}.} \bibinfo{year}{2024}\natexlab{}.
\newblock \bibinfo{title}{{Irish Data Protection Commission Fines Meta \texteuro{}251 Million Following 2018 Personal Data Breach}}.
\newblock \bibinfo{howpublished}{\url{https://www.dataprotection.ie/en/news-media/press-releases/irish-data-protection-commission-fines-meta-eu251-million}}.
\newblock


\bibitem[Deng et~al\mbox{.}(2024)]%
        {pentestgpt}
\bibfield{author}{\bibinfo{person}{Gelei Deng}, \bibinfo{person}{Yi Liu}, \bibinfo{person}{V{\'\i}ctor Mayoral-Vilches}, \bibinfo{person}{Peng Liu}, \bibinfo{person}{Yuekang Li}, \bibinfo{person}{Yuan Xu}, \bibinfo{person}{Tianwei Zhang}, \bibinfo{person}{Yang Liu}, \bibinfo{person}{Martin Pinzger}, {and} \bibinfo{person}{Stefan Rass}.} \bibinfo{year}{2024}\natexlab{}.
\newblock \showarticletitle{{PentestGPT}: Evaluating and Harnessing Large Language Models for Automated Penetration Testing}. In \bibinfo{booktitle}{\emph{33rd USENIX Security Symposium (USENIX Security)}}.
\newblock


\bibitem[Dharmaadi et~al\mbox{.}(2025)]%
        {bacfuzz}
\bibfield{author}{\bibinfo{person}{I~Putu~Arya Dharmaadi}, \bibinfo{person}{Mohannad Alhanahnah}, \bibinfo{person}{Van-Thuan Pham}, \bibinfo{person}{Fadi Mohsen}, {and} \bibinfo{person}{Fatih Turkmen}.} \bibinfo{year}{2025}\natexlab{}.
\newblock \bibinfo{title}{{BACFuzz}: Exposing the Silence on Broken Access Control Vulnerabilities in Web Applications}.
\newblock \bibinfo{howpublished}{arXiv:2507.15984}.
\newblock


\bibitem[Ferrag et~al\mbox{.}(2025)]%
        {ferrag2025securefalcon}
\bibfield{author}{\bibinfo{person}{Mohamed~Amine Ferrag}, \bibinfo{person}{Ammar Battah}, \bibinfo{person}{Norbert Tihanyi}, \bibinfo{person}{Ridhi Jain}, \bibinfo{person}{Diana Maimu{\c{t}}}, \bibinfo{person}{Fatima Alwahedi}, \bibinfo{person}{Thierry Lestable}, \bibinfo{person}{Narinderjit~Singh Thandi}, \bibinfo{person}{Abdechakour Mechri}, \bibinfo{person}{Merouane Debbah}, {et~al\mbox{.}}} \bibinfo{year}{2025}\natexlab{}.
\newblock \showarticletitle{SecureFalcon: Are we there yet in automated software vulnerability detection with LLMs?}
\newblock \bibinfo{journal}{\emph{IEEE Transactions on Software Engineering}} (\bibinfo{year}{2025}).
\newblock


\bibitem[Fong(2011)]%
        {fong2011rebac}
\bibfield{author}{\bibinfo{person}{Philip W.~L. Fong}.} \bibinfo{year}{2011}\natexlab{}.
\newblock \showarticletitle{Relationship-Based Access Control: Protection Model and Policy Language}. In \bibinfo{booktitle}{\emph{ACM Conf. on Data and Application Security and Privacy (CODASPY)}}. \bibinfo{pages}{191--202}.
\newblock
\href{https://doi.org/10.1145/1943513.1943539}{doi:\nolinkurl{10.1145/1943513.1943539}}


\bibitem[{Forum of Incident Response and Security Teams (FIRST)}(2026)]%
        {first2026}
\bibfield{author}{\bibinfo{person}{{Forum of Incident Response and Security Teams (FIRST)}}.} \bibinfo{year}{2026}\natexlab{}.
\newblock \bibinfo{title}{{Mid-Year Vulnerability Forecast 2026}}.
\newblock \bibinfo{howpublished}{\url{https://www.first.org/newsroom/releases/20260615}}.
\newblock


\bibitem[Fowler(2002)]%
        {fowler2002eaa}
\bibfield{author}{\bibinfo{person}{Martin Fowler}.} \bibinfo{year}{2002}\natexlab{}.
\newblock \bibinfo{booktitle}{\emph{Patterns of Enterprise Application Architecture}}.
\newblock \bibinfo{publisher}{Addison-Wesley}.
\newblock


\bibitem[Fu et~al\mbox{.}(2022)]%
        {fu2022vulrepair}
\bibfield{author}{\bibinfo{person}{Michael Fu}, \bibinfo{person}{Chakkrit Tantithamthavorn}, \bibinfo{person}{Trung Le}, \bibinfo{person}{Van Nguyen}, {and} \bibinfo{person}{Dinh Phung}.} \bibinfo{year}{2022}\natexlab{}.
\newblock \showarticletitle{{VulRepair}: a {T5}-based automated software vulnerability repair}. In \bibinfo{booktitle}{\emph{ACM Joint European Software Engineering Conference and Symposium on the Foundations of Software Engineering (ESEC/FSE)}}.
\newblock


\bibitem[Gamblin(2026)]%
        {gamblin2026cve}
\bibfield{author}{\bibinfo{person}{Jerry Gamblin}.} \bibinfo{year}{2026}\natexlab{}.
\newblock \bibinfo{title}{2025 CVE Data Review}.
\newblock \bibinfo{howpublished}{\url{https://jerrygamblin.com/2026/01/01/2025-cve-data-review/}}.
\newblock
\newblock
\shownote{48,185 CVEs published in 2025, counted from the CVE Program's official CVE List}.


\bibitem[Gu et~al\mbox{.}(2025)]%
        {gu2025epscan}
\bibfield{author}{\bibinfo{person}{Yue Gu}, \bibinfo{person}{Xin Tan}, \bibinfo{person}{Yuan Zhang}, \bibinfo{person}{Siyan Gao}, {and} \bibinfo{person}{Min Yang}.} \bibinfo{year}{2025}\natexlab{}.
\newblock \showarticletitle{{EPScan}: Automated Detection of Excessive {RBAC} Permissions in {Kubernetes} Applications}. In \bibinfo{booktitle}{\emph{IEEE Symposium on Security and Privacy (S\&P)}}.
\newblock


\bibitem[Gyori et~al\mbox{.}(2015)]%
        {gyori2015poldet}
\bibfield{author}{\bibinfo{person}{Alex Gyori}, \bibinfo{person}{August Shi}, \bibinfo{person}{Farah Hariri}, {and} \bibinfo{person}{Darko Marinov}.} \bibinfo{year}{2015}\natexlab{}.
\newblock \showarticletitle{Reliable Testing: Detecting State-Polluting Tests to Prevent Test Dependency}. In \bibinfo{booktitle}{\emph{Int. Symp. on Software Testing and Analysis (ISSTA)}}. \bibinfo{pages}{223--233}.
\newblock


\bibitem[Hossain et~al\mbox{.}(2024)]%
        {hossain2024deep}
\bibfield{author}{\bibinfo{person}{Soneya~Binta Hossain}, \bibinfo{person}{Nan Jiang}, \bibinfo{person}{Qiang Zhou}, \bibinfo{person}{Xiaopeng Li}, \bibinfo{person}{Wen-Hao Chiang}, \bibinfo{person}{Yingjun Lyu}, \bibinfo{person}{Hoan Nguyen}, {and} \bibinfo{person}{Omer Tripp}.} \bibinfo{year}{2024}\natexlab{}.
\newblock \showarticletitle{A deep dive into large language models for automated bug localization and repair}.
\newblock \bibinfo{journal}{\emph{Proceedings of the ACM on Software Engineering}} \bibinfo{volume}{1}, \bibinfo{number}{FSE} (\bibinfo{year}{2024}), \bibinfo{pages}{1471--1493}.
\newblock


\bibitem[Hu et~al\mbox{.}(2014)]%
        {nist_abac}
\bibfield{author}{\bibinfo{person}{Vincent~C. Hu}, \bibinfo{person}{David Ferraiolo}, \bibinfo{person}{Rick Kuhn}, \bibinfo{person}{Adam Schnitzer}, \bibinfo{person}{Kenneth Sandlin}, \bibinfo{person}{Robert Miller}, {and} \bibinfo{person}{Karen Scarfone}.} \bibinfo{year}{2014}\natexlab{}.
\newblock \bibinfo{booktitle}{\emph{Guide to Attribute Based Access Control ({ABAC}) Definition and Considerations}}.
\newblock \bibinfo{type}{NIST Special Publication} 800-162. \bibinfo{institution}{National Institute of Standards and Technology (NIST)}.
\newblock
\href{https://doi.org/10.6028/NIST.SP.800-162}{doi:\nolinkurl{10.6028/NIST.SP.800-162}}


\bibitem[Huang et~al\mbox{.}(2024)]%
        {huang2024bola}
\bibfield{author}{\bibinfo{person}{Yongheng Huang}, \bibinfo{person}{Chenghang Shi}, \bibinfo{person}{Jie Lu}, \bibinfo{person}{Haofeng Li}, \bibinfo{person}{Haining Meng}, {and} \bibinfo{person}{Lian Li}.} \bibinfo{year}{2024}\natexlab{}.
\newblock \showarticletitle{Detecting Broken Object-Level Authorization Vulnerabilities in Database-Backed Applications}. In \bibinfo{booktitle}{\emph{ACM SIGSAC Conference on Computer and Communications Security (CCS)}}.
\newblock


\bibitem[Ji et~al\mbox{.}(2026)]%
        {ji2026firmagent}
\bibfield{author}{\bibinfo{person}{Jiangan Ji}, \bibinfo{person}{Chao Zhang}, \bibinfo{person}{Shuitao Gan}, \bibinfo{person}{Lin Jian}, \bibinfo{person}{Hangtian Liu}, \bibinfo{person}{Tieming Liu}, \bibinfo{person}{Lei Zheng}, {and} \bibinfo{person}{Zhipeng Jia}.} \bibinfo{year}{2026}\natexlab{}.
\newblock \showarticletitle{FirmAgent: Leveraging Fuzzing to Assist LLM Agents with IoT Firmware Vulnerability Discovery.}. In \bibinfo{booktitle}{\emph{NDSS}}.
\newblock


\bibitem[Jiang et~al\mbox{.}(2021)]%
        {jiang2021cure}
\bibfield{author}{\bibinfo{person}{Nan Jiang}, \bibinfo{person}{Thibaud Lutellier}, {and} \bibinfo{person}{Lin Tan}.} \bibinfo{year}{2021}\natexlab{}.
\newblock \showarticletitle{{CURE}: Code-Aware Neural Machine Translation for Automatic Program Repair}. In \bibinfo{booktitle}{\emph{IEEE/ACM International Conference on Software Engineering (ICSE)}}.
\newblock


\bibitem[Jimenez et~al\mbox{.}(2024)]%
        {swebench}
\bibfield{author}{\bibinfo{person}{Carlos~E. Jimenez}, \bibinfo{person}{John Yang}, \bibinfo{person}{Alexander Wettig}, \bibinfo{person}{Shunyu Yao}, \bibinfo{person}{Kexin Pei}, \bibinfo{person}{Ofir Press}, {and} \bibinfo{person}{Karthik Narasimhan}.} \bibinfo{year}{2024}\natexlab{}.
\newblock \showarticletitle{{SWE-bench}: Can Language Models Resolve Real-World {GitHub} Issues?}. In \bibinfo{booktitle}{\emph{International Conference on Learning Representations (ICLR)}}.
\newblock


\bibitem[Kim et~al\mbox{.}(2025)]%
        {atlantis2025}
\bibfield{author}{\bibinfo{person}{Taesoo Kim}, \bibinfo{person}{HyungSeok Han}, \bibinfo{person}{Soyeon Park}, \bibinfo{person}{Dae~R Jeong}, \bibinfo{person}{Dohyeok Kim}, \bibinfo{person}{Dongkwan Kim}, \bibinfo{person}{Eunsoo Kim}, \bibinfo{person}{Jiho Kim}, \bibinfo{person}{Joshua Wang}, \bibinfo{person}{Kangsu Kim}, {et~al\mbox{.}}} \bibinfo{year}{2025}\natexlab{}.
\newblock \showarticletitle{ATLANTIS: AI-driven Threat Localization, Analysis, and Triage Intelligence System}.
\newblock \bibinfo{journal}{\emph{arXiv preprint arXiv:2509.14589}} (\bibinfo{year}{2025}).
\newblock


\bibitem[Le~Goues et~al\mbox{.}(2011)]%
        {legoues2012genprog}
\bibfield{author}{\bibinfo{person}{Claire Le~Goues}, \bibinfo{person}{ThanhVu Nguyen}, \bibinfo{person}{Stephanie Forrest}, {and} \bibinfo{person}{Westley Weimer}.} \bibinfo{year}{2011}\natexlab{}.
\newblock \showarticletitle{Genprog: A generic method for automatic software repair}.
\newblock \bibinfo{journal}{\emph{Ieee transactions on software engineering}} \bibinfo{volume}{38}, \bibinfo{number}{1} (\bibinfo{year}{2011}), \bibinfo{pages}{54--72}.
\newblock


\bibitem[Li et~al\mbox{.}(2026)]%
        {li2026detecting}
\bibfield{author}{\bibinfo{person}{Penghui Li}, \bibinfo{person}{Hong~Yau Chong}, \bibinfo{person}{Yinzhi Cao}, {and} \bibinfo{person}{Junfeng Yang}.} \bibinfo{year}{2026}\natexlab{}.
\newblock \showarticletitle{Detecting Privilege Escalation in Polyglot Microservices via Agentic Program Analysis}. In \bibinfo{booktitle}{\emph{2026 IEEE Symposium on Security and Privacy (SP)}}. IEEE, \bibinfo{pages}{1747--1765}.
\newblock


\bibitem[Liu et~al\mbox{.}(2025)]%
        {bacscan2025}
\bibfield{author}{\bibinfo{person}{Fengyu Liu}, \bibinfo{person}{Yuan Zhang}, \bibinfo{person}{Enhao Li}, \bibinfo{person}{Wei Meng}, \bibinfo{person}{Youkun Shi}, \bibinfo{person}{Qianheng Wang}, \bibinfo{person}{Chenlin Wang}, \bibinfo{person}{Zihan Lin}, {and} \bibinfo{person}{Min Yang}.} \bibinfo{year}{2025}\natexlab{}.
\newblock \showarticletitle{{BACScan}: Automatic Black-Box Detection of Broken-Access-Control Vulnerabilities in Web Applications}. In \bibinfo{booktitle}{\emph{ACM SIGSAC Conference on Computer and Communications Security (CCS)}}.
\newblock


\bibitem[Liu(2026)]%
        {github}
\bibfield{author}{\bibinfo{person}{Mingrui Liu}.} \bibinfo{year}{2026}\natexlab{}.
\newblock \bibinfo{title}{BAC-BENCH}.
\newblock \bibinfo{howpublished}{https://github.com/TianMaXingK/BAC-BENCH}.
\newblock


\bibitem[Liu et~al\mbox{.}(2024)]%
        {liu2024exploring}
\bibfield{author}{\bibinfo{person}{Peiyu Liu}, \bibinfo{person}{Junming Liu}, \bibinfo{person}{Lirong Fu}, \bibinfo{person}{Kangjie Lu}, \bibinfo{person}{Yifan Xia}, \bibinfo{person}{Xuhong Zhang}, \bibinfo{person}{Wenzhi Chen}, \bibinfo{person}{Haiqin Weng}, \bibinfo{person}{Shouling Ji}, {and} \bibinfo{person}{Wenhai Wang}.} \bibinfo{year}{2024}\natexlab{}.
\newblock \showarticletitle{Exploring $\{$ChatGPT's$\}$ capabilities on vulnerability management}. In \bibinfo{booktitle}{\emph{33rd USENIX Security Symposium (USENIX Security 24)}}. \bibinfo{pages}{811--828}.
\newblock


\bibitem[Lu et~al\mbox{.}(2022)]%
        {lu2022detecting}
\bibfield{author}{\bibinfo{person}{Jie Lu}, \bibinfo{person}{Haofeng Li}, \bibinfo{person}{Chen Liu}, \bibinfo{person}{Lian Li}, {and} \bibinfo{person}{Kun Cheng}.} \bibinfo{year}{2022}\natexlab{}.
\newblock \showarticletitle{Detecting missing-permission-check vulnerabilities in distributed cloud systems}. In \bibinfo{booktitle}{\emph{Proceedings of the 2022 ACM SIGSAC Conference on Computer and Communications Security}}. \bibinfo{pages}{2145--2158}.
\newblock


\bibitem[Luo et~al\mbox{.}(2014)]%
        {luo2014flaky}
\bibfield{author}{\bibinfo{person}{Qingzhou Luo}, \bibinfo{person}{Farah Hariri}, \bibinfo{person}{Lamyaa Eloussi}, {and} \bibinfo{person}{Darko Marinov}.} \bibinfo{year}{2014}\natexlab{}.
\newblock \showarticletitle{An Empirical Analysis of Flaky Tests}. In \bibinfo{booktitle}{\emph{ACM SIGSOFT Int. Symp. on Foundations of Software Engineering (FSE)}}. \bibinfo{pages}{643--653}.
\newblock


\bibitem[{MITRE}(2025)]%
        {mitre2025cwetop25}
\bibfield{author}{\bibinfo{person}{{MITRE}}.} \bibinfo{year}{2025}\natexlab{}.
\newblock \bibinfo{title}{2025 CWE Top 25 Most Dangerous Software Weaknesses}.
\newblock \bibinfo{howpublished}{\url{https://cwe.mitre.org/top25/archive/2025/2025_key_insights.html}}.
\newblock


\bibitem[{Mondoo}(2026)]%
        {mondoo2026}
\bibfield{author}{\bibinfo{person}{{Mondoo}}.} \bibinfo{year}{2026}\natexlab{}.
\newblock \bibinfo{title}{2026 State of Vulnerabilities}.
\newblock \bibinfo{howpublished}{\url{https://mondoo.com/vulnerability-intelligence/state-of-vulnerabilities-2026}}.
\newblock


\bibitem[Monperrus(2018)]%
        {monperrus2018}
\bibfield{author}{\bibinfo{person}{Martin Monperrus}.} \bibinfo{year}{2018}\natexlab{}.
\newblock \showarticletitle{Automatic software repair: A bibliography}.
\newblock \bibinfo{journal}{\emph{ACM Computing Surveys (CSUR)}} \bibinfo{volume}{51}, \bibinfo{number}{1} (\bibinfo{year}{2018}), \bibinfo{pages}{1--24}.
\newblock


\bibitem[Monshizadeh et~al\mbox{.}(2016)]%
        {monshizadeh2016patching}
\bibfield{author}{\bibinfo{person}{Maliheh Monshizadeh}, \bibinfo{person}{Prasad Naldurg}, {and} \bibinfo{person}{VN Venkatakrishnan}.} \bibinfo{year}{2016}\natexlab{}.
\newblock \showarticletitle{Patching logic vulnerabilities for web applications using LogicPatcher}. In \bibinfo{booktitle}{\emph{Proceedings of the Sixth ACM Conference on Data and Application Security and Privacy}}. \bibinfo{pages}{73--84}.
\newblock


\bibitem[Monshizadeh et~al\mbox{.}(2014)]%
        {mace2014}
\bibfield{author}{\bibinfo{person}{Maliheh Monshizadeh}, \bibinfo{person}{Prasad Naldurg}, {and} \bibinfo{person}{V.~N. Venkatakrishnan}.} \bibinfo{year}{2014}\natexlab{}.
\newblock \showarticletitle{{MACE}: Detecting Privilege Escalation Vulnerabilities in Web Applications}. In \bibinfo{booktitle}{\emph{ACM SIGSAC Conference on Computer and Communications Security (CCS)}}.
\newblock


\bibitem[{National Institute of Standards and Technology}(2022)]%
        {cve20222730}
\bibfield{author}{\bibinfo{person}{{National Institute of Standards and Technology}}.} \bibinfo{year}{2022}\natexlab{}.
\newblock \bibinfo{title}{{CVE-2022-2730}: Authorization Bypass Through User-Controlled Key in OpenEMR}.
\newblock \bibinfo{howpublished}{\url{https://nvd.nist.gov/vuln/detail/CVE-2022-2730}}.
\newblock
\newblock
\shownote{National Vulnerability Database, published 9 August 2022}.


\bibitem[Nong et~al\mbox{.}(2025)]%
        {appatch}
\bibfield{author}{\bibinfo{person}{Yu Nong}, \bibinfo{person}{Haoran Yang}, \bibinfo{person}{Long Cheng}, \bibinfo{person}{Hongxin Hu}, {and} \bibinfo{person}{Haipeng Cai}.} \bibinfo{year}{2025}\natexlab{}.
\newblock \showarticletitle{$\{$APPATCH$\}$: Automated adaptive prompting large language models for $\{$Real-World$\}$ software vulnerability patching}. In \bibinfo{booktitle}{\emph{34th USENIX Security Symposium (USENIX Security 25)}}. \bibinfo{pages}{4481--4500}.
\newblock


\bibitem[{OASIS}(2013)]%
        {xacml30}
\bibfield{author}{\bibinfo{person}{{OASIS}}.} \bibinfo{year}{2013}\natexlab{}.
\newblock \bibinfo{title}{e{X}tensible Access Control Markup Language ({XACML}) Version 3.0}.
\newblock \bibinfo{howpublished}{OASIS Standard}.
\newblock
\newblock
\shownote{\url{http://docs.oasis-open.org/xacml/3.0/xacml-3.0-core-spec-os-en.html}}.


\bibitem[{OpenAI}(2024)]%
        {openai2024swebenchverified}
\bibfield{author}{\bibinfo{person}{{OpenAI}}.} \bibinfo{year}{2024}\natexlab{}.
\newblock \bibinfo{title}{Introducing {SWE-bench} Verified}.
\newblock \bibinfo{howpublished}{\url{https://openai.com/index/introducing-swe-bench-verified/}}.
\newblock


\bibitem[{OpenEMR Community}(2023)]%
        {openemr2946forum}
\bibfield{author}{\bibinfo{person}{{OpenEMR Community}}.} \bibinfo{year}{2023}\natexlab{}.
\newblock \bibinfo{title}{On demographic ``PRESCRIPTION NOT AUTHORIZED'' for Physician}.
\newblock \bibinfo{howpublished}{OpenEMR Community Forum}.
\newblock
\urldef\tempurl%
\url{https://community.open-emr.org/t/physician-cant-fill-up-prescriptions/20458}
\showURL{%
Retrieved August 27, 2026 from \tempurl}


\bibitem[{OpenEMR Project}(2022)]%
        {openemr4505initial}
\bibfield{author}{\bibinfo{person}{{OpenEMR Project}}.} \bibinfo{year}{2022}\natexlab{}.
\newblock \bibinfo{title}{{OpenEMR} Commit 235b1910: Initial Repair Examined for {CVE-2022-4505}}.
\newblock
\urldef\tempurl%
\url{https://github.com/openemr/openemr/commit/235b1910ffe5296187667277d4e197a0c3a9ac33}
\showURL{%
Retrieved August 27, 2026 from \tempurl}


\bibitem[{OpenEMR Project}(2023a)]%
        {openemr2946initial}
\bibfield{author}{\bibinfo{person}{{OpenEMR Project}}.} \bibinfo{year}{2023}\natexlab{a}.
\newblock \bibinfo{title}{OpenEMR commit 81832ac}.
\newblock
\urldef\tempurl%
\url{https://github.com/openemr/openemr/commit/81832acc14207e577e76c4175967c99ae7e3d3f4}
\showURL{%
Retrieved August 27, 2026 from \tempurl}


\bibitem[{OpenEMR Project}(2023b)]%
        {openemr2946followup}
\bibfield{author}{\bibinfo{person}{{OpenEMR Project}}.} \bibinfo{year}{2023}\natexlab{b}.
\newblock \bibinfo{title}{OpenEMR commit a8ed9cd}.
\newblock
\urldef\tempurl%
\url{https://github.com/openemr/openemr/commit/a8ed9cde9e1ada229fc4638e1cd44d768527833f}
\showURL{%
Retrieved August 27, 2026 from \tempurl}


\bibitem[{OpenEMR Project}(2024)]%
        {openemr4505followup}
\bibfield{author}{\bibinfo{person}{{OpenEMR Project}}.} \bibinfo{year}{2024}\natexlab{}.
\newblock \bibinfo{title}{{OpenEMR} Commit 210a63b4: Follow-up Repair for {CVE-2022-4505}}.
\newblock
\urldef\tempurl%
\url{https://github.com/openemr/openemr/commit/210a63b4fcf0d25dee6eae9f957a31d23d4dfca5}
\showURL{%
Retrieved August 27, 2026 from \tempurl}


\bibitem[{OWASP Foundation}(2021)]%
        {owasp2021}
\bibfield{author}{\bibinfo{person}{{OWASP Foundation}}.} \bibinfo{year}{2021}\natexlab{}.
\newblock \bibinfo{title}{{A01:2021 -- Broken Access Control, OWASP Top 10:2021}}.
\newblock \bibinfo{howpublished}{\url{https://owasp.org/Top10/2021/A01_2021-Broken_Access_Control/}}.
\newblock


\bibitem[{OWASP Foundation}(2023a)]%
        {owaspapi2023}
\bibfield{author}{\bibinfo{person}{{OWASP Foundation}}.} \bibinfo{year}{2023}\natexlab{a}.
\newblock \bibinfo{title}{{API1:2023 -- Broken Object Level Authorization, OWASP API Security Top 10 2023}}.
\newblock \bibinfo{howpublished}{\url{https://owasp.org/API-Security/editions/2023/en/0xa1-broken-object-level-authorization/}}.
\newblock


\bibitem[{OWASP Foundation}(2023b)]%
        {owaspapi2023bopla}
\bibfield{author}{\bibinfo{person}{{OWASP Foundation}}.} \bibinfo{year}{2023}\natexlab{b}.
\newblock \bibinfo{title}{{API3:2023 -- Broken Object Property Level Authorization}}.
\newblock \bibinfo{howpublished}{\url{https://owasp.org/API-Security/editions/2023/en/0xa3-broken-object-property-level-authorization/}}.
\newblock


\bibitem[{OWASP Foundation}(2025)]%
        {owasp2025}
\bibfield{author}{\bibinfo{person}{{OWASP Foundation}}.} \bibinfo{year}{2025}\natexlab{}.
\newblock \bibinfo{title}{{A01:2025 -- Broken Access Control, OWASP Top 10:2025}}.
\newblock \bibinfo{howpublished}{\url{https://top10.owasp.org/2025/A01_2025-Broken_Access_Control/}}.
\newblock


\bibitem[{\"O}zkan et~al\mbox{.}(2023)]%
        {ozkan2023ddd}
\bibfield{author}{\bibinfo{person}{Ozan {\"O}zkan}, \bibinfo{person}{{\"O}nder Babur}, {and} \bibinfo{person}{Mark van~den Brand}.} \bibinfo{year}{2023}\natexlab{}.
\newblock \showarticletitle{Refactoring with Domain-Driven Design in an Industrial Context}.
\newblock \bibinfo{journal}{\emph{Empirical Software Engineering}}  \bibinfo{volume}{28} (\bibinfo{year}{2023}), \bibinfo{pages}{94}.
\newblock
\href{https://doi.org/10.1007/s10664-023-10310-1}{doi:\nolinkurl{10.1007/s10664-023-10310-1}}


\bibitem[Pearce et~al\mbox{.}(2023)]%
        {pearce2023examining}
\bibfield{author}{\bibinfo{person}{Hammond Pearce}, \bibinfo{person}{Benjamin Tan}, \bibinfo{person}{Baleegh Ahmad}, \bibinfo{person}{Ramesh Karri}, {and} \bibinfo{person}{Brendan Dolan-Gavitt}.} \bibinfo{year}{2023}\natexlab{}.
\newblock \showarticletitle{Examining Zero-Shot Vulnerability Repair with Large Language Models}. In \bibinfo{booktitle}{\emph{IEEE Symposium on Security and Privacy (S\&P)}}.
\newblock


\bibitem[Qi et~al\mbox{.}(2015)]%
        {qi2015analysis}
\bibfield{author}{\bibinfo{person}{Zichao Qi}, \bibinfo{person}{Fan Long}, \bibinfo{person}{Sara Achour}, {and} \bibinfo{person}{Martin Rinard}.} \bibinfo{year}{2015}\natexlab{}.
\newblock \showarticletitle{An Analysis of Patch Plausibility and Correctness for Generate-and-Validate Patch Generation Systems}. In \bibinfo{booktitle}{\emph{Int. Symp. on Software Testing and Analysis (ISSTA)}}. \bibinfo{pages}{24--36}.
\newblock


\bibitem[Rein et~al\mbox{.}(2023)]%
        {rein2023gpqa}
\bibfield{author}{\bibinfo{person}{David Rein}, \bibinfo{person}{Betty~Li Hou}, \bibinfo{person}{Asa~Cooper Stickland}, \bibinfo{person}{Jackson Petty}, \bibinfo{person}{Richard~Yuanzhe Pang}, \bibinfo{person}{Julien Dirani}, \bibinfo{person}{Julian Michael}, {and} \bibinfo{person}{Samuel~R. Bowman}.} \bibinfo{year}{2023}\natexlab{}.
\newblock \bibinfo{title}{{GPQA}: A Graduate-Level Google-Proof {Q\&A} Benchmark}.
\newblock \bibinfo{howpublished}{arXiv:2311.12022}.
\newblock


\bibitem[Sandhu et~al\mbox{.}(1996)]%
        {sandhu1996rbac}
\bibfield{author}{\bibinfo{person}{Ravi~S. Sandhu}, \bibinfo{person}{Edward~J. Coyne}, \bibinfo{person}{Hal~L. Feinstein}, {and} \bibinfo{person}{Charles~E. Youman}.} \bibinfo{year}{1996}\natexlab{}.
\newblock \showarticletitle{Role-Based Access Control Models}.
\newblock \bibinfo{journal}{\emph{Computer}} \bibinfo{volume}{29}, \bibinfo{number}{2} (\bibinfo{year}{1996}), \bibinfo{pages}{38--47}.
\newblock
\href{https://doi.org/10.1109/2.485845}{doi:\nolinkurl{10.1109/2.485845}}


\bibitem[Serebryany et~al\mbox{.}(2012)]%
        {serebryany2012asan}
\bibfield{author}{\bibinfo{person}{Konstantin Serebryany}, \bibinfo{person}{Derek Bruening}, \bibinfo{person}{Alexander Potapenko}, {and} \bibinfo{person}{Dmitriy Vyukov}.} \bibinfo{year}{2012}\natexlab{}.
\newblock \showarticletitle{{AddressSanitizer}: A Fast Address Sanity Checker}. In \bibinfo{booktitle}{\emph{USENIX Annual Technical Conference (USENIX ATC)}}. \bibinfo{pages}{309--318}.
\newblock


\bibitem[Shi et~al\mbox{.}(2026)]%
        {shi2025llmjudge}
\bibfield{author}{\bibinfo{person}{Sherry Shi}, \bibinfo{person}{Renyao Wei}, \bibinfo{person}{Michele Tufano}, \bibinfo{person}{Jos{\'e} Cambronero}, \bibinfo{person}{Runxiang Cheng}, \bibinfo{person}{Franjo Ivancic}, {and} \bibinfo{person}{Pat Rondon}.} \bibinfo{year}{2026}\natexlab{}.
\newblock \showarticletitle{Towards a Human-in-the-Loop Framework for Reliable Patch Evaluation Using an LLM-as-a-Judge}. In \bibinfo{booktitle}{\emph{Proceedings of the 1st International Workshop on AI for Software Quality Evaluation-Judgment, Metrics, Benchmarks, and Beyond}}. \bibinfo{pages}{19--28}.
\newblock


\bibitem[Shi et~al\mbox{.}(2025)]%
        {shi2025facilitating}
\bibfield{author}{\bibinfo{person}{Youkun Shi}, \bibinfo{person}{Fengyu Liu}, \bibinfo{person}{Guangliang Yang}, \bibinfo{person}{Yuan Zhang}, \bibinfo{person}{Yinzhi Cao}, \bibinfo{person}{Enhao Li}, \bibinfo{person}{Xin Tan}, \bibinfo{person}{Xiapu Luo}, \bibinfo{person}{Min Yang}, {and} \bibinfo{person}{Siyi Chen}.} \bibinfo{year}{2025}\natexlab{}.
\newblock \showarticletitle{Facilitating Access Control Vulnerability Detection in Modern Java Web Applications With Accurate Permission Check Identification}.
\newblock \bibinfo{journal}{\emph{IEEE Transactions on Information Forensics and Security}} (\bibinfo{year}{2025}).
\newblock


\bibitem[Smith et~al\mbox{.}(2015)]%
        {smith2015overfitting}
\bibfield{author}{\bibinfo{person}{Edward~K. Smith}, \bibinfo{person}{Earl~T. Barr}, \bibinfo{person}{Claire Le~Goues}, {and} \bibinfo{person}{Yuriy Brun}.} \bibinfo{year}{2015}\natexlab{}.
\newblock \showarticletitle{Is the Cure Worse Than the Disease? {Overfitting} in Automated Program Repair}. In \bibinfo{booktitle}{\emph{Joint Meeting on Foundations of Software Engineering (ESEC/FSE)}}. \bibinfo{pages}{532--543}.
\newblock


\bibitem[Son et~al\mbox{.}(2011)]%
        {rolecast2011}
\bibfield{author}{\bibinfo{person}{Sooel Son}, \bibinfo{person}{Kathryn~S McKinley}, {and} \bibinfo{person}{Vitaly Shmatikov}.} \bibinfo{year}{2011}\natexlab{}.
\newblock \showarticletitle{Rolecast: finding missing security checks when you do not know what checks are}. In \bibinfo{booktitle}{\emph{Proceedings of the 2011 ACM international conference on Object oriented programming systems languages and applications}}. \bibinfo{pages}{1069--1084}.
\newblock


\bibitem[Son et~al\mbox{.}(2013)]%
        {fixmeup2013}
\bibfield{author}{\bibinfo{person}{Sooel Son}, \bibinfo{person}{Kathryn~S. McKinley}, {and} \bibinfo{person}{Vitaly Shmatikov}.} \bibinfo{year}{2013}\natexlab{}.
\newblock \showarticletitle{Fix Me Up: Repairing Access-Control Bugs in Web Applications}. In \bibinfo{booktitle}{\emph{Network and Distributed System Security Symposium (NDSS)}}.
\newblock


\bibitem[Spitzner(2003)]%
        {spitzner2003honeytokens}
\bibfield{author}{\bibinfo{person}{Lance Spitzner}.} \bibinfo{year}{2003}\natexlab{}.
\newblock \showarticletitle{Honeypots: Catching the Insider Threat}. In \bibinfo{booktitle}{\emph{Annual Computer Security Applications Conference (ACSAC)}}. \bibinfo{pages}{170--179}.
\newblock


\bibitem[Sun et~al\mbox{.}(2011)]%
        {sun2011static}
\bibfield{author}{\bibinfo{person}{Fangqi Sun}, \bibinfo{person}{Liang Xu}, {and} \bibinfo{person}{Zhendong Su}.} \bibinfo{year}{2011}\natexlab{}.
\newblock \showarticletitle{Static detection of access control vulnerabilities in web applications}. In \bibinfo{booktitle}{\emph{20th USENIX Security Symposium (USENIX Security 11)}}.
\newblock


\bibitem[Tan et~al\mbox{.}(2008)]%
        {tan2008autoises}
\bibfield{author}{\bibinfo{person}{Lin Tan}, \bibinfo{person}{Xiaolan Zhang}, \bibinfo{person}{Xiao Ma}, \bibinfo{person}{Weiwei Xiong}, {and} \bibinfo{person}{Yuanyuan Zhou}.} \bibinfo{year}{2008}\natexlab{}.
\newblock \showarticletitle{{AutoISES}: Automatically Inferring Security Specifications and Detecting Violations}. In \bibinfo{booktitle}{\emph{USENIX Security Symposium}}.
\newblock


\bibitem[Ullah et~al\mbox{.}(2026)]%
        {ullah2026cve}
\bibfield{author}{\bibinfo{person}{Saad Ullah}, \bibinfo{person}{Praneeth Balasubramanian}, \bibinfo{person}{Wenbo Guo}, \bibinfo{person}{Amanda Burnett}, \bibinfo{person}{Hammond Pearce}, \bibinfo{person}{Christopher Kruegel}, \bibinfo{person}{Giovanni Vigna}, {and} \bibinfo{person}{Gianluca Stringhini}.} \bibinfo{year}{2026}\natexlab{}.
\newblock \showarticletitle{CVE-Genie: An LLM-Based Multi-Agent Framework for Reproducing CVEs}.
\newblock  (\bibinfo{year}{2026}).
\newblock


\bibitem[Wan et~al\mbox{.}(2026)]%
        {wan2026commitsuite}
\bibfield{author}{\bibinfo{person}{Zirui Wan}, \bibinfo{person}{Zhaonan Wu}, \bibinfo{person}{Xinyi Hou}, {et~al\mbox{.}}} \bibinfo{year}{2026}\natexlab{}.
\newblock \bibinfo{title}{{CommitSuite}: A Comprehensive Benchmark for Commit Classification and Message Generation}.
\newblock
\showeprint[arxiv]{2605.02256}~[cs.SE]


\bibitem[Wang et~al\mbox{.}(2026)]%
        {wang2026cybergym}
\bibfield{author}{\bibinfo{person}{Zhun Wang}, \bibinfo{person}{Tianneng Shi}, \bibinfo{person}{Jingxuan He}, \bibinfo{person}{Matthew Cai}, \bibinfo{person}{Jialin Zhang}, {and} \bibinfo{person}{Dawn Song}.} \bibinfo{year}{2026}\natexlab{}.
\newblock \showarticletitle{CyberGym: Evaluating {AI} Agents' Real-World Cybersecurity Capabilities at Scale}. In \bibinfo{booktitle}{\emph{The Fourteenth International Conference on Learning Representations}}.
\newblock
\urldef\tempurl%
\url{https://openreview.net/forum?id=2YvbLQEdYt}
\showURL{%
\tempurl}


\bibitem[Winters et~al\mbox{.}(2020)]%
        {winters2020swe}
\bibfield{editor}{\bibinfo{person}{Titus Winters}, \bibinfo{person}{Tom Manshreck}, {and} \bibinfo{person}{Hyrum Wright}} (Eds.). \bibinfo{year}{2020}\natexlab{}.
\newblock \bibinfo{booktitle}{\emph{Software Engineering at Google: Lessons Learned from Programming Over Time}}.
\newblock \bibinfo{publisher}{O'Reilly Media}.
\newblock


\bibitem[Xia et~al\mbox{.}(2025)]%
        {xia2025agentless}
\bibfield{author}{\bibinfo{person}{Chunqiu~Steven Xia}, \bibinfo{person}{Yinlin Deng}, \bibinfo{person}{Soren Dunn}, {and} \bibinfo{person}{Lingming Zhang}.} \bibinfo{year}{2025}\natexlab{}.
\newblock \showarticletitle{{Agentless}: Demystifying {LLM}-based Software Engineering Agents}. In \bibinfo{booktitle}{\emph{ACM International Conference on the Foundations of Software Engineering (FSE)}}.
\newblock


\bibitem[Xia et~al\mbox{.}(2023)]%
        {xia2023apr}
\bibfield{author}{\bibinfo{person}{Chunqiu~Steven Xia}, \bibinfo{person}{Yuxiang Wei}, {and} \bibinfo{person}{Lingming Zhang}.} \bibinfo{year}{2023}\natexlab{}.
\newblock \showarticletitle{Automated Program Repair in the Era of Large Pre-Trained Language Models}. In \bibinfo{booktitle}{\emph{IEEE/ACM International Conference on Software Engineering (ICSE)}}.
\newblock


\bibitem[Xia and Zhang(2024)]%
        {xia2024chatrepair}
\bibfield{author}{\bibinfo{person}{Chunqiu~Steven Xia} {and} \bibinfo{person}{Lingming Zhang}.} \bibinfo{year}{2024}\natexlab{}.
\newblock \showarticletitle{Automated Program Repair via Conversation: Fixing 162 out of 337 Bugs for \$0.42 Each using {ChatGPT}}. In \bibinfo{booktitle}{\emph{ACM SIGSOFT International Symposium on Software Testing and Analysis (ISSTA)}}.
\newblock


\bibitem[Xu et~al\mbox{.}(2025)]%
        {xu2025dracv}
\bibfield{author}{\bibinfo{person}{Ke Xu}, \bibinfo{person}{Bing Zhang}, \bibinfo{person}{Jingyue Li}, \bibinfo{person}{Haitao He}, \bibinfo{person}{Rong Ren}, {and} \bibinfo{person}{Jiadong Ren}.} \bibinfo{year}{2025}\natexlab{}.
\newblock \showarticletitle{DRacv: Detecting and auto-repairing vulnerabilities in role-based access control in web application}.
\newblock \bibinfo{journal}{\emph{Journal of Network and Computer Applications}}  \bibinfo{volume}{240} (\bibinfo{year}{2025}), \bibinfo{pages}{104191}.
\newblock


\bibitem[Yang et~al\mbox{.}(2024)]%
        {yang2024sweagent}
\bibfield{author}{\bibinfo{person}{John Yang}, \bibinfo{person}{Carlos~E. Jimenez}, \bibinfo{person}{Alexander Wettig}, \bibinfo{person}{Kilian Lieret}, \bibinfo{person}{Shunyu Yao}, \bibinfo{person}{Karthik Narasimhan}, {and} \bibinfo{person}{Ofir Press}.} \bibinfo{year}{2024}\natexlab{}.
\newblock \showarticletitle{{SWE-agent}: Agent-Computer Interfaces Enable Automated Software Engineering}. In \bibinfo{booktitle}{\emph{Advances in Neural Information Processing Systems (NeurIPS)}}.
\newblock


\bibitem[Yavatkar et~al\mbox{.}(2000)]%
        {rfc2753}
\bibfield{author}{\bibinfo{person}{Raj Yavatkar}, \bibinfo{person}{Dimitrios Pendarakis}, {and} \bibinfo{person}{Roch Guerin}.} \bibinfo{year}{2000}\natexlab{}.
\newblock \bibinfo{title}{A Framework for Policy-based Admission Control}.
\newblock \bibinfo{howpublished}{RFC 2753, IETF}.
\newblock
\newblock
\shownote{\url{https://www.rfc-editor.org/rfc/rfc2753}}.


\bibitem[Yu et~al\mbox{.}(2025)]%
        {patchagent}
\bibfield{author}{\bibinfo{person}{Zheng Yu}, \bibinfo{person}{Ziyi Guo}, \bibinfo{person}{Yuhang Wu}, \bibinfo{person}{Jiahao Yu}, \bibinfo{person}{Meng Xu}, \bibinfo{person}{Dongliang Mu}, \bibinfo{person}{Yan Chen}, {and} \bibinfo{person}{Xinyu Xing}.} \bibinfo{year}{2025}\natexlab{}.
\newblock \showarticletitle{$\{$PATCHAGENT$\}$: A Practical Program Repair Agent Mimicking Human Expertise}. In \bibinfo{booktitle}{\emph{34th USENIX Security Symposium (USENIX Security 25)}}. \bibinfo{pages}{4381--4400}.
\newblock


\bibitem[Yuan et~al\mbox{.}(2024)]%
        {mqttactic2024}
\bibfield{author}{\bibinfo{person}{Bin Yuan}, \bibinfo{person}{Zhanxiang Song}, \bibinfo{person}{Yan Jia}, \bibinfo{person}{Zhenyu Lu}, \bibinfo{person}{Deqing Zou}, \bibinfo{person}{Hai Jin}, {and} \bibinfo{person}{Luyi Xing}.} \bibinfo{year}{2024}\natexlab{}.
\newblock \showarticletitle{MQTTactic: Security analysis and verification for logic flaws in MQTT implementations}. In \bibinfo{booktitle}{\emph{2024 IEEE Symposium on Security and Privacy (SP)}}. IEEE, \bibinfo{pages}{2385--2403}.
\newblock


\bibitem[Zhang et~al\mbox{.}(2026a)]%
        {zhang2026aixccsok}
\bibfield{author}{\bibinfo{person}{Cen Zhang}, \bibinfo{person}{Younggi Park}, \bibinfo{person}{Fabian Fleischer}, \bibinfo{person}{Yu-Fu Fu}, \bibinfo{person}{Jiho Kim}, \bibinfo{person}{Dongkwan Kim}, \bibinfo{person}{Youngjoon Kim}, \bibinfo{person}{Qingxiao Xu}, \bibinfo{person}{Andrew Chin}, \bibinfo{person}{Ze Sheng}, \bibinfo{person}{Hanqing Zhao}, \bibinfo{person}{Brian~J. Lee}, \bibinfo{person}{Joshua Wang}, \bibinfo{person}{Michael Pelican}, \bibinfo{person}{David~J. Musliner}, \bibinfo{person}{Jeff Huang}, \bibinfo{person}{Jon Silliman}, \bibinfo{person}{Mikel Mcdaniel}, \bibinfo{person}{Jefferson Casavant}, \bibinfo{person}{Isaac Goldthwaite}, \bibinfo{person}{Nicholas Vidovich}, \bibinfo{person}{Matthew Lehman}, {and} \bibinfo{person}{Taesoo Kim}.} \bibinfo{year}{2026}\natexlab{a}.
\newblock \bibinfo{title}{{SoK}: {DARPA}'s {AI} Cyber Challenge ({AIxCC}): Competition Design, Architectures, and Lessons Learned}.
\newblock
\showeprint[arxiv]{2602.07666}~[cs.CR]


\bibitem[Zhang et~al\mbox{.}(2025)]%
        {acfix2025}
\bibfield{author}{\bibinfo{person}{Lyuye Zhang}, \bibinfo{person}{Kaixuan Li}, \bibinfo{person}{Kairan Sun}, \bibinfo{person}{Daoyuan Wu}, \bibinfo{person}{Ye Liu}, \bibinfo{person}{Haoye Tian}, {and} \bibinfo{person}{Yang Liu}.} \bibinfo{year}{2025}\natexlab{}.
\newblock \showarticletitle{{ACFix}: Guiding {LLMs} With Mined Common {RBAC} Practices for Context-Aware Repair of Access Control Vulnerabilities in Smart Contracts}.
\newblock \bibinfo{journal}{\emph{IEEE Transactions on Software Engineering}} (\bibinfo{year}{2025}).
\newblock


\bibitem[Zhang et~al\mbox{.}(2023)]%
        {zhang2023critical}
\bibfield{author}{\bibinfo{person}{Quanjun Zhang}, \bibinfo{person}{Tongke Zhang}, \bibinfo{person}{Juan Zhai}, \bibinfo{person}{Chunrong Fang}, \bibinfo{person}{Bowen Yu}, \bibinfo{person}{Weisong Sun}, {and} \bibinfo{person}{Zhenyu Chen}.} \bibinfo{year}{2023}\natexlab{}.
\newblock \showarticletitle{A critical review of large language model on software engineering: An example from chatgpt and automated program repair}.
\newblock \bibinfo{journal}{\emph{arXiv preprint arXiv:2310.08879}} (\bibinfo{year}{2023}).
\newblock


\bibitem[Zhang et~al\mbox{.}(2019)]%
        {zhang2019pex}
\bibfield{author}{\bibinfo{person}{Tong Zhang}, \bibinfo{person}{Wenbo Shen}, {and} \bibinfo{person}{Ahmed~M. Azab}.} \bibinfo{year}{2019}\natexlab{}.
\newblock \showarticletitle{{PeX}: A Permission Check Analysis Framework for {Linux} Kernel}. In \bibinfo{booktitle}{\emph{USENIX Security Symposium}}.
\newblock


\bibitem[Zhang et~al\mbox{.}(2024)]%
        {zhang2024autocoderover}
\bibfield{author}{\bibinfo{person}{Yuntong Zhang}, \bibinfo{person}{Haifeng Ruan}, \bibinfo{person}{Zhiyu Fan}, {and} \bibinfo{person}{Abhik Roychoudhury}.} \bibinfo{year}{2024}\natexlab{}.
\newblock \showarticletitle{{AutoCodeRover}: Autonomous Program Improvement}. In \bibinfo{booktitle}{\emph{ACM SIGSOFT International Symposium on Software Testing and Analysis (ISSTA)}}.
\newblock


\bibitem[Zhang et~al\mbox{.}(2026b)]%
        {zhang2026fixing}
\bibfield{author}{\bibinfo{person}{Yuntong Zhang}, \bibinfo{person}{Jiawei Wang}, \bibinfo{person}{Dominic Berzin}, \bibinfo{person}{Martin Mirchev}, {and} \bibinfo{person}{Abhik Roychoudhury}.} \bibinfo{year}{2026}\natexlab{b}.
\newblock \showarticletitle{Fixing Security Vulnerabilities with Agentic AI in OSS-Fuzz}. In \bibinfo{booktitle}{\emph{Proceedings of the IEEE/ACM 48th International Conference on Software Engineering: Software Engineering in Practice}}. \bibinfo{pages}{374--384}.
\newblock


\bibitem[Zuo et~al\mbox{.}(2017)]%
        {authscope}
\bibfield{author}{\bibinfo{person}{Chaoshun Zuo}, \bibinfo{person}{Qingchuan Zhao}, {and} \bibinfo{person}{Zhiqiang Lin}.} \bibinfo{year}{2017}\natexlab{}.
\newblock \showarticletitle{{AuthScope}: Towards Automatic Discovery of Vulnerable Authorizations in Online Services}. In \bibinfo{booktitle}{\emph{ACM SIGSAC Conf. on Computer and Communications Security (CCS)}}. \bibinfo{pages}{799--813}.
\newblock
\href{https://doi.org/10.1145/3133956.3134089}{doi:\nolinkurl{10.1145/3133956.3134089}}


\end{thebibliography}

\clearpage
\appendix
\section{Examples of Patch-Application Failures}
\label{app:patch-failures}
\begin{figure}
    \centering
    \includegraphics[width=0.6\linewidth]{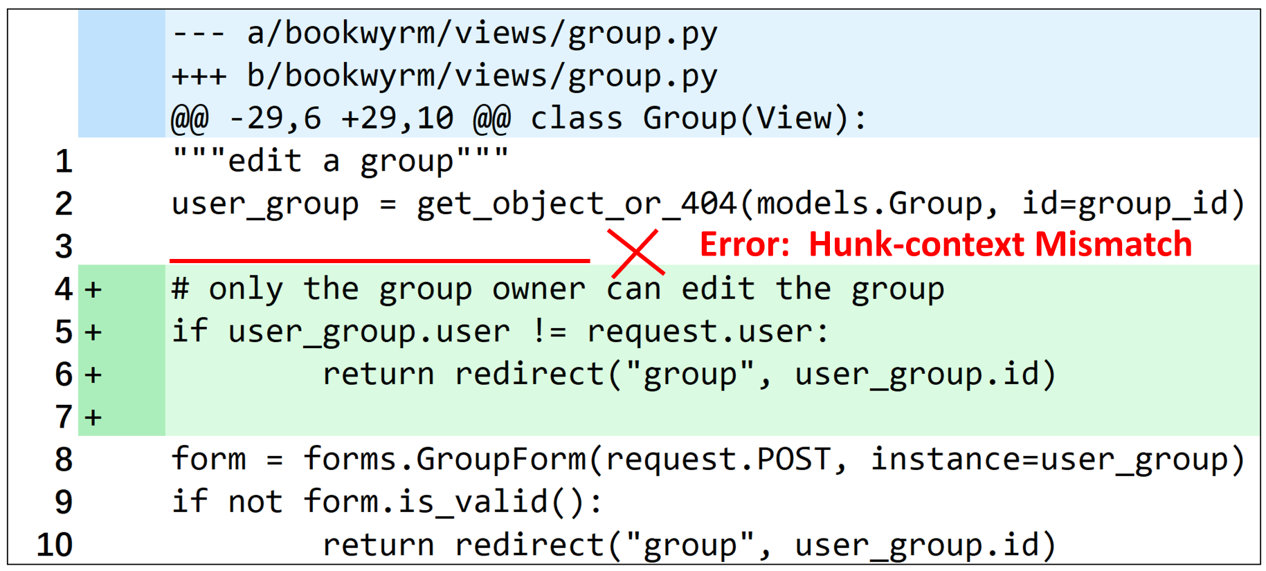}
    \caption{Case C: a patch generated by Kimi K2.7-Code under Base condition for BookWyrm's BW-2 vulnerability. The patch includes an extra blank line that is absent from the target source, resulting in a hunk-context mismatch.}
    \label{fig:figure-5-3}
\end{figure}

\begin{figure}
    \centering
    \includegraphics[width=\linewidth]{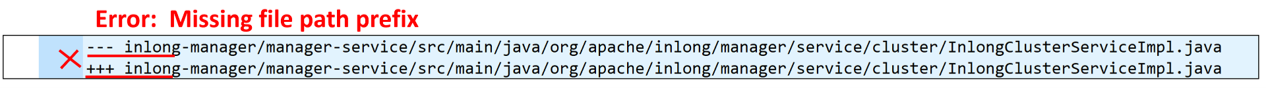}
    \caption{Case D: a patch generated by Qwen3.7-Max under Base condition for Apache InLong's CVE-2023-31454. The patch omits the \texttt{a/} and \texttt{b/} path prefixes, resulting in a patch-path resolution failures under the \texttt{git apply -p1} rule.}
    \label{fig:figure-5-4}
\end{figure}

As illustrated in Figure~\ref{fig:figure-5-3}, Case C exemplifies the hunk-context mismatch pattern.
The Kimi K2.7-Code patch for BookWyrm's BW-2 vulnerability contains an extra blank line near the \(\texttt{user\_group}\) assignment that is absent from the target source version. Although the discrepancy is visually minor, \texttt{git apply} performs strict hunk-context matching and therefore reports \texttt{patch failed}.
Similar failures arise when a model adds or omits comments, decorators, blank lines, or nearby statements while reconstructing the target source context.
Case D corresponding to Figure~\ref{fig:figure-5-4} presents a patch-path resolution failure. The Qwen3.7-Max patch for Apache InLong's CVE-2023-31454 omits the conventional \(\texttt{a/}\) and \(\texttt{b/}\) prefixes from the file headers.
Under the \(\texttt{git apply -p1}\) path-stripping rule used in the experiment, the resulting path does not resolve to the target repository file, producing a \texttt{No such file or directory} error.
The patch identifies the intended file in a human-readable form, but its path representation is incompatible with the application procedure.

\section{Examples of Semantic Failures}
\label{app:semantic-failures}
\begin{figure}
    \centering
    \includegraphics[width=0.6\linewidth]{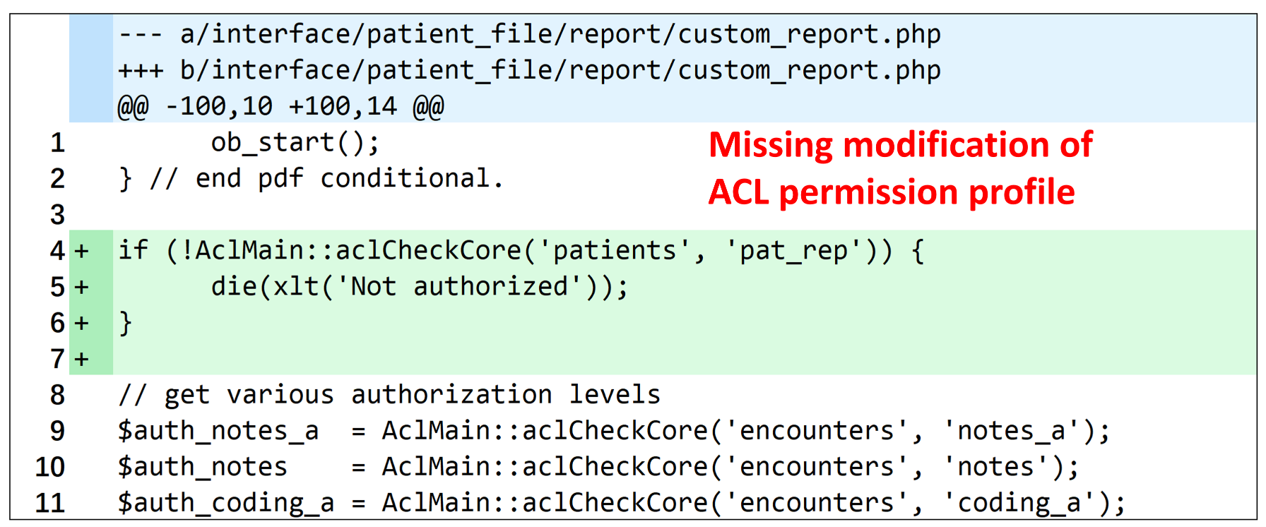}
    \caption{Case F: a patch generated by DeepSeek-V4 Pro under the Apply-Feedback condition for OpenEMR's CVE-2022-1177. The patch adds the \texttt{patients/pat\_rep} entry check but omits the ACL configuration updates that remove this permission from the Front Office and Accounting groups, resulting in an incomplete ACL-policy update.}
    \label{fig:figure-5-6}
\end{figure}

\begin{figure}
    \centering
    \includegraphics[width=0.7\linewidth]{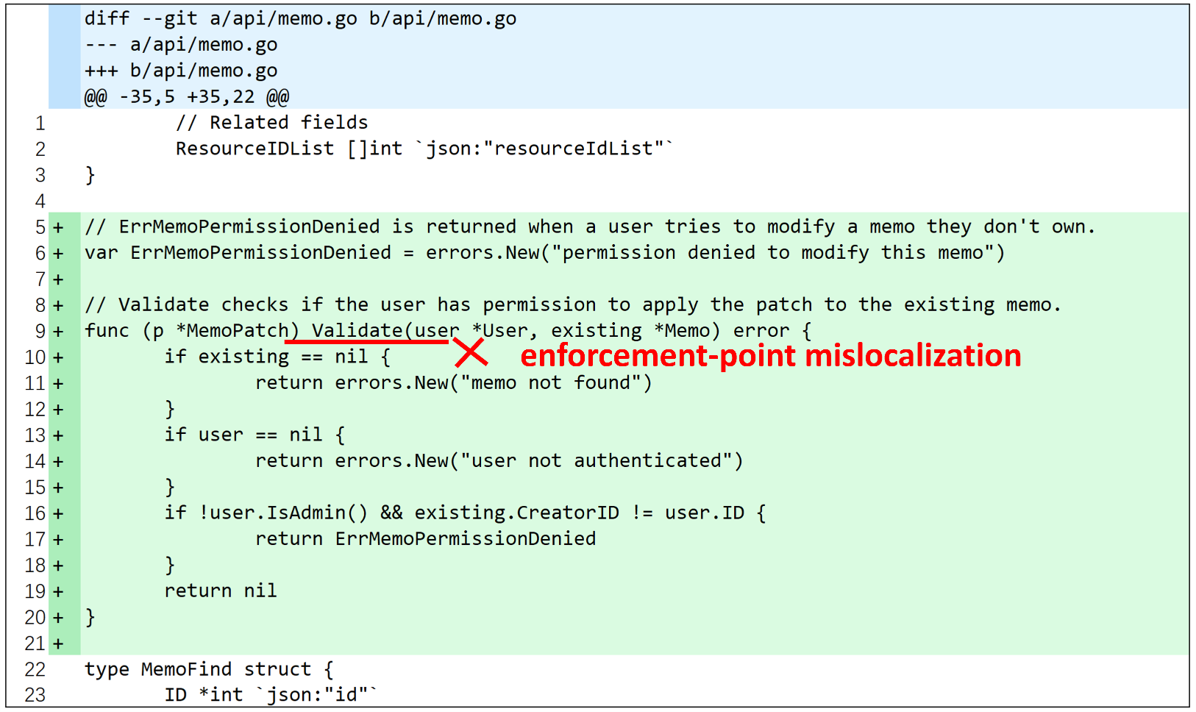}
    \caption{Case G: a patch generated by DeepSeek-V4 Pro under the Apply-Feedback condition for Usememos' CVE-2022-4684. The patch adds \texttt{MemoPatch.Validate}, but the vulnerable \texttt{PATCH /memo/:memoId} handler never invokes it, leaving the authorization check outside the effective request path.}
    \label{fig:figure-5-7}
\end{figure}

In Case F, shown in Figure~\ref{fig:figure-5-6}, the DeepSeek-V4 Pro patch for OpenEMR's CVE-2022-1177 exemplifies incomplete authorization-state reconstruction, specifically an incomplete ACL-policy update.
The patch passes application-stage validation and adds an entry check for \texttt{patients/pat\_rep} in \texttt{custom\_report.php}.
However, the complete repair also requires changing the ACL configuration state so that the existing \texttt{patients/pat\_rep} permission is removed from the Front Office and Accounting ACL groups.
The official patch implements this configuration change through installation or upgrade paths in files such as \texttt{acl\_upgrade.php}, \texttt{Installer.class.php}, and \texttt{AclExtended.php}. Without these ACL configuration updates, the two groups continue to retain the relevant permission.
Although the patch adds the appropriate authorization check, it does not update the underlying ACL state. Under the restricted source context, the LLM had no opportunity to fully explore the relevant parts of the OpenEMR repository and understand its complete ACL authorization mechanism.
As a result, it produced a locally plausible but incomplete repair.

Figure~\ref{fig:figure-5-7} presents Case G, an instance of enforcement-point mislocalization.
The DeepSeek-V4 Pro patch for Usememos' CVE-2022-4684 adds \texttt{MemoPatch.Validate(user, existing)} in \texttt{api/memo.go}.
However, the vulnerable \texttt{PATCH /memo/:memoId} handler in \texttt{server/memo.go} never invokes this method.
The proposed ownership check therefore lies outside the effective execution path of the vulnerable request and cannot protect the operation.

\section{Frozen Vulnerability Unit Counts in \textsc{BAC-Bench}}
\label{app:vuln-list}

  \begin{table}[t]
  \caption{The value of \(n\) (the number of frozen vulnerability units) for each vulnerability in \textsc{BAC-Bench}.}
  \label{tab:vuln-full}
  \small
  \setlength{\tabcolsep}{4pt}
  \renewcommand{\arraystretch}{1.15}
  \begin{tabular}{@{}lp{11.3cm}@{}}
  \toprule
  \textbf{Application} & \textbf{Vulnerability ($n$)}\\ \midrule
  OpenEMR (PHP) & CVE-2022-2730 ($n{=}1$); CVE-2022-2824 ($n{=}4$); CVE-2022-4505 ($n{=}1$); CVE-2022-4567 ($n{=}2$); CVE-2023-2942 ($n{=}1$); CVE-2023-2944 ($n{=}3$); CVE-2023-2945 ($n{=}1$); CVE-2023-2946 ($n{=}3$); (OE-1) ($n{=}1$); CVE-2022-1177 ($n{=}1$)\\
  \midrule
  BookWyrm (Python) & BW-1 ($n{=}1$); BW-2 ($n{=}1$); BW-3 ($n{=}1$); BW-4 ($n{=}1$); BW-5 ($n{=}1$)\\ \midrule
  Apache InLong (Java) & CVE-2023-31066 ($n{=}3$); CVE-2023-31064 ($n{=}1$); CVE-2023-31453 ($n{=}1$); CVE-2023-31454 ($n{=}2$)\\ \midrule
  XWiki (Java) & CVE-2022-23621 ($n{=}1$); CVE-2022-23615 ($n{=}2$); CVE-2021-32620 ($n{=}1$); CVE-2022-41929 ($n{=}2$); CVE-2021-32731 ($n{=}1$)\\ \midrule
  Usememos (Go) & CVE-2022-4684 ($n{=}1$); CVE-2022-4689 ($n{=}1$); CVE-2022-4796 ($n{=}1$); CVE-2022-4797 ($n{=}1$); CVE-2022-4798 ($n{=}1$); CVE-2022-4799 ($n{=}1$); CVE-2022-4801 ($n{=}1$); CVE-2022-4802 ($n{=}3$); CVE-2022-4803 ($n{=}2$); CVE-2022-4804 ($n{=}1$);
  CVE-2022-4805 ($n{=}3$); CVE-2022-4806 ($n{=}1$); CVE-2022-4807 ($n{=}2$); CVE-2022-4808 ($n{=}1$); CVE-2022-4809 ($n{=}1$); CVE-2022-4810 ($n{=}1$); CVE-2022-4811 ($n{=}2$); CVE-2022-4812 ($n{=}1$); CVE-2022-4813 ($n{=}1$); CVE-2022-4814 ($n{=}1$)\\ \midrule
  Nextcloud (PHP) & CVE-2021-32688 ($n{=}2$); CVE-2024-52514 ($n{=}1$)\\
  \bottomrule
  \end{tabular}
  \end{table}

Table~\ref{tab:vuln-full} illustrate the value of \(n\) (the number of frozen vulnerability units) for each vulnerability in \textsc{BAC-Bench}.

\end{document}